\documentclass[12pt,draftclsnofoot,onecolumn]{IEEEtran}
\usepackage[pagewise]{lineno}
\usepackage[type=none,unit=in]{fgruler}
\usepackage[shortlabels]{enumitem}
\usepackage{balance}
\usepackage{amsmath,amssymb,amsfonts}
\usepackage{mathtools}
\usepackage[ruled,vlined]{algorithm2e}
\usepackage{mathabx}
\usepackage{dsfont}
\usepackage{mathrsfs}
\usepackage{eufrak}
\usepackage[mathscr]{eucal}
\usepackage{nomencl}

\ifCLASSOPTIONcompsoc
  \usepackage[caption=false,font=normalsize,labelfont=sf,textfont=sf]{subfig}
\else
  \usepackage[caption=false,font=footnotesize]{subfig}
\fi
\usepackage[dvipsnames]{xcolor}
\usepackage{titletoc} 
\usepackage{amsthm,thmtools} 
\newtheorem{thm}{Theorem}[section]
\newtheorem{dfn}{Definition}[section]
\newtheorem{crl}{Corollary}[section]
\newtheorem{prp}{Preposition}[section]

\newtheorem{exm}{Example}[section]
\newtheorem{lm}{Lemma}[section]
\newtheorem{rmk}{Remark}[section]

\newtheorem{obs}{Observation}[section]

\DeclareMathOperator{\conv}{conv}

\DeclareMathOperator{\supp}{supp}

\DeclareMathOperator{\LIM}{LIM}
\makenomenclature
\begin{document}
\title{Compressible Topological Vector Spaces}
\author{Mohammadreza Robaei,
        and~Robert Akl
\thanks{M. Robaei is with the Department
of Computer Science and Engineering, University of North Texas, Denton,
Tx, 76207 USA (e-mail:mohammadrezarobaei@my.unt.edu.)}
\thanks{R. Akl is with the Department
of Computer Science and Engineering, University of North Texas, Denton,
Tx, 76207 USA (e-mail:robert.akl@unt.edu.)}
}

\maketitle
\vspace{-16mm}
\begin{abstract}
The optimum subspace decomposition of the infinite-dimensional compressible random processes in the locally convex Hausdorff space has been propose and its dimension has been measured. For this purpose, we conduct extensively topological analysis of finite- and infinite-dimensional compressible vector spaces. We prove that if there are a sufficient number of separating points in compressible topological vector space, then Banach Limit coincides the minimal linear functional. Then, optimum orthogonal decomposition of the compressible topological vector space can be formulated as a limit for which minimal linear functional occurs. Accordingly, we show that the purposed topological analysis method decomposes signal space into separable and inseparable subspace. It has been shown that the separable space, also referred to as an optimum subspace, is the subset of the compressible vector space that contains its extreme points. The inseparable subspace is characterized using a novel absorbing null space property through Minkowski functional and Lebesgue measure. We prove that the absorbing null space is a connected subspace of the given locally convex space. We purpose reflexive homomorphism that establishes a relation between signal space and double dual space. It has been shown that the Banach limit can be determined by the compact convergence of the Cauchy nets on the left and right hands sides of a reflexive homomorphism. Finally, we propose to measure the optimum subspace dimension using the Fr\'echet distance metric. To apply the Fr\'echet distance metric to the compressible topological vector space, the K\"othe sequence sampled from a continuous function has been proposed. The sampled K\"othe sequence generates the Borel $\sigma$-algebra proportional to the dimensional growth of the compressible topological vector space. Briefly, the proposed approach measures the sufficient number of separating points of the finite- and infinite-dimensional compressible vector space for the given undersampled operator with respect to Hahn-Banach and Daniell-Kolmogorov theorems. The numerical analysis have been presented for (1) finite-dimensional millimeter-wave channel estimation with and without beam squint, and (2) infinite-dimensional signal with local fluctuation.
\end{abstract}
%
\IEEEpeerreviewmaketitle
\section{Introduction}
The main idea in the compressed sensing is to reconstruct sparse signal $X\in{\mathds{K}^{d}}$ by finding the infimum subspace. For this purpose, compressed sensing provides a mathematical framework to represent $X$ in much lower dimension $X\in{\mathds{K}^{n}}$, $n \ll d$ and $d$ is finite. The continuity condition with respect to vector space operations indicates that the compressibility factor is chosen from a continuous domain $\Omega$. The original idea recommended by Donoho \cite{donoho2006compressed} and others \cite{foucart2010gelfand,candes2006near} tries to find the size of the infimum subspace and the required number of measurements to construct infimum subspace of $X$ as the weighted combination of linear functionals. However, both the size and the number of required measurements are determined loosely, such that in many practical applications, the compressibility factor is an ad-hoc prameter. While Gelfand's $n$-wdith and compressive $m$-width provides a close approximation for $n$ asymptotically, they are far from the optimum subspace. This has been shown in the \cite{robaeiP7CSKLE} by examining variety of finite and infinite-dimensional signals.

The main question we answer in the current work is to find the optimum compact subset of $X$ whose its closure is equal to the closure of convex hull of the compressible topological vector space $X$. This is equal to the original question of Restricted Invertibility Property (RIP) recommended by Bourgain and others \cite{bourgain1987invertibility,rudelson1999random,aubrun2007sampling} in the Banach space and Restricted Isometry Property (RIP) of compressed sensing initiated by Candes and Tao \cite{candes2006near,rauhut2007random}. The aforementioned studies rely on more probabilistic approaches, particularly, Markov and Bernstein inequalities, to determine the optimum number of samples adequate to recover an unknown convex body using random sampling theory. The original groundbreaking works by Donoho \cite{donoho2006compressed} and Foucart \cite{foucart2010gelfand} in $L^{p}$-spaces are also more focused on characterizing the volume of finite-dimensional compressible signals by considering Kolmogorv and Gelfand's n-width. None of the studies above, neither in Banach space nor in $L^{p}$-space, has been considered the topological structure of the compressible vector spaces directly. In this work, we formulate the random sampling of locally convex body $X$ in the Fr\'echet space to measure the optimum subspace $X_{n}$ of $X$ that satisfies both RIP of Bourgain and also RIP of Candes-Tao.
\subsection{Compressed Sensing From Functional Space Point of View}
Compressed sensing aims to solve a problem
\begin{equation}
    \centering
    \overline{y} = \Gamma X+\overline{e}
    \label{bkg:mthd_cs1100}
\end{equation}
and estimate unknown signal $X\in{\mathds{K}^{d}}$ using underdetermined sampling operator $\Gamma\in{\mathds{K}^{M \times d}}$. A sampling operator $\Gamma$ can be written in the form of $\Gamma = \Phi\Psi$, where $\Psi\in{\mathds{K}^{d \times d}}$ is a dictionary matrix, and $\Phi\in{\mathds{K}^{M\times d}}$ is a random sensing matrix.

As explained in \cite{robaeiP7CSKLE}, the randomness of the signal space $X$ (resp. $\Gamma\subset{X^{\ast}}$; $X^{\ast}$ is dual space of $X$) is a function of the random process path that governs compressibility. Stochastically, the underlying randomness can be explained by the path of process $t\rightarrow \left(\psi \circ \left(\omega\right)\right)\left(t\right)$ where $t$ is a longitudinal variable, $\omega$ is a random variable, and $\psi$ is a characteristic function in dual space $X^{\ast}$. Consequently, one can consider the random process path as an unknown distribution law of the compressed sensing problem that has to be estimated. The underlying distribution law that the compressible vector space $X$ takes its values is the unknown of this work. In particular, we aim to study the uniform convergence of the $X^{\ast}$ (resp. $X$) and determine the optimum subspace $X^{\ast}_{n}$ of $X^{\ast}$ (resp. $X_{n}$ of $X$) by introducing topological vector space on $X^{\ast}$ (resp. $X$).

While the random process path determines the randomness of the dual space $X^{\ast}$ (resp. $X$), the randomness of the undersampled operator $\Gamma$ and its construction is not clearly defined, and it is subjective. The general approach in compressed originally recommended by Donoho \cite{donoho2006compressed}, is to sample the dual space uniformly from the dual space $X^{\ast}$. For some applications, such as directional communication, $\Phi$ can be constructed using columns that are randomly selected from the row space of the dictionary $\Psi$. However, for many applications such as image processing, the concepts of being compressible, consequently randomness, may not follow the uniform sampling law. For image processing and continuous spectrum approximation of the linear operator, one also needs to design a random mask that is imposed by the nature of the problem \cite{lustig2007sparse,adcock2016generalized,robaeiP7CSKLE}.  

We mainly focuses on Fourier basis as a universal basis for spectrum approximation problems. Concerning $\Psi$, $X$ has a restricted representation $X_{n}$ with respect to its dual space $X^{\ast}_{n}$ such that
\begin{equation}
    \centering
     \|X-X_{n}\|_{p}<\delta \text{ iff } \|X^{\ast}\left(X\right)-X^{\ast}_{n}\left(X\right)\|_{p}<\epsilon
    \label{bkg:mthd_cs1101}
\end{equation}
for some unknown $n < d$ and $\delta,\epsilon\in{\mathds{R}^{+} \cup \{0\}}$. $L_{p}$ is a norm-space where the $X_{n}$ is an optimal space with respect to $\ell_{p}$-norm error. \eqref{bkg:mthd_cs1101} indicates that if $X$ has no \textit{exact sparse representation}, it can be compressible. That is, $X$ can be approximated by subset $X_{n}\subset{X}$ such that \eqref{bkg:mthd_cs1101} is satisfied. Since \eqref{bkg:mthd_cs1101} suppresses the tails of the spread of the sparse representation of $X$, it leads to \textit{sparse representation error} $\epsilon$. Note that $\epsilon$ is equivalent to the $\ell_{p}$-error of the best $s$-term approximation of compressed sensing. The expression \textit{exact sparse representation} is mathematically fuzzy because the entries of the actual sparse representation may decay exponentially or sub-exponentially, letting the range of subsets of $X^{\ast}_{n}$ (resp. $X_{n}$) converge sequentially to $X$. Here, an important question arises: \textit{What is the $n$ for which $X^{\ast}_{n}$ (resp. $X_{n}$) is an optimum sparse representation of $X^{\ast}$ (resp. $X$)?} Notice that we have used the phrase \textit{optimum} instead of \textit{exact}, which is more consistent with the concept of the uniform convergence in \eqref{bkg:mthd_cs1101}.
\subsection{Compressed Sensing From Kolmogorov Extension Theorem Point of View\label{eq:dfn:wos:1406}}
We can think of compressible random process as the direct sum of i.i.d. product spaces generated with respect to some distribution law \cite{robaeiP7CSKLE}. For this purpose, let $X=\{f\left(t_{0}\right)\in{\mathcal{B}_{0}\left(\mathds{R}\right)},f\left(t_{1}\right)\in{\mathcal{B}_{1}\left(\mathcal{R}\right)},\cdots f\left(t_{d-1}\right)\in{\mathcal{B}_{d-1}\left(\mathds{R}\right)}\}$ be a random vector space takes value with respect to some distribution $\mu_{X}$ from $\mathcal{B}\left(\mathds{R}^d\right)$ where $\mathcal{B}\left(R^{d}\right)$ is a corresponding $\sigma$-algebra Borel set. This problem can formulated as $\left(\Omega,f\left(t\right),\mu_{X}\right) \coloneqq \left(\Omega,\mathcal{B}\left(\mathds{R}^d\right),\mu_{X}\right)$. The corresponding random process path is obtained as $t\rightarrow f\left(\omega,t\right)$. Then, the measurement space $\left(\Omega,\mu_{X}\right)$ for all $f\left(t_{i}\right)$ is determined by evaluating $\mu_{X}$ on all $f\left(t_{i}\right)$, $i\in{\mathds{N}}$, as
\begin{equation}
	\centering
	\mu_{X_{i}}\left(\mathcal{B}\left(\mathds{R}^{d}\right)\right) = \mu_{X}\left(\mathcal{B}_{i}\left(\mathds{R}\right)\right) \text{, } \forall{i}\in{\left[0,d-1\right]}
	\label{bkg:mthd_cs1102}
\end{equation}
According to \eqref{bkg:mthd_cs1102} a measurement on subset $X_{i}$ of $X$ with respect to $\sigma$-algebra is equivalent to the measurement on $X$ with respect to a subset of the $\sigma$-algebra. 
The consistency here is expressed by Kolmogorov Extension Theorem in the following.
\begin{thm}[Kolmogorov Extension Theorem] Let $\mu_{X} = \{\mu_{x_{t_{0}}},\mu_{x_{t_{1}}},\cdots,\mu_{x_{t_{d-1}}}\}$ be a family of consistent distribution measurements for all positive values of $d\in{\mathds{N}}$ and a directed set $t_{i}$ for all $i\in{\left[0,d-1\right]}$. Then, there is random process $X = \{x_{t_{0}},x_{t_{1}},\cdots,x_{t_{d-1}}\}$ consistence with the family of given distribution measurements $\mu_{X}$. 
	\label{bkg:thm_cs1100}
\end{thm}

\begin{dfn}[Projective Probability Measure] \cite[Page 115]{kallenbergfoundations2002} Let $\widehat{T}$ and $\overline{T}$ be classes of finite and countable subsets of $T$. A family of probability measures $\mu_{I}$, $I\in{\widehat{T}}$ or $I\in{\overline{T}}$, is said to be projective if the following regularity is satisfied
\begin{equation}
	\centering
	\mu_{J}\left(\cdot\times \mathcal{B}_{J \backslash I}\right) = \mu_{I} \text{, } \left(I\subset{J}\right)\in{\widehat{T} \text{ or }\overline{T}} 	
	\label{eq:lm_cs1500}
\end{equation}

\end{dfn}

\begin{thm}[Kolmogorov Existence of Random Process] \cite[Theorem 6.16]{kallenbergfoundations2002} Let $\mathcal{B}_{t}$, $t\in{T}$ is a directed set, be a set of Borel set. Consider a family of projective probability measures $\mu_{I}$ on $\mathcal{B}_{I}$, $I\in{\widehat{T}}$. Then, there is a random sequence $X_{t}$ in $\mathcal{B}_{t}$, $t\in{T}$, such that
\begin{equation}
	\centering
	\mathcal{L}\left(X_{I}\right) = \mu_{I} \text{, } \forall{I\in{\widehat{T}}}
	\label{eq:lm_cs1501}
\end{equation}
where $\mathcal{L}\left(X_{I}\right)$ is the distribution of random sequence $X_{I}\subset{X_{t}}$.
	\label{thm:lm_cs1500}
\end{thm}

It is more helpful to rephrase \eqref{eq:lm_cs1501} as
\begin{equation}
	\centering
	\mu\left(\pi_{t_{1}}\in{\mathcal{B}_{1}},\pi_{t_{2}}\in{\mathcal{B}_{2}},\cdots,\pi_{t_{n}}\in{\mathcal{B}_{n}}\right) = \mu_{t_{1},t_{2},\cdots,t_{n}}\left(\mathcal{B}_{1} \times \mathcal{B}_{2} \times \cdots \times \mathcal{B}_{n}\right)
	\label{eq:lm_cs1502}
\end{equation}
which indicates the consistency between marginal spaces (given by the sequence of projective operators $\pi_{t_{i}}\in{\mathcal{B}_{i}}$) and product spaces (given by $\mathcal{B}_{1} \times \mathcal{B}_{2} \times \cdots \times \mathcal{B}_{n}$). The continuous extension of the Kolmogorov existence theorem in \ref{thm:lm_cs1500} from $X_{I}$ to $X_{t}$ is inherited from Daniel Extension Theorem. 
\begin{thm}[Daniell Extension Theorem] \cite[Theorem 27]{shalizi200736} Let $\mathcal{B}_{t}$, $t\in{T}$ is a directed set, be a set of Borel set. Also, let $\mu_{n}$ be a family of projective probability measures on a countably finite measure on product space $\mathcal{B}_{1} \times \mathcal{B}_{2} \times \cdots \times \mathcal{B}_{n}$. Then, there is a finite random sequence $X_{I}$ that takes values from $\mathcal{B}_{I}$, $I\subset{\mathds{N}}$, such that
\begin{equation}
	\centering
	\mathcal{L}\left(X_{I}\right) = \mu_{n} \text{, } \forall{n\in{I}}
	\label{eq:lm_cs1503}
\end{equation} 
and continuous measure $\mu$ such that $\mu_{n}$ is equal to the projection of $\mu$ onto product space $\mathcal{B}_{1} \times \mathcal{B}_{2} \times \cdots \times \mathcal{B}_{n}$.
	\label{thm:lm_cs1501}
\end{thm}

It is more meaningful to rephrase \eqref{eq:lm_cs1503} as
\begin{equation}
	\centering
	\mu_{t_{1},t_{2},\cdots,t_{n}}\left(\mathcal{B}_{1} \times \mathcal{B}_{1} \times \cdots \times \mathcal{B}_{n-1} \times \mathds{R}\right) = \mu_{t_{1},t_{2},\cdots,t_{n-1}}\left(\mathcal{B}_{1} \times \mathcal{B}_{1} \times \cdots \times \mathcal{B}_{n-1}\right)
	\label{thm:lm_cs1502}
\end{equation}

Consequently, we can ask what is the $n$ that satisfies the regularity\footnote{Radon regularity will be discussed in section \ref{m:mth:rdm}.} in \eqref{thm:lm_cs1502}. Developing a systematic approach to defining $\mathcal{B}_{\epsilon}\left(X_{I}\right)$ is one of our main goals in this paper. 

In a closely related problem of Kolmogorov and Daniell extension theorems, in compressed sensing, one aims to estimate unknown $X^{\ast}_{n}$ (resp. $X_{n}$) corresponding to $X^{\ast}$ (resp. $X_{n}$) from an undersampled measurement vector $\overline{y}$. While neither are known the $X$ and $n$, how can we estimate $X_{n}$? This question partially is answered in $\ell_{1}$ convex and greedy algorithms by optimizing the subset selection regarding the expected error. However, it is still unknown what the optimum $n$ is, or more precisely, how to measure optimum $n$ from undersampled measurement $\overline{y}$. In some applications, including millimeter-wave channel estimation, the compressibility factor is assumed to be known priori. This is an unfeasible assumption. Especially, considering the ambiguity in definition of compressible signal, it is reasonable to consider the optimum $n$ as unknown priori. 

\section{Contribution}
\subsection{Functional and Stochastic Analysis}
Considering the random process path $t\rightarrow X\left(t,\omega\right)$, we prove that the Daniell- Kolmogorov continuous extension theorem in stochastical analysis coincides with the Hahn-Banach continuous extension in functional analysis. Consequently, one can treat the compressible random process in stochastical space as a locally convex topological vector space in functional analysis.

\subsection{Separability and metrizability}
What kind of metrizability are we going to define? Equivalently, what property of an object is going to be measured? It is common in compressed sensing and sparse problem solutions to measure the number of non-zero components of some sparse vector. The sparsity level of data has been discussed by a groundbreaking paper by Donoho \cite{donoho2006compressed} and others in the Banach space as the volume of convex bodies \cite{bourgain1987invertibility,rudelson1999random}. The Kolmogorov and Gelfand’s $n$-width are used to provide asymptotically the sense of metrizability \cite{donoho2006compressed,foucart2010gelfand}. In order to define a topology on a given sparse structure, a more generalized definition of sparsity that allows continuous extension is required. First of all, we generalize optimum subspace as a mathematical structure with a closed convex hull. Second, we define a metrizable parameter of an object as the optimum Borel set that maximizes a Radon inner regularity defined on the object. This is equivalent to finding an optimum compact subset of compressible topological vector space. Naturally, one expects that the optimum Borel set also maximizes other metrics inherent to the application. For example, we will show that the optimum Borel set also maximizes the beamforming gain in directional millimeter-wave massive MIMO application.

On the one hand, the metrizability of the given compressible vector space is an important property of decomposing compressible topological vector space in its dual space. This can be obtained by the $\sigma$-finite Borel set as described above. On the other hand, the separability is essential to be able to separate points in compressible topological vector space and, even more important, in the dual space. The condition for this can be obtained if the compressible vector space satisfies the Hahn-Banach separability. Furthermore, if the dual space is compact with respect to Banach-Agao\u{g}lu theorem, then dual space is also separable. We will prove that provided there are sufficient numbers of separating points, with respect to Hahn-Banach continuous extension theorem and Riesz extension theorem in general, one can optimally decompose the compressible topological vector space by defining weak-compact and weak$^{\ast}$-compact topologies on the compressible random process and its dual space, respectively.

\subsection{Absorbing Null Space}
In any attempt to measure the compressibility factor of the compressible topological space, one first needs to clarify the null-space property of the compressed sensing. Although the null-space property has implicitly included in the RIP and mutual-coherency formulation, its explicit formulation is not well-defined in the compressed-sensing literature. A candidate formulation based on Grassmannian has been recommended by Xu et al. \cite{xu2010compressive}. The Grassmannian analysis in \cite{xu2010compressive} is based on the $d$-polytope model, which has been recommended by Donoho \cite{donoho2006high}. However, the recommended model is far from estimating the optimum subspace of sparse signals. In addition, it relies on an ad-hoc threshold with no sharp definition. We propose a systematic approach to extend the definition of null-space property to Hausdorff topological vector space. For this purpose, we introduce an absorbing null space property based on Minkowski functional and connected topological subspaces. 

Knowing that the compressible topological vector space can be decomposed into discrete separable and solid inseparable subspaces and the fact that the minimal linear functional determines the optimum decomposition points, we propose absorbing null space as an extension of null space property in the compressed sensing. The absorbing null space intrinsically can be constructed using the Lebesgue measure. In particular, the general philosophy of the Lebesgue measure that suggests assigning zero to subsets with sufficiently small measurement provides a powerful tool to construct absorbing null space.

Absorbing null space is a mathematically inseparable structure that coincides with the nontrivial subspace of the compressible topological vector space. From a topological point of view, such an object has to be a connected subspace of the compressible topological vector space that behaves like Martingale stochastically. In this thesis, the mean Lebesgue measurement of this object approaches becomes sufficiently small at some point. The corresponding points is determined by the Banach limit of the compressible topological vector space. The connected solid mathematical object allows the compressible topological vector space to satisfy the homeomorphism. As a result, one can tackle the invertibility of the solution space without involving in the difficulty of the nontrivial null space. Readers should note that the solid absorbing null space agrees the RIP condition.

\subsection{Optimum Decomposition of Compressible Topological Vector Spaces}
Most practical signals are compressible up to certain factor. Using the Riesz extension theorem and minimal linear functional (min-max theorem), we prove that there is a point for which there are sufficient numbers of separating points such that the given compressible vector space can be decomposed optimally. We define the compressibility for a minimal linear function for which there is a Banach limit that divides a given compressible topological vector space into sublinear and linear functional spaces. 

In order to find a minimal linear functional space of a compressible topological vector space that satisfies the Banach limit, the given data should be well-ordered with respect to some invariant property. The question we must answer is What kind of invariant topological deformation can be defined on compressible topological vector space? Here, deformation is defined in the sense of shape. Ironically, it is invariant in the sense that it preserves the compressibility factor, i.e., information rate and corresponding Lebesgue measurements. Alternatively, one can ask what the symmetry will be corresponding to the invariant property? The invariant property brings us to rearrangement-invariant spaces. The invariant property of the compressible topological vector space is the volume of the locally convex space. 

Knowing an invariant property of compressible topological vector space, then what sequence space does this invariant property define? In order to define the proper invariant space, we consider two important properties of a system, (1), local properties, and (2) global properties. Here, we utilize the following observation from \cite[Lecture Notes 1, Section 5]{Tao}
    \begin{obs}
    \begin{quote}[$L^{p}$ Sequence Spaces] 
    	\begin{itemize}
    		\item $L^{p}$-space with large $p$: Control in $L^{p}$-space with large $p$ rules out sever local singularities. In other words, $L^{p}$ for large $p$ reveals global properties of a given vector space.
    		\item $L^{q}$-space with small $q$: Control in $L^{q}$-space with small $q$ rules out insufficiently rapid decay at infinity. In other words, it preserves the local properties for sufficiently rapid decay.
    	\end{itemize}
    \end{quote}
    \label{quote:cont:1100}
    \end{obs}
    From observation \ref{quote:cont:1100}, for sufficiently rapid decay $L^{q,p}$-space, where $p$ is sufficiently large and $q$ is sufficiently small, the global and local properties of the compressible topological vector spaces are measurable provided that there is $\sigma$-compact that can separate sufficient number of points in compressible vector space (resp. in dual space). The $L^{q,p}$-space, also known as Lorentz space, is the rearrangement-invariant space. We adopt $L^{1,\infty}$-space to study the uniform convergence behavior of the compressible topological vector space. The idea has been shown in Fig. \ref{fig:fdos:1100}, where the compressible topological vector space $X$ is decomposed into discrete separable subspace $V\subset{X}$ and solid inseparable subspace $U\subset{X}$.

Considering the non-increasing rearrangement-invariant space, one need to figure out how the sufficient number of separating points can be determined. We should emphasize that a sufficient number of points are obtained with respect to the Riesz extension theorem in general, and Hahn-Banach
continuous extension theorem, in particular for the vector space. Then, the following solutions are equivalent
    \begin{itemize}
    	\item The sufficient number of separating points is obtained for the Banach limit for which the Martingale condition is satisfied. 
    	\item The sufficient number of separating points is obtained for vanishing total variation defined using family of semi-norms. For a sufficiently rapidly decaying scenario, the compressible topological vector space satisfies the Schwartz space definition. Then, one can define distribution $\mathcal{D}_{K}\left(\Omega\right)$ for the optimum compact subset $K_{n}$ of the compressible topological vector space. 
    	\item The sufficient number of separating points provides the closed convex hull of the compressible topological vector space. Also, Krein-Milman theorem indicates that the sufficient number of separating points provides the smallest compact subset of the compressible topological vector that contains its extreme points. While the idea here is partially addressed by Hansen \cite[Theorem 8.2]{hansen2008approximation} where author recommend $n$-pseudospectrum to approximate the spectrum of the a continuous operator, the fact that the optimum compact subset of the compressible topological vector space has to contains its extreme points is a novel and important contribution in this work.
    	Considering (1) Riesz extension theorem and minimal linear, (2) solid inseparable absorbing null space, and (3) optimum compact subset of compressible topological vector space, we propose reflexive homeomorphism of compressible topological vector space in Fr\'echet space in Theorem \ref{thm:mth:ta:h:1099}. It has been shown that the Cauchy nets obtained through forward and inverse homeomorphism enable us to compute the Banach limit that separates the compressible topological vector space into separable (optimum subspace) and inseparable (absorbing null space) subsets.  
    \end{itemize}

We prove that the smallest compact subset has to contain its extreme points in order for the Cauchy net defined on the non-increasing rearrangement-invariant of the compressible topological vector space (resp. on its dual space) to be convergent. Then, the given optimum subset is complete.

Finally, the strong sparsity/simplicity condition claimed in \cite{donoho2010precise} is not a strong condition as expected. We prove that the compressibility factor obtained with respect to the compact subset of the compressible topological vector space, which contains its extreme points, so it is a closed convex hull with respect, provides the actual strong condition. Then, the strong condition decomposes compressible topological vector space into separable and inseparable subspaces. Considering this, the optimum dimension is the indicator of the compactness of the compressible topological vector space. And, in this way, the topological analysis of the compressible topological vector space crystallizes the ambiguity in the definition of compressiblility.

\subsection{Fr\'echet Metric}
Considering that (1) bases of topological vector spaces are locally convex at some fundamental neighborhood and (2) broaden null space of the compressible topological vector space, we prove that the compressible vector space can be analyzed in locally convex Hausdorff space, also known as Fr\'echet space. In fact, Fr\'echet space is the most natural space to analyze the uniform convergence behavior of the compressible vector space.
The contributions in the following summarize the achievements in this chapter (not necessarily in the order followed in this chapter).

We leverage the metric space defined on the Fr\'echet space to measure the optimum dimension $n$ of the given compressible topological vector space. $n$ is the index of a point in $L^{1,\infty}$-space that decomposes the compressible topological vector space into discrete separable and solid inseparable subspace. In order to measure the optimum dimension $n$ using Fr\'echet space directly from compressed sensing measurement vector $\overline{y}$ in \eqref{bkg:mthd_cs1100}, we define K\"othe sequence that inherits the internal structure of the compressible topological vector space $X$. As we explain in section \ref{m:mth:fd}, the K\"othe sequence is necessary to measure the optimum dimension $n$ of a given compressible topological vector space.

From measurement theory point of view, the Borel set $\mathcal{B}_{X}$, i.e., $\sigma$-finite, defined on the compressible topological vector space $X$, is the bridge between the sampling operator $\Gamma$ and the metric space $\left(X,\mathcal{B}_{X},\mu_{X}\right)$ such that both $X$ and $\mu_{X}$ take their values from Borel set $\mathcal{B}_{X}$. This fact has been used to define K\"othe sequence for the compressible topological vector space $X$.

It is shown that that the transition point, recommended in \cite{robaeiP7CSKLE}, occurs for the sequentially complete locally convex space. The transition point occurs at a neighborhood of a point in compressible topological vector space for which all Cauchy sequences and Cauchy nets converge to their limit. The compact subset that contains the limits is the optimum subspace of a given compressible topological vector space.

\subsection{Algorithm}
It has been shown that the reflexive homeomorphic relation of compressible topological vector space in Theorem \ref{thm:mth:ta:h:1099}, decays to a class of orthogonal projection greedy algorithms, in particular, Orthogonal Matching Pursuit (OMP) for suboptimum subspace estimation. We also discover the relation between reflexive homeomorphic relation of compressible topological vector space in Theorem \ref{thm:mth:ta:h:1099} and Green's function. We prove that the relation between Green's function and reflexive homeomorphism leads to CS-KLE relation recommended in \cite{robaeiP7CSKLE}.

\subsection{Approximate Infinite-Dimensional Compressible Topological Vector Spaces}
Since the local and global parameters are preserved through a Lorentz $L^{q,p}$-space, then reflexive homeomorphic relation of compressible topological vector space in Theorem \ref{thm:mth:ta:h:1099} can approximate the continuous spectrum of infinite-dimensional vector spaces. Accordingly, one is able to compute the local spectral parameters of a joint-distribution signal without using bilinear transforms, such as Short Time Fourier Transform (STFT), and quadratic Cohen class methods, such as Wigner distribution. For example, the Gabor transform has been conventionally used to estimate the local spectral components of the radar signal. This has already been evaluated in \cite{robaeiP7CSKLE} using CS-KLE algorithm. In section \ref{m:mth:dt}, we will prove that the reflexive homeomorphic relation of compressible topological vector space in Theorem \ref{thm:mth:ta:h:1099} decays to CS-KLE through Green’s function.
\section{Compressible Topological Vector Spaces}
\subsection{Compressible Topological Vector Space}
In the following section, we summarize the definitions and preliminaries for the compressible topological vector space $X\in{\mathds{K}^{d}}$, where $d \leq \infty$.

\begin{dfn}[Dual Space of Topological Vector Space] \cite[Definition 3.1]{rudin1962fourier} The dual space of a topological vector space $X$ is a vector space $X^{\ast}$ with members $\Lambda\in{X^{\ast}}$ that are continuous\footnote{With respect to addition and multiplication on vector space \cite[Section 1.6]{rudin1962fourier}.} linear functionals acting on $X$ as $\Lambda: X\rightarrow Y$. Equivalently, we can write 
	\begin{equation}
			Y \coloneqq X^{\ast}\left(X\right) = \Lambda X
		\label{eq:bck:cs:cv:1098}
	\end{equation}
	\label{dfn:bck:cs:cv:1098} 
\end{dfn}

\begin{dfn}[Compressible Topological Vector Space and Dual Space] A vector space $X$ is called compressible under a linear map $\Lambda:X\rightarrow Y$ where $\Lambda\in{X^{\ast}}$, if the null space of $\Lambda$ is non-trivial but absorbing with respect to Minkowski functional.
\end{dfn}

\begin{dfn}[Support and Null Space Property of Compressible Topological Vector Space]
The support of compressible topological vector space $X$ is the index of the entries of $X^{\ast}$ that satisfies
    \begin{equation}
        \centering
        \supp\left(\Lambda X\right) = \{\Lambda_{i} \vert \forall{\Lambda_{i}\in{X^{\ast}}} \text{, } \forall{x\in{X}} \text{ such that } \left|\Lambda_{i} x\right|\geq \alpha\}
        \label{eq:bck:cs:cv:1099}
    \end{equation}
where $\alpha$ is a threshold that determines the order of compressibility. \textit{Note that the value of $\alpha$ determines how well the null space absorbs the points in $\left|\Lambda x\right|<\alpha$, for all $x\in{X}$.} Then, the null space property with respect to \eqref{eq:bck:cs:cv:1099} is obtained using Minkowski Functional
	\begin{equation}
		\mathcal{N}\left(\Lambda\right) = \lim_{r \rightarrow 0} \Lambda^{-1} \left(r\epsilon \right) = \lim_{r \rightarrow 0} \sup_{\Lambda} \{x\in{X} \vert \Lambda x< r\epsilon \text{, } r\epsilon\in{Y}\}		
		\label{eq:bck:cs:cv:1099b}
	\end{equation}
where $r\epsilon < \alpha$. 
\label{dfn:bck:cs:cv:1099}
\end{dfn}

\begin{dfn}[$\ell_{p}$-error of best $n$-term approximation of compressible topological vector space $X\in{\mathds{K}^{d}}$]
For some $p>0$, the $\ell_{p}$-error of best $n$-term approximation of compressible topological vector space $X\in{\mathds{K}^{d}}$ is the error corresponding to $X_{n}\in{\mathds{K}^{n}}$ such that the entries of $\Lambda X_{n}$ are the set consisting of $n$-largest non-zero entries of $\Lambda X$, or equivalently,
    \begin{equation}
        \centering
        \sigma_{n}\left(\Lambda x\right) = \inf_{X_{n}} \|\Lambda X - \Lambda X_{n}\|_{p} \text{, } \Lambda^{-1}_{n}\left(Y\right) = X_{n}
        \label{eq:bck:cs:cv:1100}
    \end{equation}
for $n\ll d$. $\Lambda: X\rightarrow Y$ is a linear functional on $X$ with a quotient map defined as $X^{\ast}\slash \mathcal{N}\left(\Lambda\right)$. Alternatively, we can write $\Lambda_{n}\in{X^{\ast}}$ such that $\Lambda = \Lambda_{n} \oplus \Lambda^{\perp}_{n}$.
    \label{dfn:bck:cs:cv:1100}
\end{dfn}

Annihilators are the $L^{p}$-space generalization of the orthogonality in $L^{2}$-space. They are useful tools to define subspaces with respect to the intrinsic structure of the compressible topological vector space $X$. In particular, annihilators allow to separate compressible topological vector space $X$ (resp. $X^{\ast}$) to the sublinear subspace $V\subset{X}$ and linear subspace $U\in{X}$ with a functional behavior
\begin{equation}
\centering
f\left(x+y\right) \leq f\left(x\right)+f\left(y\right) \text{, } x,y\in{V}
\label{eq:dfn:wos:1500}
\end{equation}
\begin{equation}
\centering
f\left(x+y\right) = f\left(x\right)+f\left(y\right) \text{, } x,y\in{U}
\label{eq:dfn:wos:1501}
\end{equation}
As we will see later in sections \ref{m:mth:mlf} and \ref{m:bck:fdos}, the decomposition of compressible signal $X$ into sublinear spaces $V$ and $U$ will enable to find optimum subspaces $X_{n}$ and $X^{\ast}_{n}$. In the following brief discussion, we develop operator subspaces for compressible topological vector space $X$ and its separating dual space $X^{\ast}$. For this purpose, we define the following relation between the compressible topological vector space $X$, dual space $X^{\ast}$, and double dual space $\left(X^{\ast}\right)^{\ast}$
\begin{equation}
	\centering
	\langle x,x^{\ast} \rangle = \langle x^{\ast},\phi x\rangle
	\label{eq:dfn:wos:1502}
\end{equation}
where $\phi:X \rightarrow \left(X^{\ast}\right)^{\ast}$ is an isometric isomorphism of $X$ onto a closed subspace of $\left(X^{\ast}\right)^{\ast}$ \cite[Section 4.5]{rudin1991functional}. The notation $\langle x, x^{\ast}\rangle$ evaluates the values of the linear functional $x^{\ast}\in{X^{\ast}}$ at $x\in{X}$.

\begin{dfn}[Annihilator] Let $X$ be a Banach space with linear functional $\Lambda: X \rightarrow Y$ and dual space $X^{\ast}$. Also, let $M\subset{X}$ and $N\subset{X^{\ast}}$, then annihilators are defined as 
\begin{equation}
	\centering
	M^{\perp} = \{\Lambda\in{X^{\ast}} \vert \Lambda x = 0 \text{, }  \forall{x\in{M}}\}\subset{X^{\ast}}
	\label{eq:dfn:wos:1400}
\end{equation}
\begin{equation}
	N_{\perp} = \{x\in{X} \vert \Lambda x = 0 \text{, }  \forall{\Lambda\in{N}}\}
	\label{eq:dfn:wos:1401}
\end{equation}
\label{dfn:wos:1100}
\end{dfn}

\begin{dfn}[Compressible Topological Vector Space] The vector $X\in{\mathds{K}^{d}}$ is called compressible topological vector space if there is $n\ll d$ such that $X$ can be approximated in a $\ell_{p}$-ball of radius $1$, or equivalently
    \begin{equation}
        \centering
        \mathfrak{B}_{\epsilon>0} = \{X_{n}\in{\mathds{K}^{n}} \vert \epsilon^{-1} \| \Lambda X - \Lambda X_{n}\|_{p} \leq 1\}
        \label{eq:bck:cs:cv:1101}
    \end{equation}

\textit{Equivalently, there exist sequentially compact $K\in{\mathfrak{B}}$ of cardinality $n$ such that $X_{n}$ uniformly converges to $X$. As we will see later, for $\|\cdot\|_{1,\infty}$ the topology defined on locally convex Hausdorff space $X$ is the weakest topology that makes dual space $X^{\ast}$ continuous.}
    \label{dfn:bck:cs:cv:1101}
\end{dfn}
Considering orthogonal decomposition in the form of 
\begin{equation}
	\centering
	X=X_{n} \oplus X^{\perp}_{n}	
	\label{eq:dfn:wos:1406}
\end{equation} 
the orthogonal complement of $X_{n}$ is given as $X^{\perp}_{n}$. For dual space, we can derive a similar decomposition with respect to quotient norm. Using annihilator $M^{\perp}$ in Definition \ref{dfn:wos:1100}, the dual spaces of $M$ and $X\slash M$ are defined as $M^{\ast} = X^{\ast}\slash M^{\ast}$ and $\left(X\slash M\right) = M^{\perp}$. Then, from Hahn-Banach Continuous Extension Theorem \ref{thm:bck_lca_1100}, $M^{\ast}$ extends to $X^{\ast}_{M}$ as
\begin{equation}
	\centering
	X^{\ast}_{\sigma} = X^{\ast}_{M} + M^{\perp}
	\label{eq:dfn:wos:1407}
\end{equation}
such that $X^{\ast}_{\sigma}$ is an isometric isomorphism of $M^{\ast}$ onto $X^{\ast}/M^{\perp}$ \textit{with respect to the Hahn-Banach continuous extension theorem}. Inasmuch as $X^{\ast}$ is a vector space, \eqref{eq:dfn:wos:1407} can be rephrased as an internal direct sum
\begin{equation}
	\centering
	X^{\ast}_{\sigma} = X^{\ast}_{M} \oplus M^{\perp}
	\label{eq:dfn:wos:1408}
\end{equation}

Two main questions must be answered here. First, the orthogonal decomposition of compressible sensing is not unique since the null space is non-trivial. In order to address this issue, we purpose to \textit{estimate the whole signal space and then decompose the signal space to optimum subspace $X_{n}$ (resp. $X^{\ast}_{n}$) and $X^{\perp}_{n}$ (resp. $\left(X^{\ast}\right)^{\perp}_{n}$) with respect to optimum subspace dimension measured through Fr\'echet distance metric.} 

Second, the relation between internal direct sum in \ref{eq:dfn:wos:1406} and external direct sums, i.e., product space, can be obtained as an optimum subspace for finite-dimensional signals. The extension of the finite-dimensional optimum orthogonal decomposition to infinite-dimensional space can be explained using Daniell-Kolmogorov theorem of stochastical analysis and the Hahn-Banach extension theorem in functional space. For more clarification, we refer to the theorem in the following.

\begin{thm}[Finite-Dimensional Distribution of Random Process] \cite[Theorem 23]{shalizi200736} Let $X$ and $Y$ be two random processes. Then, $X$ and $Y$ have the same distribution, if and only if, all their finite-dimensional distributions agree.  
	\label{thm:dfn:fdd_rpd:1100}
\end{thm}

Comparing Theorem \ref{thm:dfn:fdd_rpd:1100} with an optimum orthogonal decomposition proposed above, we notice that $X^{\ast}_{n}$ will have a distribution that uniformly converges to $X$, and for certain $n$, we can approximate infinite-dimensional space. A grasp understanding of the underlying concept is important here. For this purpose, let $M^{\ast} = X^{\ast}\slash M^{\perp}$ be an estimated distribution, then the Hahn-Banach extension theorem extends the distribution $M^{\ast}$ to $X^{\ast}_{M}$ such that the isometric isomorphism $X^{\ast}_{\sigma}$ in \eqref{eq:dfn:wos:1408} is satisfied. A more pragmatic realization is obtained if we assume a loose isometric condition. Watchful readers should already notice that the isometric isomorphism obtained through Hahn-Banach continuous extension theorem and Daniell-Kolmogorov extension theorem is equivalent to the Restricted Isometric Property. Accordingly, we made the following observation:

\begin{quote} In the topological analysis of the compressible random process, the Hahn-Banach continuous extension theorem of functional analysis and Daniell-Kolmogorov extension theorem of stochastical analysis concur at the optimum orthogonal decomposition.
\end{quote}\label{m:ctvs}
\subsubsection{Absorbing Null Space Property} \label{m:ansp1}
For a linear functional to be continuous, the null space needs to be closed, as discussed in \cite[Lemma 3.4.2]{kriegl2015advanced} \cite{rudin1991functional}. Otherwise, how can linear functional be continuous and isomorphic? The null space of compressed sensing problems, either finite or infinite-dimensional compressible signal, is an unknown non-trivial space and should be treated as an open coset of the compressible signal $X$. 

A potential approach is to apply the Lebesgue measurement and assign zero to all linear functionals in absorbing null space. Null space property of compressed sensing implies that the null space of \eqref{bkg:mthd_cs1100} cannot be too tight. In fact, this agree with the studies in the literature where they indicate that for given sparse recovery algorithm, successful recovery probability decreases as the ratio $\rho = \frac{n}{M}$ increases. Donoho \cite{donoho2006high} has shown a sharp theoretical threshold, also called transition phase, for $\ell_{1}$ convex recovery such that the probability of sparse solution only exists for a certain range of sparsity ratio $\rho$ and undersampling ration $\delta = n/d$. 

In order to find optimum subspace $n$, one needs to expand null space as much as possible. Since null space is unknown, topologically, it has to be absorbing. The absorbing null space property can be formulated as Minkowski functional for Lebesgue measurement on subsets of $X$ that are sufficiently small. To how extent the null space must be expanded is unsolved problem of the compressed sensing. If we can find subspace $n$, then $\dim \mathcal{N}\left(\Lambda\right) = d-n$. For infinite-dimensional compressible signal, obviously, $\dim \mathcal{N}\left(\Lambda\right) \rightarrow \infty$.
\begin{dfn}[Absorbing Null Space] Let $X = X_{n} \oplus X^{\perp}_{n}$ be a compressible topological vector space decomposed as an optimum restricted orthogonal representation $X_{n}$ and orthogonal complement $X^{\perp}_{n}$. By applying Lebesgue measurement to $X^{\perp}_{n}$, we define the absorbing null space as a Lebesgue measurement
	\begin{equation}
    	\mu_{X^{\perp}_{n}}\left(X\right) = \mu\left(K\bigcap X^{\perp}_{n}\right)
        = \inf_{r}\{r>0 \vert x\in{r\left(K\bigcap X^{\perp}_{n}\right)} \text{, } f\left(x\right) <\epsilon \}
    \label{bkg:mthd_nsp1100}
	\end{equation}
where $K\subset{X}$, $f\left(x\right)$ is a family of semi-norms and $\sum^{d-n}_{i=0}\epsilon= \left(d-n\right)\epsilon=\|X^{\perp}_{n}\|_{1}$, for sufficiently small positive $\epsilon\rightarrow 0$. $K$ can be constructed sequentially growing from $X$ such that ultimately $X_{n}\subseteq{K}$. In order to definition to be valid for $n \ll d$, $\|X_{n}\|_{p}\gg \|X^{\perp}_{n}\|_{p}$ has to be satisfied. Equivalently, one can say $X$ has to decay rapidly enough in $L^{p}$-space to be able to define absorbing null space properly. 
\label{dfn:mthd_nsp1100}
\end{dfn}

Note that a definition of absorbing null space in \eqref{bkg:mthd_nsp1100} includes the Gelfand's $n$-width, and at the same time, it keeps the measurement on the absorbing null space infimum. Although it may not be obvious at first glance, the measurement $\mu_{X^{\perp}_{n}}\left(X\right)$ satisfies inequality 
\begin{equation}
    \centering
    d_{n}\left(X_{n}\vert X\right) \leq \sup_{x\in{K \cap X^{\perp}_{n}}} \|x\|
    \label{bkg:mthd_nsp1099}
\end{equation}
which can be generalized using the absorbing null space property and $\mu_{X^{\perp}_{n}}\left(X\right) \leq d_{n}\left(X_{n}\vert X\right)$ as
\begin{equation}
    \centering
    \mu_{X^{\perp}_{n}}\left(X\right) \leq \sup_{x\in{K \cap X^{\perp}_{n}}} \mu\left(x\right)
    \label{bkg:mthd_nsp1098}
\end{equation}
where $\mu\left(x\right)$ is a local measurement at a neighborhood of $x\in{X}$.

Definition \ref{dfn:mthd_nsp1100} leads to the open null space for infinite-dimensional compressible vector space. If $X$ is a finite-dimensional signal, then absorbing null space property leads to compact optimum subspace $X_{n}$, for $n=d-n_{\perp} = d-\dim r \left(K\bigcap X^{\perp}_{n}\right)$.

Alternatively, $\mu_{X^{\perp}_{n}\left(X\right)}$ can be formulated using the Lebesgue outer measure as
\begin{equation}
    \begin{split}
    \mu_{X^{\perp}_{n}\left(X\right)} & = \inf_{\bigcup^{d}_{i=n+1}\mathcal{B}_{i}\subset U} \sum^{d}_{i=n+1}\left|\mathcal{B}_{i}\right|\\
    & = \inf_{\bigcup^{n^{\perp}}_{i=1}\mathcal{B}_{i}\subset U} \sum^{n^{\perp}}_{i=1}\left|\mathcal{B}_{i}\right|
    \end{split}
    \label{bkg:mthd_nsp1102}
\end{equation}
where $\mathcal{B}_{1}\subset \mathcal{B}_{2}\cdots \subset \mathcal{B}_{n^{\perp}}\subset U\subset X$ are the $\sigma$-algebra and $U$ is an open subset of $X$. The complement of $U$ is denoted as $V = X \backslash U$ is a closed compact subset of $X$ such that $X = V \bigcup U$. Decomposition of $X$ into $V$ and $U$ is equivalent to orthogonal decomposition of compressible topological vector space as $X=X_{n}\bigcup X_{n}^{\perp}$ such that there is a $\sigma$-finite set $V$ corresponding to $X_{n}$ with the Lebesgue inner measure denoted as $\mu_{X_{n}}\left(X\right)=\bigcup^{n}_{i=1}\mu\left(V_{i}\right)$. Similarly, $U$ is a $\sigma$-algebra corresponding to $X_{n}^{\perp}$ with a Lebesgue measurement given by $\mu_{X^{\perp}_{n}}\left(X\right) = \bigcup^{n^{\perp}}_{i=1}\mu\left(U_{i}\right)$.

Although $\mu_{X^{\ast}_{n^{\perp}}}$ will not be calculated directly in this paper, detection of a transition point naturally reveals the absorbing null space for the compressible signal. Two issues occur when we want to extend the formulation to infinite-dimensional scenarios. First, $\|X^{\perp}_{n}\|_{1} = \sum^{n^{\perp}}_{i=1}\left(d-n\right)\epsilon$ is unbounded for $d\rightarrow \infty$. Second, for finite-dimensional signal $\mu_{X_{n^{\perp}}}$ is the union of finitely many $\sigma$-finite subsets of $U_{i}$. If $d \rightarrow \infty$, then $U$ is not $\sigma$-finite, so not Lebesgue measurable. So, the formulation in \eqref{bkg:mthd_nsp1102} collapses. Consequently, absorbing null space in the current formulation cannot be extended to infinite-dimensional problems. However, a potential solution can be obtained using first by assuming that $\epsilon$ in Definition \ref{dfn:mthd_nsp1100} is very small such that it approaches zero faster than the dimension growth of $X$, then $\|X^{\perp}_{n}\|_{1}$ is asymptotically bounded and vanishes at a certain coordinate. A better solution to address the infinite-dimensional compressible random process follows from the fact that the asymptotic behavior of the infinite-dimensional distribution is determined by their finite-dimensional sub-distributions. According to the Daniell-Kolmogorov extension theorem, the finite-dimensional distribution of a random process describes the process. Also, the Hahn-Banach theorem guarantees the continuous functional extension of the given distribution to the infinite dimension. In \cite{robaeiP7CSKLE}, it has been shown that the infinite-dimensional signal can be induced with a $\sigma$-algebra by designing a proper measurement mask. We will show in section \ref{m:mth:fd} that the topology induced by the derived K\"othe sequence accurately reveals the optimum subspace $X_{n}$.
\section{Regularity Measure}
\subsection{Radon Measure}\label{m:mth:rdm}
This section describes the outer and inner regular measures. The inner regular measure, in particular, recommends an approach to measure the interior volume of a given topological vector space. In order to formulate the outer and inner measure, the Borel $\sigma$-algebra will be defined. Finally, we will explain the Lebesgue outer and inner measurements.

\begin{dfn}[Lebesgue Outer Measure and Lebesgue Measurability] \cite[Page 16, and 17]{tao2011introduction} Let $E$ be a bounded subset of $\mathds{R}^{d}$ equipped with elementary set $\mathcal{B}_{1},\mathcal{B}_{2},\cdots$. Also, let $E\subset{\bigcup^{n}_{i=1}\mathcal{B}_{n}}$. Then, the Lebesgue outer measure on the set of volume $\mathcal{B}_{i}$, $i\in{\mathds{N}}$, is defined as
    \begin{enumerate}
    \item Lebesgue measurability: Let $E\subseteq{U}\subset{\mathds{R}^{d}}$, then $E$ is called Lebesgue measurable, if there is $\epsilon\in{\mathds{R}^{+}}\cup \{0\}$ such that $m^{*}\left(U \backslash E\right)\leq \epsilon$.
        \item Lebesgue outer measure: $m^{*}\left(E\right) = \inf_{\substack{\bigcup^{\infty}_{n=1}}\\B_{n}\supset{E}}\sum^{\infty}_{n=1}\left|B_{n}\right|$.
    \end{enumerate}
    \label{dfn:mth_reg_meas_intro_1103}
\end{dfn}

The fundamental problem of the measuring topological vector space is to answer the following questions: 
\begin{quote}
(1) What does it mean to perform measurements on compressible topological vector space $X$? 
(2) what are the measurable of the of compressible topological vector space $X$? 
(3) What is the proper Borel $\sigma$-algebra defined over $X$? 
(4) How can we measure the desired properties using the a Borel $\sigma$-algebra of $X$?
\end{quote}

Constructing Borel set from open subsets is to make sure linear functional from $X$ to $Y$ has inverse image $f^{-1}: Y \rightarrow X$.

\begin{dfn}[Radon Measure] Let $X$ be a $\sigma$-compact metric space with a Borel $\sigma$-algebra $\mathcal{B}\left(X\right)$ defined on $X$. Also, let $K\subset{U}\subset{E}$ such that $K$ and $E$ are compact and open subsets of $X$. The measurement $\mu:\mathcal{B} \rightarrow \mathds{R}^{+}$ on $X$ is called Radon measure if
    \begin{enumerate}[(a)]
        \item Local Finiteness: $\mu\left(K\right) < \infty$, $\forall \ \text{compact} \ K\subset{X}$
        \item Outer Regularity: $\mu^{*}\left(E\right) = \inf_{U\supset{E}}\mu\left(U\right)$ for every $E\subset{\mathcal{B}\left(X\right)}$
        \item Inner Regularity: $\mu_{*}\left(E\right) = \sup_{K\subset{E}}\mu\left(K\right)$ for every $E\subset{\mathcal{B}\left(X\right)}$
    \end{enumerate}
    \label{dfn:mth_reg_meas_intro_1100}
\end{dfn}
 
Applying Radon measure to the Lebesgue inner measure over the collection of sequentially compact open sets on $X$ (also referred to as Borel $\sigma$-algebra), $\mathcal{B}_{i}$ for all $i\in{\mathds{N}}$, such that $\mathcal{B}_{1}\subset\mathcal{B}_{2}\cdots$, we obtain
\begin{equation}
    \centering
    \mu_{*}(E) = \sup_{K\subset{E}}\mu\left(K\right) = \sup_{\bigcup^{K}_{i=1}\mathcal{B}_{i}\supset{E}}\sum^{K}_{i=1}\left|\mathcal{B}_{i}\right|
    \label{eq:mth_reg_meas_intro_1100}
\end{equation}
In compressible topological vector space, we aim to find the smallest subset $K$ of $X$ such that the following regularity (compatible with \eqref{eq:lm_cs1502}) is satisfied
\begin{equation}
	\centering
	\left|\mu_{*}(E)-\mu(X)\right|<\epsilon \text{, \quad } \epsilon\in{\mathds{R}^{+}\cup \{0\}}
	\label{eq:mth_reg_meas_intro_1100_1}
\end{equation}

\eqref{eq:mth_reg_meas_intro_1100_1} denotes regularity for which the supremium energy is recovered from the infimum number of basis that span $n$-dimensional topological space $K\subset{X}$. 

Now, we can answer the first two questions. First, by measuring compressible topological vector space $X$, we mean to determine the closed convex hull of $X$. Another, property we are interested in measuring is the volume of the compressible topological vector space $X$ for which the regularity in \eqref{eq:mth_reg_meas_intro_1100} is satisfied. With this discussion, it is clear now that the measurable of compressible topological vector space $X$ is the smallest compact subset of $\sigma$-algebra of the $X$ that contains its extreme points. In this sense, compactness is the indication of optimum dimension of $X$.

\begin{dfn}[Absorbing Lebesgue Inner Measure] Since $K$ is unknown, the best strategy to compute the Lebesgue inner measure is to employ the absorbing method. In this method, we need to optimize the Borel $\sigma$-algebra by including components from $E$ into the $K$. In this way, a measurement on $X$ is an absorbing because if $V$ is the neighborhood of $0$, then
\begin{equation}
	\begin{split}
	K \subset \bigcup^{n}_{i=1} V_{i} 
	& \equiv \bigcup^{n}_{i=1} \mathcal{B}_{i} \text{, \quad } n\in{\mathds{N}}
	\end{split}
	\label{eq:mth_reg_meas_intro_1200}
\end{equation}
\label{dfn:mth_reg_meas_intro_1200}
\end{dfn}

In order to discuss the connection between the Lebesgue measure and Fr\'echet space developed in section \ref{m:bck:fs}, we define Lebesgue measurable spaces in the following definition. We find the Lebesgue measurable space in Definition \ref{dfn:mth_reg_meas_intro_1104} helpful to describe the measurement behavior of $X$ and its non-increasing rearrangement-invariant $X^{-}$ in Fr\'echet space.

\begin{dfn}[Lebesgue Measurable Space] Let $\left(X,\mathfrak{B},\mu_{X}\right)$ be a measurable space equipped with Borel $\sigma$-algebra. Also, let $E_{i}\in{\mathcal{B}}$ and $i\in{I}$ where $I$ is index set. The measurable space defined by Lebesgue $\sigma$-algebra has the following properties: 
    \begin{enumerate}
        \item Empty set: $m\left(\emptyset\right) = 0$.
        \item Monotonocity: If $E_{1}\subset{E_{2}}$, then $m\left(E_{1}\right)\leq m\left(E_{2}\right)$.
        \item Countable subadditivity: $\mu\left(\bigcup^{\infty}_{n=1}E_{n}\right) \leq \sum^{\infty}_{n=1}\mu\left(E_{n}\right)$.
        \item Downwards monotone convergence: If $E_{1}\supset{E_{2}}\supset{}\cdots$, then $\mu\left(\bigcap^{\infty}_{n=1}E_{n}\right) \leq \lim_{n\rightarrow \infty}\mu\left(E_{n}\right) = \inf_{n}\mu\left(E_{n}\right)$.
        \item Upwards monotone convergence: If $E_{1}\subset{E_{2}}\subset{}\cdots$, then $\mu\left(\bigcup^{\infty}_{n=1}E_{n}\right) \leq \lim_{n\rightarrow \infty}\mu\left(E_{n}\right) = \sup_{n}\mu\left(E_{n}\right)$.
    \end{enumerate}
    \label{dfn:mth_reg_meas_intro_1104}
\end{dfn}

Options (4) and (5) are important when applied to rearrangement-invariant of $X$. We are dealing with non-increasing set $X^{-}$ with a family of Cauchy nets $a^{-}$ that follows downward monotone converges almost surely. Accordingly, family of measurements $p^{-}_{n} = \{p^{-}_{1},p^{-}_{2},\cdots,p^{-}_{n}\}$ corresponding to $X^{-} = \{X^{-}_{1},X^{-}_{2},\cdots,X^{-}_{n}\}$. The local dynamics of $X^{-}$ is given by $\nabla X^{-}$ monotonically converges to zero. Furthermore, giving epsilon of room \footnote{As recommended in \cite[Section 2.1.2]{tao2011introduction}}, one can assume that there is an optimum index $n\in{\mathds{N}}$, for which $X^{-}_{n} = \{X^{-}_{1},X^{-}_{2},\cdots,X^{-}_{n}\}$ uniformly converges over compact subsets $E_{i}$ to $X^{-}$. Then, by Vitali convergence theorem, a family of non-decreasing semi-norms $P^{-}$ converges uniformly to its limit as $n \rightarrow \infty$. By giving epsilon of room again, $n\in{\mathds{N}}$ can be obtained as a point for which the total variation of $P^{-}$ falls below some $\epsilon>0$.
\subsection{Riesz Representation Theorem}
\begin{thm}[Riesz Representation Theory]\footnote{Called after Frigyes Riesz} Let $X$ be a $\sigma$-compact metric space, that is, $X$ can be expressed as a union of countable compact subset of $X$. Also, let $\Lambda:X\rightarrow \mathds{R}$ be a continuous linear functional such that $\Lambda\in{C_{c}\left(X\right)}$. Then, there is uniquely defined measurement $\mu$ on $X$ such that
    \begin{equation}
        \centering 
        \Lambda\left(f\left(x\right)\right) = \int_{X} f\left(x\right)d\mu\left(x\right)
        \label{eq:mth_reg_meas_intro_1104}
    \end{equation}
    \label{thm:mth_reg_meas_intro_1100}
\end{thm}

Theorem \ref{thm:mth_reg_meas_intro_1100} indicates that there is a family of measurement on $X$ that are uniquely correspondent to the linear functional $\Lambda$. We end this section with the following theorem which indicates the linear functional from compact space to compact continuous function $C_{c}\left(\Omega\right)$. It is direct result of Krein-Milman Theorem and applying Riesz representation theorem to  weak$^{\ast}$-compact topology in $X$.

\begin{thm}[Dual space $X^{\ast}$ Replaced With $C_{c}\left(\Omega\right)$] Let $X$ be a compact topological vector space with dual space $X^{\ast}$ that separates points on $X$. The dual space $X^{\ast}$ can be replaced by $C_{c}\left(\Omega\right)$ with respect to (1) separating $\|\cdot\|_{1,\infty}$ defined by the Radon regularity on $X$ as a finite union of semi-norms, and (2) Riesz representation theorem.
    \label{thm:mth:reg:reg:1100}
\end{thm}

Theorem \ref{thm:mth:reg:reg:1100} is critical for the development of metric space in Fr\'echet space as we discuss in section \ref{m:mth:vcb_fr}. This answer the question 4) where the Reisz representation theorem is used as a tool to generate the Bore $\sigma$-compact to measure the compactness of the topological vector space $X$. Question 3), however, is answered in section \ref{m:mth:vcb_fr} where the sequentially compact subsets of dual space will be generated using $\|\cdot\|_{\infty}$.\label{m:mth:rg}
\subsection{Connectedness of Absorbing Null Space} \label{m:mth:cans}
The absorbing null space in \eqref{bkg:mthd_nsp1100}, considers the null space as a connected inseparable object with respect to some invariant property. In the following, we show that the Lebesgue measure in \eqref{bkg:mthd_nsp1102} is constant over connected area and it equals to zero. 
\begin{dfn}[Connected Topological Space] \cite{nlabConnected} A topological space $\left(X,\mathscr{T}\right)$ is connected if $\mathscr{T}$ is irreducible, in other words, $X$ cannot be represented as the union of two disjoint and non-empty subspaces. 
	\label{dfn:mth:rgm:cabns:1100} 
\end{dfn}

\begin{dfn}[Equivalence Relation] The equivalence relation $x \overset{c}{\thicksim} x^{\prime}$ implies that $x$ and $x^{\prime}$  belongs to the same subspace of $X$. 
	\label{dfn:mth:rgm:cabns:1101}
\end{dfn}

\begin{dfn}[Equivalence Classes With Respect To Invariance Property] Two linear functional $\Lambda_{1}\in{\Lambda}$ and $\Lambda_{2}\in{\Lambda}$ belong to the same equivalence class if they satisfy certain invariant property of that class.
\label{dfn:mth:rgm:cabns:1101b} 
\end{dfn}

The invariance property of interest in this letter is the null space of an operator $\Lambda$. Then, null space defines an equivalence class of linear functionals coincide with the absorbing null space property. The following definition introduces the connectedness with respect to certian invariance property. 

\begin{dfn}[Connected Components for Equivalence Class] Two points $x,x^{\prime}$ are the connected components of the same equivalence classes $x \overset{c}{\sim} x^{\prime}$ if there is connected subset of $U\subset{X}$ with respect to certain invariant property such that $x,x^{\prime}\in{U}$.
	\label{dfn:mth:rgm:cabns:1102}
\end{dfn}

\begin{prp}[Connected Co-domain] \cite{ivankhachatourian2018} Let $\Lambda:X \rightarrow Y$ be a continuous surjective map. Then, if $U\subset{X}$ is connected subspace of $X$, so the $V\subset{Y}$.
\label{prp:mth:rgm:cabns:1100} 
\end{prp}

\begin{crl}[Connected Absorbing Null Space] Let $\Lambda: X\rightarrow Y$ be a linear functional between topological spaces $X$ and $Y$. Also, let $\underline{\Lambda}\in{\Lambda}$ for all $i\in{\left[1,n^{\perp}\right]}$ be a set of linear functional satisfies null space property in the form of Minkowski functional in \eqref{eq:bck:cs:cv:1099b}. Then, for all $x,x^{\prime}\in{X_{n^{\perp}}}$, there are open sets $U\in{X}$ and $V\in{Y}$. Then, $\underline{\Lambda}\in{\Lambda}$ constructs connected components of $\Lambda$ and connectedness of absorbing null space is followed.
\label{crl:mth:rgm:cabns:1100}  
\end{crl}

\begin{IEEEproof}
The existence of connected subspace $V\in{Y}$ follows directly from the Proposition \ref{prp:mth:rgm:cabns:1100}. Then, the surjective continuity implies that there is an open map from $\underline{U}:U \rightarrow V$. Then, letting the null space property in \eqref{eq:bck:cs:cv:1099b}, the set of all $\underline{\Lambda}\in{\Lambda}$ provides the equivalence class; i,e., $\underline{\Lambda}_{i} \overset{c}{\sim} \underline{\Lambda}_{j}$ for all $i,j\in{\left[0,d-n-1\right]}$. Consequently, $\underline{\Lambda}$ provides a connected space in $\Lambda$ with respect to null space property. For all $x \in{X^{\perp}_{n}}$, the connectedness of absorbing null space is followed.
\end{IEEEproof}

Obviously, the equivalence class in Definition \ref{dfn:mth:rgm:cabns:1101b} implies that considering the connectedness of absorbing null space, the Lebesgue measurement $\mu\left(\mathcal{B}_{i}\right) = \mu\left(\mathcal{B}_{j}\right) = const.$ and $\mathcal{B}_{i},\mathcal{B}_{j}\subset{U}$ where $U$ is the connected subspace of $X$. $\mathcal{B}_{i},\mathcal{B}_{j}$ for all $i,j\in{I}$ belongs to the Borel $\sigma$-algebra  coincide with the $x\in{X^{\perp}_{n}}$. Considering, the null space in the form of Minlowski functional, we have $\mu\left(\mathcal{B}_{i}\right) = \mu\left(\mathcal{B}_{j}\right) < \epsilon$ for $\epsilon \rightarrow 0$. As a result, the absorbing null space $X^{\perp}_{n}$ is a connected subspace of $X$. 
However, the optimums subspace $X_{n}\subset{X}$ is not connected necessarily because it can be decomposed as the union of restricted subspaces.
\section{Topological Analysis Of Compressibility} \label{s:ta}
The rearrangement-invariant space and its properties for totally-ordered directed nets play a critical role in the topological analysis of the sparsity. In the following section, we define rearrangement-invariant space with respect to a certain invariant property. We select the $\|X_{i}\|_{1,\infty}$, for all $X_{i}\subset{X}$ and $i\in{\left[0,d-1\right]}$, as an invariant parameter of interest. In Theorem \ref{thm:mth:ta:h:1099}, we prove that formulating a problem with respect to invariant parameters enable establishing a symmetry in the form of reflexive homeomorphism. We prove the resultant reflexive homeomorphism as forward and inverse homeomorphism. We develop a systematic approach using locally convex space to define a weak-compact topology on compressible topological vector space $X$ (resp. weak$^{\ast}$-compact on dual space $X^{\ast}$ of $X$). Then, we prove the reflexive homeomorphism by showing that the Cauchy nets defined on LHS and RHS of the \eqref{eq:mth:ta:h:1098} are equivalent. Considering the convergence of the Cauchy nets, we prove that the approach proposed in Theorem \ref{thm:mth:ta:h:1099} clearly provides approximated space $X_{n}$ (resp. $X^{\ast}$) of $X$ (resp. $X^{\ast}$). Finally, we give the examples for such a reflexive homeomprphism in section \ref{m:mth:sgca}.
\subsection{Rearrangement-Invariant Theory} \label{m:bck:rit}
\begin{dfn}[Measurable Function and Level Set]\cite[Section 1.5]{lieb2001analysis} Let $\left(X,\mathcal{B},\mu\right)$ be measure space defined on compressible topological vector space $X$. We say that semi-norm $f:X\rightarrow \mathds{R}^{+} \bigcup \{0\}$ is a measurable function if the level set 
\begin{equation}
	\centering
	L_{f}\left(X\right) = \{x\in{X} \vert f\left(x\right)<t \}
	\label{eq:bck_rit_1400}
\end{equation}
is measurable, that is $L_{f}\left(X\right)\in{\mathcal{B}}$.
\end{dfn}

\begin{dfn}[Rearragenment of Sets] Let $\left(X,\mathcal{B},\mu\right)$ be a measure space defined on compressible topological vector space $X$, where $\mathcal{B}$ is a Borel set on $X$ and $\mu$ is the Lebesgue measure. Then, a set $X^{-}$ is a non-increasing rearrangement-invariant of $X$ as
	\begin{equation}
	X^{-} = \{ x\in{X} \vert \left|x\right|<r \}
	\label{eq:bck_rit_1401}
	\end{equation}
\end{dfn}
In particular, we define a non-increasing rearrangement-invariant of $X$ with respect to entries from directed set $D$ as
\begin{equation}
	\centering
	X^{-} = \{ \left|x_{i}\right|>\left|x_{j}\right| \vert \forall{x_{i},x_{j}}\in{X} \text{, } i<j\in{\mathds{N}} \}
	\label{eq:bck_rit_1402}
\end{equation}

\begin{dfn}[Character Function of Rearrangement Set] \cite[Section 3.3]{lieb2001analysis} Let $\mathcal{X}$ to be a character function of set $X$, then the following equality is satisfied
	\begin{equation}
		\centering
		\mathcal{X}^{-}_{\mathcal{B}} = \mathcal{X}_{\mathcal{B}^{-}}
		\label{eq:bck_rit_1403}
	\end{equation}
That is, the nonn-increasing rearrangement-invariant character function $\mathcal{X}^{-}$ of $X$ is the same as the character function $\mathcal{X}$ obtained for rearranged Borel set $\mathcal{B}^{-}$.
\end{dfn}
Rearrangement-invariant transform manipulates a geometry of data while preserving its volume.

\begin{figure}[!t]
    \centering
    \includegraphics[width = 3.0in]{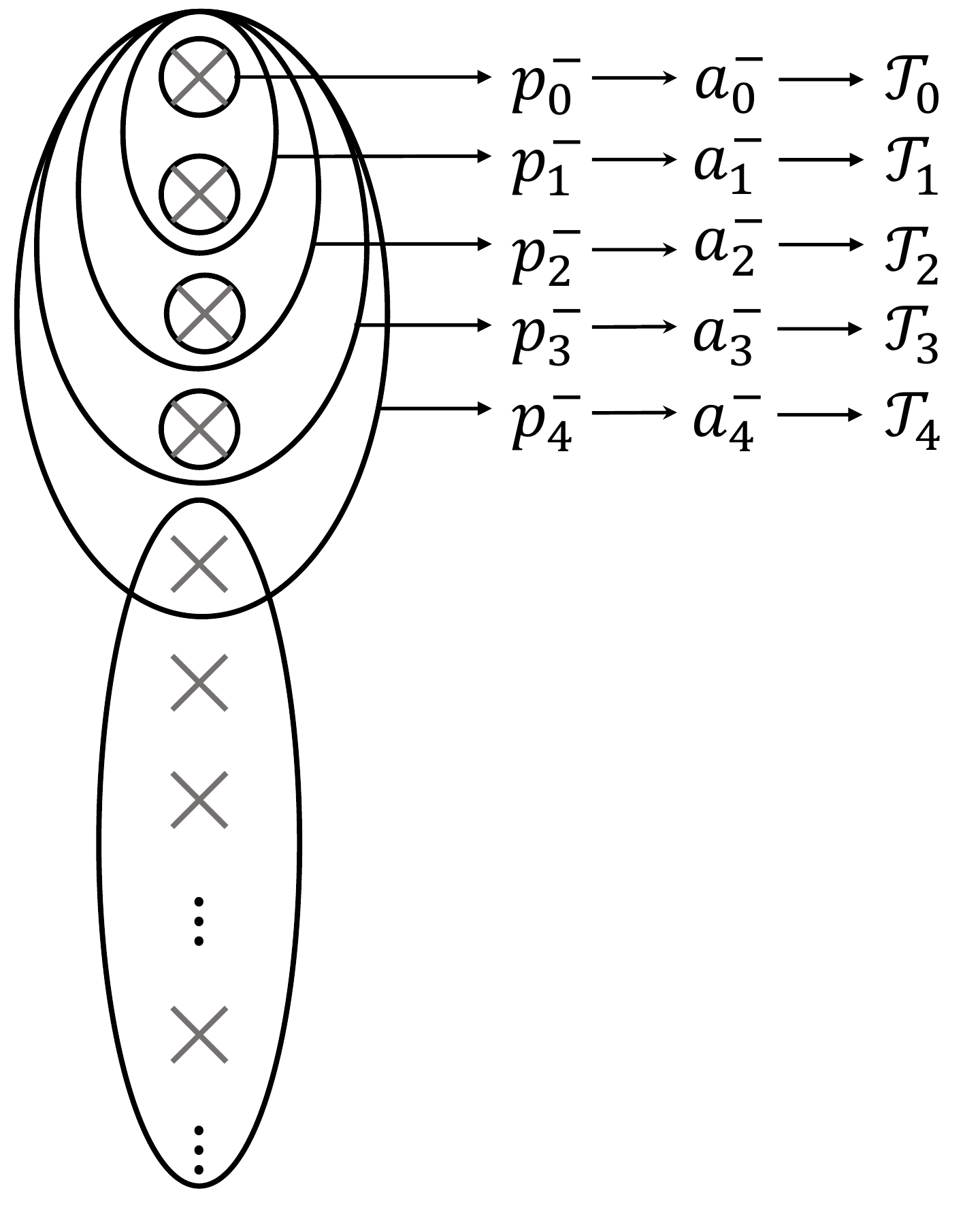}\label{fig:bck_rit_1102a}
    \caption{Topological vector space defined on non-increasing rearrangement-invariant space. $\mathscr{T}_{4}$ is the coarsest topology on $X^{-}$ corresponding to the Radon inner regularity measure. The link between topological vector space and measurable space is established by Borel $\sigma$-algebra.}
    \label{fig:bck_rit_1102}
\end{figure}

The relation between rearrangement-invariant transform and Radon measure has been shown in Fig. \ref{fig:bck_rit_1102}. One can define a topology on inner regularity as shown in Fig. \ref{fig:bck_rit_1102}, where $\mathscr{T}_{4}$ induced on the optimum subspace $X_{n}$ is the coarsest topology on $X$. Note that each topology on $X^{-}$ is defined based on a directed Cauchy net $a^{-}_{i}$ for all $i\in{D}$\footnote{For $i,j\in{D}$, $i \succ j$, if $X^{-}_{i} \succ X^{-}_{j}$, then $f^{-}\left(i\right) \prec f^{-}\left(j\right)$.}. One can define $\nabla f^{-}$ and directed Cauchy net $a^{-}$ on $\nabla f^{-}$ such that $a^{-}_{i} \prec a^{-}_{j}$. For a ball $\mathcal{B}_{\epsilon}\left(X\right)$, then we can define a volume $n$ on $X$ and $X^{-}$ with respect to $\left|a^{-}_{i+1}-a^{-}_{i}\right| \leq \epsilon$. Equivalently, we say that $a^{-}$ vanishes in a closed ball $\mathcal{B}_{\epsilon}\left(X\right)$.

The converging behavior of the directed Cauchy net with entries from $f^{-}\left(x\right)$ is crucial to formulate Radon regularity for the optimum dimension $n$. As shown in Fig. \ref{fig:bck_rit_1102}, due to the smaller size of the optimum subspace, given a non-increasing rearrangement-invariant $X^{-}$ of $X$, it is simpler to measure inner the regularity to determine optimum volume $n$.

In order to formulate rearrangement-invariant, let $\left(X,\mathcal{B}_{X},\mu_{X}\right)$ be a measurable space induced with semi-norms $p^{-}$ that vanishes at a certain coordinate. In other words, Lebesgue measurement goes to zero at a specific finite coordinate $n\in{D}$, such that for some $i,j\in{D}$, $i \succ j \succ n$ is satisfied. Then, $X$ is a locally convex space with a volume
\begin{equation}
    \centering
    vol\left(\{x\in{X} \vert \left|f\left(x\right)\right|>\alpha\}\right) < \infty
    \label{eq:bck_rit_1100}
\end{equation}
where $\alpha$ is a threshold proportional to the optimum volume of $X$. The measurable function $f$ and its rearranged form $f^{-}$ are unsinged measurements and conventionally are defined point-wise for all $x\in{X}$
\begin{equation}
	\centering
	\left|f\left(x\right)\right| = \int^{\infty}_{0} \mathcal{X}_{\left|f\right|>\alpha}\left(x\right)d\alpha
	\label{eq:bck_rit_1404a}
\end{equation}
\begin{equation}
	\centering
	f^{-}\left(x\right) = \int^{\infty}_{0} \mathcal{X}^{-}_{\left|f\right|>\alpha}\left(x\right)d\alpha
	\label{eq:bck_rit_1404b}
\end{equation}

The volume in \eqref{eq:bck_rit_1100} can be represented by a distribution function with respect to measurement $\mu$ as\footnote{Alternatively, one can define $\lambda_{f}\left(a\right)$ for the compressible topological vector space as \cite{Tao}
\begin{equation}
	\centering
	\lim_{p \rightarrow 0}\|f\|^{p}_{p} = \mu\left(\supp\left(f\right)\right)
	\label{eq:bck_rit_1101a}
\end{equation}
where $p$ refers to $\ell_{p}$-norm.
}
\begin{equation}
    \centering
    \lambda_{f}\left(\alpha\right) = \mu\left(\{x\in{X} \vert f\left(x\right)>\alpha \}\right) < \infty
    \label{eq:bck_rit_1101}
\end{equation}
\eqref{eq:bck_rit_1101} gives the volume of a locally convex space $X$ with respect to $\sigma$-finite algebra.

\begin{thm}[Optimum Volume Preservation] The optimum volume $n$ of $X$ can be obtained from $X^{-}$.
\label{thm:bck_rit_1400}
\end{thm}

In order to prove Theorem \ref{thm:bck_rit_1400}, we need to show that a measurement on $X^{-}$ in $L^{p,\infty}$-space implies a measurement on $X$ in $L_{p}$-space. For this purpose, we adopt the following Lemma from \cite[Lemma 1.4]{burchard2009short}.

\begin{lm}[Rearrangement in $L_{p}$-Space] Let $f$ and $f^{-}$ be family of semi-norms defined on $X$ and $X^{-}$, then we have
    \begin{equation}
        \centering
        \|f\|_{p} = \|f^{-}\|_{p} \text{, for } 1\leq p \leq \infty
        \label{eq:bck_rit_1102}
    \end{equation}
    \label{lm:bck_rit_1100}
\end{lm}
 
\begin{IEEEproof}
We follow the proof from \cite[Lemma 1.4]{burchard2009short}, which is based on equimeasurability property. In the following proof, $p$ refers to $\ell_{p}$-norm.
\begin{equation}
    \begin{split}
        \|f\|^{p}_{p} & = \int_{\mathds{R}^d}\left|f\left(x\right)\right|^{p}dx = \int_{\mathds{R}^d}\int^{\infty}_{0} \mathcal{X}_{f\left(x\right)^{p}>\alpha} d\alpha dx \\
        & = \int^{\infty}_{0}vol\left(\{f\left(x\right)^{p}>\alpha\}\right)d\alpha \\
        & = \int^{\infty}_{0}\lambda_{f}\left(\alpha\right)d\alpha \\
    \end{split}
    \label{eq:bck_rit_1103}
\end{equation}
Since, the volume is invariant under rearrangement, then $\lambda_{f}\left(\alpha\right) = \lambda_{f^{-}}\left(\alpha\right)$ implies $\|f\|_{p} = \|f^{*}\|_{p}$.
\end{IEEEproof}

Lemma \ref{lm:bck_rit_1100} indicates an extremely important property of the rearrangement-invariant transforms that under rearrangement-invariant transform the $\ell_{p}$ norm is preserved.

\begin{crl} In spacial case for $p=1$, Lemma \ref{lm:bck_rit_1100} indicates that
    \begin{equation}
        \centering
        \mathcal{B}_{1}\left(X_{n}\right) \equiv \mathcal{B}_{1}\left(X^{-}_{n^{\prime}}\right)
        \label{eq:bck_rit_1105}
    \end{equation}
    By applying Minkowski functional theorem to \ref{eq:bck_rit_1105}
    \begin{equation}
        \centering
        \mathcal{B}_{\epsilon}\left(X_{n}\right) \equiv \epsilon \mathcal{B}_{1}\left(X_{n}\right) \equiv \epsilon \mathcal{B}_{1}\left(X^{-1}_{n^{\prime}}\right) = \mathcal{B}_{\epsilon}\left(X^{-}_{n^{\prime}}\right)
        \label{eq:bck_rit_1106}
    \end{equation}
    From Lemma \ref{lm:bck_rit_1100}, we conclude $n = n^{\prime}$. In other words, there is an isometric isomorphic map $\phi$
    \begin{equation}
        \centering
        \Rightarrow \phi: X^{-}_{n^{\prime}} \longmapsto X_{n}
    \label{eq:bck_rit_1107}
    \end{equation}
    \label{crl:bck_rit_1100}
\end{crl}

Having the ingredient, proof of Theorem \ref{thm:bck_rit_1400} is in the following.
\begin{IEEEproof}
From Corollary \ref{crl:bck_rit_1100}, the optimum subspace $X_{n}$ can be obtained equivalently from $X^{-}$.
\end{IEEEproof}
\subsection{Locally Convex Spaces} \label{m:bck:lcs}
\begin{dfn}[Locally Convex Space] A vector space $X$ equipped with semi-norm $P$ is called locally convex space such that a family of semi-norms induces a weak topology on $X$. The locally convex space satisfies the following properties.
    \begin{enumerate}
        \item Semi-norms: Every locally convex topology is induced by a family of semi-norms defined on a vector space $X$. 
        \item Locality and Convexity: Locally convex space has a set of convex $0$-neighborhood bases \cite[Section 1.4.9]{kriegl2015advanced}. In Hilbert space, the convex bases coincide with the convex basis\footnote{The bases are defined as the union of open balls with respect to Minkowski functional, which are not necessarily unique. The basis is a set of orthonormal in the Hilbert space, which are unique}. The topology of locally convex subspaces is generated by applying the absorbing Minkowski function to each convex base at the $0$-neighborhood.
        \item Uniform Convergence: A sequence $X_{n}$ converges to $X$, if $p\left(X_{n}-X\right) \rightarrow 0$. Note that this is a point-wise convergence. A more useful property is obtained if one considers Cauchy sequences \cite[Section 1.4.9]{kriegl2015advanced}. 
        \item Sequentially Completeness: A directed set $E\subset{X^{-}}$ equipped with binary operation $\succ$ is Cauchy sequence if there is $n^{\prime}$ such that for all $j \succ i \succ n^{\prime}$ all in $\mathds{N}$ and for all $p^{-}$ defined sequentially on $E$
        \begin{equation}
            \centering
            p\left(x^{-}_{i}-x^{-}_{j}\right) < \epsilon \text{, } \forall{x^{-}_{i}, x^{-}_{j}}\in{E}
            \label{eq:bck_lca_1101}
        \end{equation}
        For directed net\footnote{Directed net is a directed set takes its value from topological space,} $p^{-}$, this equivalently leads to
        \begin{equation}
        	\centering
        	p^{-}\left(x_{k}-x_{l}\right) < \epsilon \text{, } \forall{x_{k},x_{l}}\in{X}
        	\label{eq:lcs:1406}
        \end{equation}
        for $k,l\in{D}$, where $D$ is directed set that $p^{-}$ takes its index from.
        The equivalent convergence of $X^{-}$ and $X$ can be summarize as
        \begin{equation}
        	\centering
        	p_{X^{-}} = p^{-}_{X}
        	\label{eq:lcs:1406}
        \end{equation}
        
        A locally convex space $X$ (resp. $X^{-}$) is called sequentially complete if every Cauchy sequence in $X$ (resp. $X^{-}$) converges to its limits. \textit{Note that this definition leads to vanishing total variation for the given functional}.
        \item Continuity via nets: Let $X$ be a locally convex space. A linear functional $\Lambda: X \rightarrow Y$ forms a locally convex topological space, if for all convergent net $x\in{X}$, $x\rightarrow a$ implies an image net $\Lambda x \rightarrow \Lambda a$.
    \end{enumerate}
    \label{dfn:bck_lca_1103}
\end{dfn}
\subsection{Minimal Linear Functional}\label{m:mth:mlf}
It is helpful if we can compare different semi-norms defined on vector space $X$. Two semi-norms $p$ and $q$ are comparable either if $p\left(x\right) \leq q\left(x\right)$ or $q\left(x\right) \leq p\left(x\right)$. Comparable semi-norms are useful to define linear functionals from sublinear functionals over non-increasing rearrangement-invariant $X^{-}$ of $X$ with index taken from entries of direct set $D$. A discussion in the following is based on the Lemma \cite[Lemma 5.1.1] {kriegl2015advanced} that states the linear functional $l$ is the minimal of the given sublinear functional $p$. We end this section with Theorem \ref{thm:bck_lca_1100+}, which indicates that the optimum dimension $n$ coincides with the linear subspace $X_{n}$ of $X$ through minimal linear functional.

\begin{lm}[Minimal Linear Functional]\cite[Lemma 5.1.1 and Lemma 5.1.2]{kriegl2015advanced} There is a linear functional $l:X \rightarrow \mathds{R}$ that is the minimal of sublinear functional $p:X \rightarrow \mathds{R}$. Then, $l$ and $p$ are comparable in the sense that $l\leq p$.
    \label{lm:bck_lca_1100}
\end{lm}
\begin{IEEEproof}
We have adopted a proof from \cite[Lemma 5.1.2 and 5.1.1]{kriegl2015advanced}, respectively, to show that, first, there exists a linear functional that is less than other all sublinear functional. Then, we prove that there is a minimal sublinear functional such that a linear functional is less than the minimal.
The proof leads by applying Zorn's Lemma. For this purpose, let $X^{-}$ be a totally ordered set with a family of semi-norms $p$, which is bounded for all points in $X^{-}$. Also, let $S=\{q \preccurlyeq p\}$ where $q$ is sublinear semi-norm .Also, let $L\subseteq S$ be an ordered net. According to Zorn's Lemma, there is an $m\in{S}$ such that for all $l\in{L}$, $m \preccurlyeq l$. Then, there is an infimum $l_{0}\in{L}$ such that for all $l\in{L}$, $l_{0} \preccurlyeq l$, and $m \preccurlyeq l_{0}$. 

First, we need to show that $l_{0}$ is linear. For this purpose, we need to prove additive and homogenity properties over $\mathds{R}^{+}\cup \{0\}$. In order to prove additivity, we leverage the convexity of space to define $l_{y}\left(x\right) = \inf_{0 \leq \theta\leq 1}\{l\left(x+\theta y\right)-\theta l\left(y\right)\}$ for $x,y\in{X}$. Then, there is a minimal point $a\in{X}$ such that $l_{a}\preccurlyeq l$ for all $l\in{L}$, for which $l_{a}\left(b\right) \leq l\left(a+b\right)-l_{a}$. If $a$ is a minimal point, the additivity is followed by $l\left(a+b\right) = l_{a}\left(b\right)+l_{a}$. Then, $l_{a}$ is the minimal of $L$. Obviously, $l_{a}\triangleq l_{0}$. 

For $\mathds{R}^{+} \cup \{0\}$ homogenity, we have
\begin{equation}
    \begin{split}
    l_{0}\left(rx\right) & = \inf_{0 \leq \theta \leq 1}\{l\left(rx+\theta y\right)-l\left(\theta y\right)\} \\
    & = \inf_{0 \leq \theta \leq 1}\{l\left(r\left(x+\frac{\theta}{r} y\right)\right)-\theta l\left(y\right)\}\\
    & = \inf_{0 \leq \theta \leq 1}\{r\left(l\left(x+\frac{\theta}{r} y\right)\right)- \frac{\theta}{r} l\left(y\right)\}\\
    & = \inf_{0 \leq \theta^{\prime} \leq 1}\{l\left(x+\theta^{\prime} y\right)-\theta^{\prime} l\left(y\right)\}\\
    & \overset{a}{=} r \cdot l_{0}\left(x\right)
    \end{split}
\end{equation}
where (a) $\theta^{\prime} \leftarrow \frac{\theta}{r}$. And this end proof.
\end{IEEEproof}

\begin{rmk}[Minimal Linear Functional, Optimum Restricted Functional, and Weak Topology] Lemma \ref{lm:bck_lca_1100} indicates that there is some minimal $l_{0}$ beyond which a family of semi-norms defined on a totally ordered set is linear. In fact,  minimal linear functional ${l\vert}_{l_{0}}$ is the optimum restricted representation of the functional space $p$ of compressible topological vector space. Therefore, the minimal linear functional $l_{0}$ defined the boundary of the weakest topology can be defined on $X$. The formal proof will be provided in Theorem \ref{thm:bck_lca_1100+}.
	\label{rmk:bck_lca_1200}
\end{rmk}

\begin{thm}[Hahn-Banach Theorem - Continuous Extension] Let $X$ be a vector space over $\mathds{K}$ and equipped with semi-norms $p:X\rightarrow \mathds{C}$. Also, let $U$ be a linear subspace of $X$ with a linear functional $l:U\rightarrow \mathds{K}$ such that $l\leq {p\vert}_{U}$, where ${p\vert}_{U}$ denotes sublinear semi-norms restricted to $U$. Then, there is a family of linear functional $\Lambda\in{X^{\ast}}$ such that ${\Lambda\vert}_{U} \triangleq l$, and $\left|\Lambda\right|\leq p$. Then $\Lambda$ is a linear extension of $l$ to $X$.
    \label{thm:bck_lca_1100}
\end{thm}
Proof of Theorem \ref{thm:bck_lca_1100} can be found in functional analysis textbooks including \cite[Theorem 5.1.3 and Corollary 5.1.4 Page 68]{kriegl2015advanced}, and \cite[Theorem 3.6 Page 61]{rudin1991functional}.

\begin{thm}[Hahn-Banach Theorem- Separation]\cite[Theorem 3.4]{rudin1991functional} Let $X$ be a locally convex topological space with dual space $X^{\ast}$. Then, there is $\Lambda\in{X^{\ast}}$ acting on $X$ such that for all $x,y\in{X}$, the mappings $\Lambda x$ and $\Lambda y$ are distinct, indicating $\Lambda x \neq \Lambda y$.
    \label{thm:mth:ta:ts:1100}
\end{thm}

\begin{rmk} According to Theorem \ref{thm:mth:ta:ts:1100}, it is said that $X^{\ast}$ separates points on $X$. This is equal to saying that the linear function is onto. 
    \label{rmk:mth:ta:ts:1100}
\end{rmk}

The Hahn-Banach separation theorem is a powerful tool to formulate the idea of a closed linear hull that spans compressible topological vector space $X$. In particular, the linear functional can decompose $X$ as in the orthogonal decomposition if the basis are othonormal in Hilbert space.

\begin{thm}[Existence of Optimum Subsapce via Minimal Linear Functional] Let $X$ be a vector space equipped with a family of semi-norm $p$. Let $p^{-}$ be non-increasing rearrangement-invariant of $p$. Also, let $X_{n}$ be an optimum subspace of $X$. Then, there is an infimum of sublinear semi-norm denoted by $l_{n} \leq p^{-}$ that for which an optimum linear subspace $X^{\ast}_{n}$ (resp. $X_{n}$) of $X^{\ast}$ (resp. $X$).
    \label{thm:bck_lca_1100+}
\end{thm}
\begin{IEEEproof}
Considering Riesz representation theorem, The non-increasing rearrangement-invariant $p^{-}$ is convex set generated equivalent to basis $\Lambda$ \ref{thm:mth_reg_meas_intro_1100}. Obviously, ${\Lambda\vert}_{X_{n}}$ is separable with respect to the Hahn-Banach separation theorem \ref{thm:mth:ta:ts:1100}. Then, the topology induced by $X_{n}$ is the coarsest topology on $X$. From Lemma \ref{lm:bck_lca_1100}, we understand that there is a linear function $l_{0}$ that generates a topology for ${\Lambda\vert}_{X_{n}}$. Accordingly, Hahn-Banach continuous extension theorem \ref{thm:bck_lca_1100} extends ${\Lambda\vert}_{X_{n}}$ to ${\Lambda\vert}_{X}$. Finally, Lemma \ref{lm:bck_lca_1100} implies that there is an $n\in{\mathds{N}}$ such that $l_{0}\triangleq p^{-}_{n^{\prime}}$ is the minimal linear function on topological vector space $X^{-}$. As a result, there is $X_{n}\subset{X}$ such that $n^{\prime} \equiv n$, for $n\in{D}$.
\end{IEEEproof}

In conclusion, the existence of $n$ depends on minimal linear functional that may lie randomly somewhere in $X^{\ast}$. This requires $X$ to be separable and measurable. These condition follow from the Hahn-Banach separable theorem and $\sigma$-finite algebra, which allows to obtain Lebesgue measure as the countable union of open subsets of $X$.
\subsection{Further Discussion on Optimum Subspace}\label{m:bck:fdos}
By applying the Hahn-Banach extension theorem, one should answer the question if there are sufficient numbers of separating points so that the compressible topological vector space $X$ can be decomposed into optimum subspace $X_{n}$ and $X^{\ast}_{n}$. In this section, we employ the Riesz extension theorem to examine relation between minimal linear functional and sufficient separating points.

\begin{dfn}[Banach Limit]\cite[Corollary 7.14]{einsiedler2017functional} Let $c\left(\mathds{N}\right)$ be a sequence space in Banach space with a linear functional defined as
\begin{equation}
	\centering
	\LIM \coloneqq c\left(\mathds{N}\right)\ni X_{n} \rightarrow \lim\left(X_{n}\right) = \lim_{n \rightarrow \infty} X_{n} \text{, } n\in{\mathds{N}}
	\label{eq:fdos:1098}
\end{equation}
Obviously, $\LIM\in{\left(L^{\infty}\left(X_{n}\right)\right)^{\ast}}$. Then, $\LIM$ satisfies the following properties
\begin{itemize}
	\item $\LIM\left(X_{n}\right) = \lim_{n \rightarrow \infty} X_{n}$, if the latter limit exists.
	\item $\LIM\left(X_{n}\right)\in{\left[\lim \inf_{n\rightarrow \infty} X_{n}, \lim \sup_{n\rightarrow \infty} X_{n} \right]}$, for all $X_{n}\in{\mathds{R}}$ and $n \geq 1$.
	\item $\LIM\left(X_{n}\right) = \LIM\left(X_{n+1}\right)$
\end{itemize}
Then, function $\LIM$ is called Banach limit.
\label{dfn:fdos:1100}
\end{dfn}

From second property (b), we have
\begin{equation}
	\centering
	\lim \inf_{n \rightarrow \infty} X_{n} \leq \LIM \leq \lim \sup_{n \rightarrow \infty} X_{n}
	\label{eq:fdos:1097}
\end{equation}

Now, let $X_{n} = X^{-}_{n}$ to have a C\'esaro mean in the form of
\begin{equation}
	\centering
	\mathcal{M} = \left(x_{1},\frac{x_{1}+x_{2}}{2},\frac{x_{1}+x_{2}+x_{3}}{3},\cdots,\frac{x_{1}+x_{2}+\cdots+x_{i}+x_{j}}{j},\cdots\right)
	\label{eq:fdos:1096}
\end{equation}
Let $L$ be a linear functioncal with norm one as $\|L\|=\|\lim\|$, i.e., $L$ is an operator with a norm equal to the $\|\lim\|$, then,
\begin{equation}
	\centering
	\LIM\left(X_{n}\right) = L\left(\mathcal{M}\right)
	\label{eq:fdos:1095}
\end{equation}
Obviously, for $n\geq 1$
\begin{equation}
	\centering
	\left|L\left(\mathcal{M}\right)\right| \leq \|X_{n}\|_{\infty}
	\label{eq:fdos:1094}
\end{equation}

Since $X_{n}$ is monotonically downward convergent sequence, we can find $i,j\in{D}$, $j = i+1$ such that
\begin{equation}
	\begin{split}
	\frac{x_{1}+x_{2}+\cdots+x_{i}}{i} & \approx \frac{x_{1}+x_{2}+\cdots+x_{i}+x_{j}}{j} \\
	& \Rightarrow i \times \left(x_{1}+x_{2}+\cdots+x_{i}\right)+\left(x_{1}+x_{2}+\cdots+x_{i}\right) \\
	& = i \times \left(x_{1}+x_{2}+\cdots+x_{i}\right)+ i \times x_{j} \\
	& \Rightarrow \frac{x_{1}+x_{2}+\cdots+x_{i}}{i} = x_{j}+\epsilon \\
	& \Rightarrow x_{j} = \frac{1}{i}\sum^{i}_{k = 0}x_{k} 
	\end{split}	
	\label{eq:fdos:1093}
\end{equation}

Then, there is $x_{j}\in{X_{n}}$ such that for $n\rightarrow \infty$, then $\LIM\left(X_{n}\right) = x_{j}$ and
\begin{equation}
	\centering
	\lim \inf_{n \rightarrow \infty} X_{n} \leq x_{j} \leq \lim \sup_{n \rightarrow \infty} X_{n}
	\label{eq:fdos:1092}
\end{equation}

\eqref{eq:fdos:1092} indicates that, given infinite dimensional sequence $X$ in Banach space with linear function $\LIM$, there is a $x_{j}\in{X_{n}}$ that provides the sufficient number of separating points to satisfy the Hahn-Banach extension theorem. Then, we have
\begin{equation}
	\centering
	X^{-} = \{\underbrace{\{x_{1},x_{2},\cdots,x_{j}\}}_{V^{-}},\underbrace{\{x_{j+1},x_{j+2},\cdots\}}_{U^{-}}\}
	\label{eq:fdos:1091}
\end{equation}

Inspired by \eqref{eq:fdos:1092}, we provide alternative proof for the existence of an optimum subspace in the following. The proof in the following is based on the Riesz extension theorem, which is usually used to prove the Hahn-Banach extension theorem \ref{thm:bck_lca_1100}. We have found the proof in \cite{castillo2005note}, after we have developed the idea holistically in section \ref{m:mth:mlf} and, in particular, in Theorem \ref{thm:bck_lca_1100+}.

\begin{thm}[Reisz Extension Theorem]\footnote{Called after Marcel Riesz} Let $X^{-} = \{V^{-},U^{-}\}$ be a totally ordered locally convex space (in particular, compressible topological vector space). Also, let $Y$ be a linear subspace of $X$ with a linear functional $f_{0}:Y \rightarrow \mathds{R}$. Assume that for every $x\in{X}$ there exists $y\in{Y}$ such that $x \leq y$. Then, there is a sublinear functional $f:X\rightarrow \mathds{R}$ such that $f\vert_{Y} = f_{0}$.
\label{thm:fdos:1100}
\end{thm}
\begin{IEEEproof}
Let 
\begin{equation}
	\centering
	U = \{y\in{Y}\vert y\leq x_{0}\} \text{, \quad}
	V = \{y\in{Y}\vert y\geq x_{0}\}
	\label{eq:fdos:1099}
\end{equation} 
be non-empty sets, where $x_{0}\in{X\backslash Y}$. In order to extend $f_{0}$ to $f$, we need to show that $Y+tx_{0}$, $t\in{\mathds{R}}$, lies in a positive linear functional space. Also, let $y_{1}\in{U}$ and $y_{2}\in{V}$. Since, $f_{0}$ is linear, then $f_{0}\left(y_{2}\right)<f_{0}\left(y_{1}\right))$. Sine $X$ is bounded sequence with well-defined Banach limit $\LIM$, there is $a = x_{j}$ satisfies \eqref{eq:fdos:1092} as
\begin{equation}
	\centering
	\inf\{f_{0}\left(y_{2}\right)\} \leq a \leq \sup\{f_{0}\left(y_{1}\right)\} \text{, } a\in{\mathds{R^{+} \cup \{0\}}}
	\label{eq:fdos:1100}
\end{equation}
Let
\begin{equation}
	f\left(y+tx_{0}\right) \leq f_{0}\left(y\right) + ta
	\label{eq:fdos:1101}
\end{equation} 
Assume that $y+tx_{0}>0$, then $-y/t < x_{0}$. Two scenarios are possible. (1) If $t>0$, then $-y/t\in{U}$ and $f_{0}\left(-y/t\right)<a$ (with respect to \eqref{eq:fdos:1100}). (2) If $t<0$, then $y/t\in{V}$ and with respect to \eqref{eq:fdos:1100} $f_{0}\left(y/t\right) > a$. From (1) and (2), we notice that $f\left(y+tx\right)\geq 0$. And this proves the existence of $f:X \rightarrow \mathds{R}$. 

For $X\in{\mathds{C}^{d}}$, the results follows from complex-linear property as $f\left(y+tx\right) = t f\left(y/t+x\right)$.
\end{IEEEproof}

\begin{crl}[Existence of Optimum Subsapce via Riesz Extension Theorem] \cite{castillo2005note} The continuous extension of $f_{0}$ to $f$ is unique an optimum, if and only if, for all $x_{0}\in{X}$
\begin{equation}
	\centering
		\inf\{f_{0}\left(y_{2}\right)\} = \sup\{f_{0}\left(y_{1}\right)\} \text{, } a\in{\mathds{R^{+} \cup \{0\}}}
	\label{eq:fdos:1101}
\end{equation}
\label{crl:fdos:1100}
\end{crl}

\begin{crl}[Optimum Subspace of Compressible Topological Vector Space with respect to Riesz Extension Theorem] Corollary \ref{crl:fdos:1100} indicates that the continuous extension of the $f_{0}$ to $f$ allows to obtain an optimum subspace for which the whole compressible topological vector space $X$ can be divided into two subspaces, $U$ and $V$ with respect to a linear functional (resp. semi-norms with respect to Riesz representation Theorem in \ref{thm:mth_reg_meas_intro_1100}). It is important to understand what happens here briefly. The continuous extension of the $f_{0}$ to $f$ is optimum and unique if there is an external point $x_{i}\in{V}$ of $f_{0}\left(x_{i}\right)$, and internal point $x_{j}\in{U}$ with $f_{0}\left(x_{j}\right)$ such that $f_{0}\left(x_{i}\right) = f_{0}\left(x_{j}\right)$. Since $f_{0}$ is a function, then $x_{i} = x_{j}$ for $i = j$. Then, there is an optimum point which is the exterior of the $V$ and the interior of the $U$. Because it is difficult to satisfy the condition in \eqref{eq:fdos:1101}, we define the loose condition as 
\begin{equation}
	\centering
		\inf\{f_{0}\left(y_{2}\right)\} \approx \sup\{f_{0}\left(y_{1}\right)\}
	\label{eq:fdos:1102}
\end{equation}
\label{crl:fdos:11000}
\end{crl}

\begin{thm}[Optimum Subspace and Continuous Spectrum approximation] Let $X$ be a compressible topological vector space with separating dual space $X^{\ast}$. The subsets $V^{\ast}$ and $U^{\ast}$ such that $\left(X^{-}\right)^{\ast} = V^{\ast} \oplus U^{\ast}$ coincide with the diagonalization and continuous spectrum extension of the dual space $X^{\ast}$ of $X$. Then, optimum subspace $V\subset{X^{-}}$ coincides with continuous spectrum approximation of $X^{\ast}_{n}$ of $\Lambda: X\rightarrow Y$ such that $\Lambda\vert_{V} \subset{X^{\ast}}$.
	\label{thm:fdos:11010}
\end{thm}

\begin{figure}[t!]
	\centering
	\includegraphics[width = 3.0in]{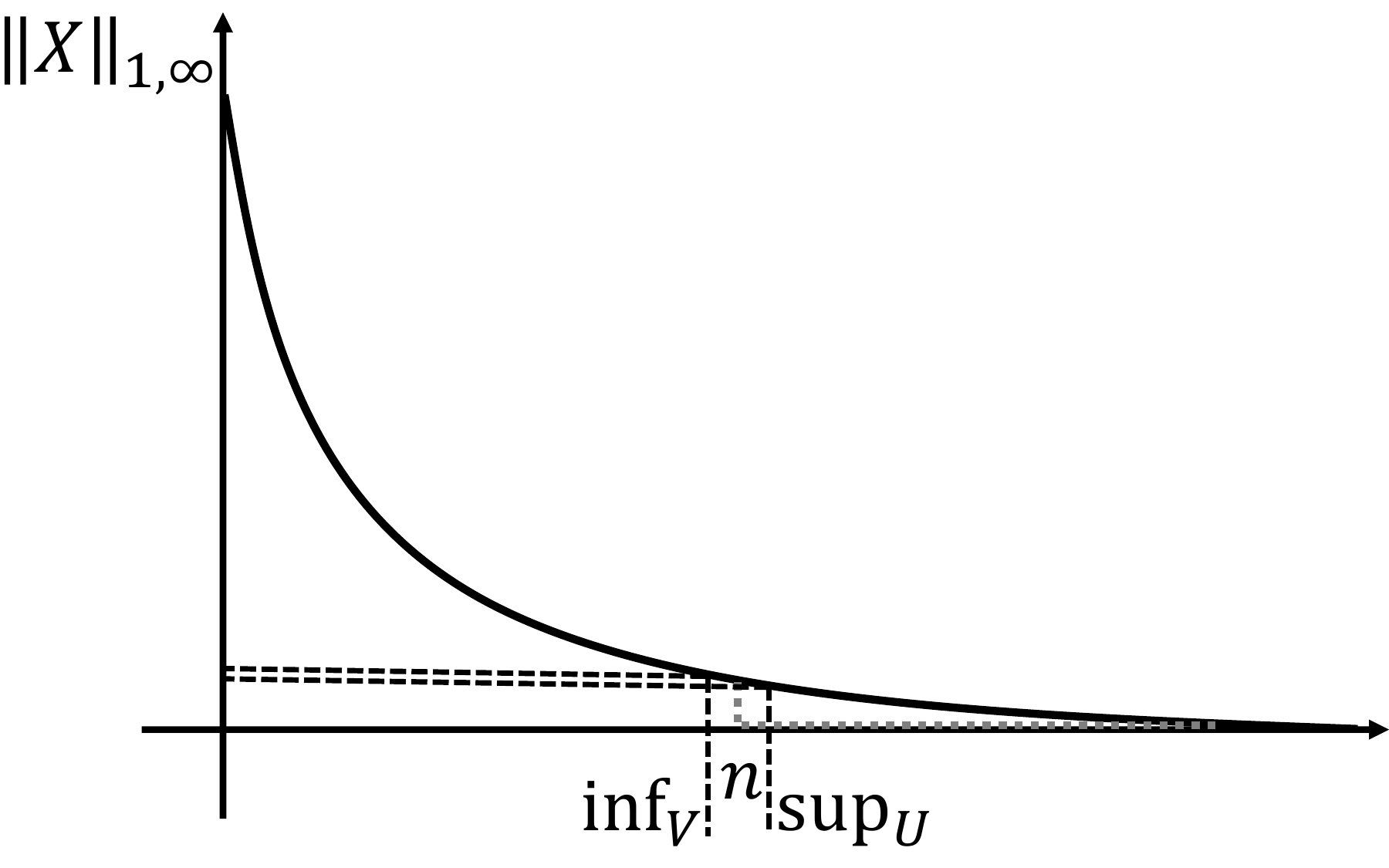}
	\caption{Optimum subspace with respect to regularity in Theorem \ref{thm:fdos:11010}.}
	\label{fig:fdos:1100}
\end{figure}

\begin{IEEEproof}
The proof follows from Kolmogorov and Daniell extension theorems in \ref{thm:lm_cs1500} and \ref{thm:lm_cs1501}. From \eqref{thm:lm_cs1502}, we have
\begin{equation}
	\begin{split}
	&\mu_{t_{1},t_{2},\cdots,t_{n}}\left(\mathfrak{B}_{1} \times \mathfrak{B}_{1} \times \cdots \times \mathfrak{B}_{n-1} \times \mathds{R}\right) = \mu_{t_{1},t_{2},\cdots,t_{n-1}}\left(\mathfrak{B}_{1} \times \mathfrak{B}_{1} \times \cdots \times \mathfrak{B}_{n-1}\right)\\
	& \Rightarrow
	\begin{cases}
	\mu\left(X\right) = \mu_{t_{1},t_{2},\cdots,t_{n}}\left(\mathfrak{B}_{1} \times \mathfrak{B}_{1} \times \cdots \times \mathfrak{B}_{n-1} \times \mathds{R}\right) & \text{, } \forall{y}\in{X}\\
	\mu\left(V\right) = \mu_{t_{1},t_{2},\cdots,t_{n-1}}\left(\mathfrak{B}_{1} \times \mathfrak{B}_{1} \times \cdots \times \mathfrak{B}_{n-1}\right) & \text{, } \forall{y}\leq x_{0}\\
	\mu\left(U\right) = \mu\left(X\slash V\right) & \text{, } \forall{y<x_{0}} 
	\end{cases}
	\end{split}
	\label{eq:fdos:1103}
\end{equation} 
\end{IEEEproof}
Fig. \ref{fig:fdos:1100} depicts the regularity between supremum and infimum defined on $U$ and $V$. Note that regularity is guaranteed where the total variation vanishes.
Two observations can be made from the regularity formulated in Theorem \ref{thm:fdos:11010}. First, regularity in \eqref{eq:fdos:1103} determines how many points $x\in{X}$ (resp. $\Lambda_{i}\in{X^{\ast}}$, $i\in{\mathds{N}}$) are sufficient for the compressible topological vector space to be optimally separable to $V$ and $U$ (equivalently, $X_{n}$ and $X^{\perp}_{n}$). From the point of measurement theory, this is equivalent to find the optimum $\sigma$-finite compatible with weak-compact and weak$^{\ast}$-compact topologies on $X$ and $X^{\ast}$, respectively. Obviously, this is directly related to the RIP, which expresses the condition for the existence of a solution for the compressed sensing problem. Second, the extension of the Borel set requires an integral operator that can perform a combinatorial search over $X^{\ast}$ to optimize $\Lambda_{n}$ such that $\Lambda_{n}$ uniformly converges to $X^{\ast}$. Obviously, conventional $\ell_{1}$ convex optimization and greedy projective algorithms fail to meet the specification.
\subsection{Fr\'echet Space for Compressible Topological Vector Space}\label{m:bck:fs}
\begin{figure}[t!]
	\centering
	\includegraphics[width=3.0in]{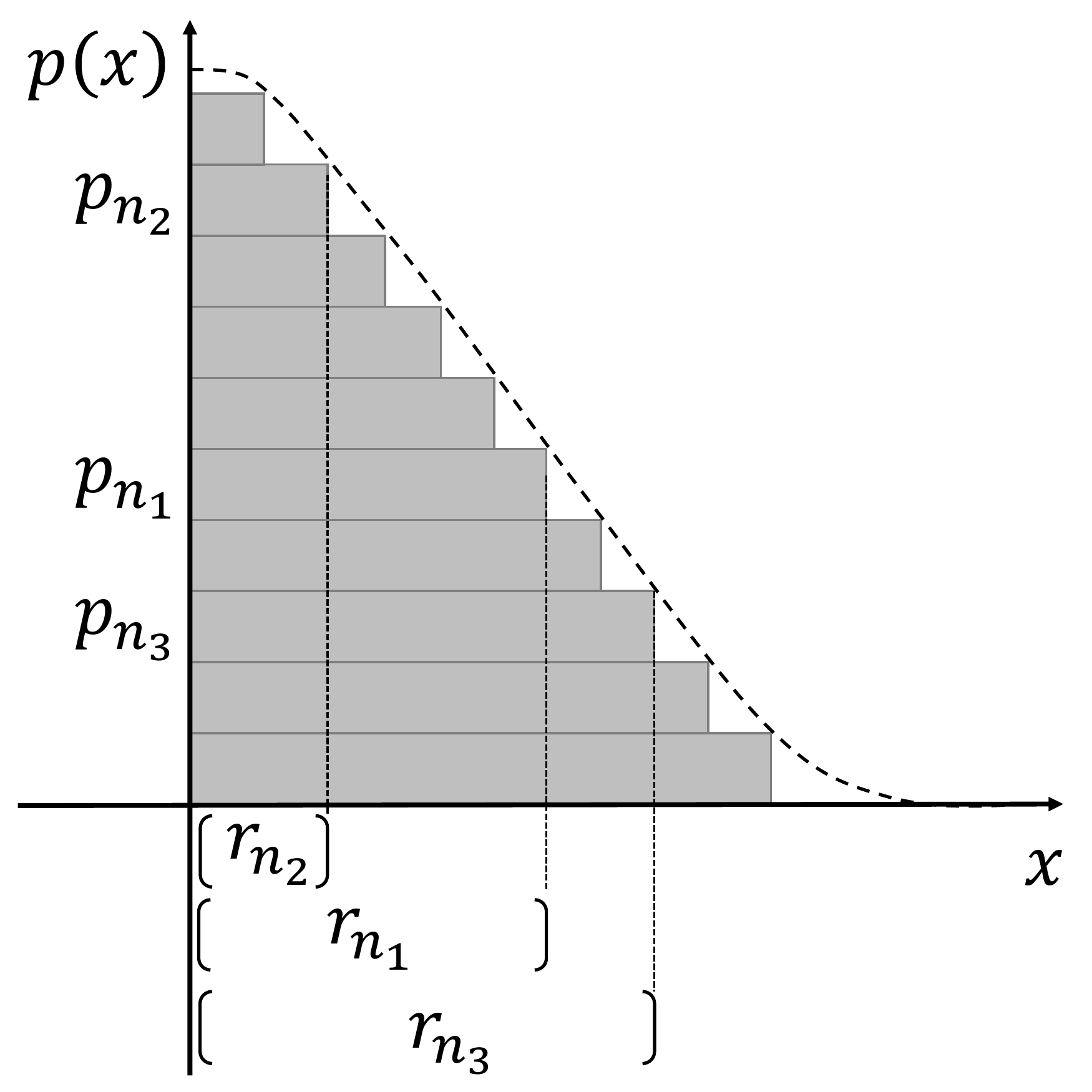}
	\caption{Rearrangement invariant distribution. The interval $r_{n_{i}}$, for all $n_{i}\in{\mathds{N}}$ and $i\in{D}$, is an ordered set. $n_{i}$ is totally ordered in space $X$ such that if $n_{i} \prec n_{j}$, then $f_{n_{i}}\left(x\right)\succ f_{n_{j}}\left(x\right)$.}
	\label{fig:fs:1100}
\end{figure}

\begin{dfn}[Topology Induced by Directed semi-norms] Let $X$ be a vector space. Also, let $P^{-}$ be a family of directed semi-norms defined on non-increasing convex  rearrangement-invariant $X^{-}$ of $X$ with a base corresponding to $p_{n}$, $n\in{\mathds{N}}$ in neighborhood of $0$ as
    \begin{equation}
        \centering
        p^{-}_{n} = \{x\in{X} \vert p^{-}\left(x\right)<2^{-n}\}
        \label{eq:bck_fs_1100}
    \end{equation}
    \label{dfn:bck_fs_1100}
\end{dfn}
Fig. \ref{fig:fs:1100} shows the example non-increasing rearrangement distribution. Obviously,
the following statement are equivalent
\begin{enumerate}
	\item $n_{1} \preccurlyeq n_{2}$
	\item $r_{n_{2}} \preccurlyeq r_{n_{1}} \Rightarrow \epsilon_{n_{2}}\mathcal{B}_{1}\subset \epsilon_{n_{1}}\mathcal{B}_{1}$
	\item $p_{n_{1}} \preccurlyeq p_{n_{2}}$
\end{enumerate}

The topology induced by $P^{-}_{0}$, $P^{-}_{0}\in{P^{-}}$, on $X^{-}$ is the coarsest topology on $X$. The family of directed semi-norms $P$ is first countable. Then, we can define define family of semi-norms for non-decreasing convex rearrangement $X^{-}$ as
\begin{equation}
	\begin{split}
    	P^{-} & = \sum^{N}_{i=1} p\left(X^{-}_{i}\right)\\
    	& \{p_{1},p_{1}+p_{2},p_{1}+p_{2}+p_{3},\cdots\}
    \end{split}
    \label{eq:bck_fs_1101}
\end{equation}
$P^{-}$ is a set one-to-one to power set 
\begin{equation}
	\centering
	\{X^{-}_{1},X^{-}_{2},X^{-}_{3}\cdots\} = \{\{\left|x^{-}_{1}\right|\},\{\left|x^{-}_{1}\right|,\left|x^{-}_{2}\right|\},\{\left|x^{-}_{1}\right|,\left|x^{-}_{2}\right|,\left|x^{-}_{3}\right|,\cdots\},\cdots\}
	\label{eq:bck_fs_1401}
\end{equation}
for all $x^{-}_{i}\in{X^{-}}$. 

If $X^{-}_{n^{\prime}}$ uniformly converges to $X^{-}$ for some $n^{\prime}$ such that $\|x^{-}_{i}-x^{-}_{j}\|\rightarrow 0$, $i\succ j\succ n^{\prime}$, then a directed net $P^{-}_{n^{\prime}}$ also converges to $P^{-}$ such that $\left|P^{-}_{i}-P^{-}_{j}\right|\rightarrow 0$, or equivalently for $X_{k},X_{l}\in{X}$, $k,l\in{N}$ where $x^{-}_{k} = x_{i}$ and $x^{-}_{l} = x_{j}$, then we have $p\left(x_{i}-x_{j}\right)\rightarrow 0$. Note that $X^{-}_{n^{\prime}}\triangleq X_{n}$ contains the $n$-strongest entries of $X$. The set 
\begin{equation}
	\centering
	V_{n} = \{x^{-}\in{X^{-}} \vert \bigcup^{n^{\prime}}_{i=1}p\left(x^{-}_{i}\right)\}
	\label{eq:bck_fs_1400}
\end{equation}
forms local convex bases for $X$. We should also pay attention that the semi-norm $p\left(x^{-}_{i}\right)$ is separable.

\begin{dfn}[Initial Definition of Fr\'echet Space] \cite[p. 167]{folland1999real} A complete Hausdorff topological vector space whose topology is defined by a countable family of semi-norms is called Fr\'echet space.
    \label{dfn:mth:ta:fr:1100}
\end{dfn}

Having define ingredients, we define the Fr\'echet space in the following.
\begin{dfn}[Fr\'echet space] Let $X$ be a Hausdorff locally convex space. Then, $X$ is Fr\'echet space if any of the following equivalent property are satisfied
    \begin{enumerate}
        \item $X$ is countable and complete.
        \item $X$ is metrizable and complete.
        \item The weak topology on $X$ is compatible with complete and translation invariant metric.
        \item $X$ is complete and a weak topology on $X$ is given by a countable family of semi-norms.
    \end{enumerate}
    \label{dfn:bck_fs_1100}
\end{dfn}
Note that the countable family of semi-norms implies Borel $\sigma$-algebra set defined on $X$, which makes $X$ metrizable. In order to approximate $X^{-}$, Fr\'echet space is defined on a restricted family of semi-norms ${P^{-}\vert}_{X_{n}}$. Applying Hahn-Banach theorem, ${P^{-}\vert}_{X^{-}_{n}}$ is continuously extended to $P^{-}$. Accordingly, from Reisz representation theorem \ref{thm:mth_reg_meas_intro_1100}, there is a restricted linear functional ${\Lambda\vert}_{X^{-}_{n}}$ acting on $X^{-}_{n}$ and continuously extends to $\Lambda$ acting on $X^{-}$.

\begin{dfn}[Fr\'echet Pseudo-Metric Space] Having countable family of separable semi-norms $P^{-} = \{P^{-}_{1},P^{-}_{2},\cdots\}$ in non-increasing format where $P^{-}_{i} = p^{-}_{1}+p^{-}_{2}+\cdots+p^{-}_{i}$, a pseudo-metric\footnote{Metric space distance between two point is zero if the two points are identical, i.e., metric space is non-degenerative. Pseudo-metric space is generalization of metric space to degenerative spaces such that the distance between two \textit{distinct} points can also be zero. Obviously, Cauchy net is pseudo-metric space.} for $x^{-}_{j},x^{-}_{i}$ and $n^{\prime} \prec i \prec j$, $i,j,n^{\prime}\in{\mathds{N}}$, is defined as
\begin{equation}
    \centering
    d\left(x^{-}_{i},x^{-}_{j}\right) = \sum^{d}_{i=1} c_{i}\frac{\left|P^{-}_{i}-P^{-}_{j}\right|}{1+\left|P^{-}_{i}-P^{-}_{j}\right|}
    \label{eq:bck_fs_1102}
\end{equation}
	\label{dfn:bck_fs_1400}
\end{dfn}
In general, $c_{i}>0$ is unknown. It is common in functional analysis to set $c_{i} = 2^{-i}$. Pseudo-metric $d\left(x,y\right)$ has three important properties that are of interest in this letter: (1) it maps $\left[0,\infty\right]$ to $\left[0,1\right]$, (2) $d\left(x,y\right)$ induces a topology on $X$ the same as $P^{-}$, and (3) given an optimum subspace $X^{-}_{n}$, the total variation approaches zero as
\begin{equation}
    \centering
    V^{T}_{d} = \sup_{X^{-}}d\left(x^{-}_{i},X^{-}_{j}\right) \text{, for all } x^{-}\in{X}^{-}_{j} 
\end{equation}
such that $\left|P_{i}-P_{j}\right| \rightarrow 0$, and $X^{-}_{j} = {x^{-}_{j},x^{-}_{j+1},\cdots,x^{-}_{d}}$, Then, $X^{-}_{n^{\prime}} = \{x^{-}_{1},x^{-}_{2},\cdots,x^{-}_{n^{\prime}}\}$ is the optimal subspace corresponding to index $n^{\prime}$ in Definition \ref{dfn:bck_fs_1400} with a countable family of separable semi-norms $P^{-}_{n^{\prime}} \subset {P^{-}}$ and $P^{-}_{n^{\prime}} = \{p^{-}_{1},p^{-}_{1}+p^{-}_{2},p^{-}_{1}+p^{-}_{2}+p^{-}_{3},\cdots,p^{-}_{1}+p^{-}_{2}+\cdots+p^{-}_{n^{\prime}}\}$.

So far, we have proved that $X^{-}_{n^{\prime}}\triangleq X_{n}$ for some $n^{\prime}\in{D}$, and $n\in{\mathds{N}}$. We show that for uniform convergence of $X^{-}_{n^{\prime}} \rightarrow X^{-}$, $n^{\prime}$ is measured as a dimension for which the total variation of pseudo-metric $d\left(x^{-}_{i},x^{-}_{j}\right)$ approaches zero. Uniform convergence of $X^{-}_{n^{\prime}}$ to $X^{-}$ also proves that $d\left(x_{i},x_{j}\right)$ contains subsequences that are sequentially complete for the corresponding Cauchy nets $\{X^{-}_{1},X^{-}_{2},\cdots,X^{-}_{n}\}$. 

We end this section by defining reflexive Fr\'echet space which is essential for the proof of reflexive homeomorphism in section \ref{m:mth:vcb_fr}.
\begin{thm}[Reflexivity in Fr\'echet Space] Let $X$ and $\left(X^{\ast}\right)^{\ast}$ be Fr\'echet space. Then, the following statements are equivalent 
    \begin{enumerate}
        \item $\Delta_{\lambda_{f}\left(\alpha\right)}: X \rightarrow \left(X^{\ast}\right)^{\ast}$ is evaluation map.
        \item Cauchy nets define on $X$ and $\left(X^{\ast}\right)^{\ast}$ are equivalent in the form of uniformly convergent over compact subsets.
    \end{enumerate}
\end{thm}
\begin{IEEEproof}
Proof directly follows from \ref{thm:mth_dist_intro_1103}.
\end{IEEEproof}
\subsection{Homeomorphic Relation of Compressible Topological Vector Space}\label{m:mth:hr_cs_kle}
As we have discussed, compressibility can be considered as sparsity in the linear functional acting on the signal. In fact, the sparsity appears to be fundamental property of the distribution, e.g., as given in \eqref{eq:bck_rit_1101}. As a result, we formulate the sparse representation in the following
\begin{equation}
	\centering
	\Lambda: X \hookrightarrow X^{\ast}
	\label{eq:lcs:1400}
\end{equation}
The representation in \eqref{eq:lcs:1400} is a modification over conventional linear functional $\Lambda:X \rightarrow Y$, which provides additional information about the mapping performed with respect to a certain distribution obtained from \eqref{eq:bck_rit_1101}. We aim to show that the relation in \eqref{eq:lcs:1400} is doubly reflexive. The forward relation can be defined using an evolution map such that 
\begin{equation}
	\centering
	\Delta: X \rightarrow \left(X^{\ast}\right)^{\ast}
	\label{eq:lcs:1401}
\end{equation}

The inverse relation also can be define by identity inclusion map as
\begin{equation}
	\centering
	\Delta^{-1}: \left(X^{\ast}\right)^{\ast} \rightarrow X
	\label{eq:lcs:1405}
\end{equation}

The discussion in section \ref{m:bck:fdos} and, in particular, the conclusion in Theorem \ref{thm:fdos:11010} enables us to purpose the reflexive homeomorphic relation for the compressible topological vector space in the following.

\begin{thm}[Reflexive Homeomorphic Relation of Compressible Topological Vector Space]
Let $\Delta_{\lambda_{f}\left(\alpha\right)}:X \underset{\lambda_{f}\left(\alpha\right)}{\rightarrow} \left(X^{\ast}\right)^{\ast}$ be a linear functional on compressible topological vector space $X$, separating dual space $X^{\ast}$, and double dual space $\left(X^{\ast}\right)^{\ast}$. Also, let $X^{-}=\{V,U\}$ be a non-increasing rearrangement-invariant transform of $X$ where $V$ and $U$ are given with respect to \eqref{eq:fdos:1099}. Then, we can find $X^{\#}\left(t,\omega\right)$ and $\widehat{X}$ that take points from $\left(V^{\ast}\right)^{\ast}$ and $\left(U^{\ast}\right)^{\ast}$, respectively. Then, $\Delta_{\lambda_{f}\left(\alpha\right)}$ has to satisfy the following symmetry\footnote{Symmetry has been defined with respect to the invariant volume.},\footnote{The candidate algorithm for $\Delta_{\lambda_{f}\left(\alpha\right)}$ that satisfies the symmetry in \eqref{eq:mth:ta:h:1098a} will be proposed in section \ref{m:mth:sgca}.}

    \begin{equation}
        \begin{split}
            & \mathds{E}\{X^{\#}\left(t,\omega\right)+\widetilde{X}\}\leq
            \mathds{E}\{\underbrace{\left|X^{\#}\left(t,\omega\right)+\widetilde{X}\right|}_{X^{L}}\}\\
            & \mathds{E}\{X^{\#}\left(t,\omega\right)+\widetilde{X}\}\leq \sup_{j}\sum^{N}_{i=1} \left|X^{L}_{ij}\right| = \|X^{L}\|_{1,\infty}
            \end{split}
        \label{eq:mth:cs_kle:kle_rnd:1101b}
  \end{equation}
  Considering the relation between $\|X^{L}\|_{1,\infty}$ and a topological space $X$, we can write:
  \begin{equation}
      \centering
      \|X^{L}\|_{1,\infty} \thicksim \|X\|_{1,\infty}
      \label{eq:mth:cs_kle:kle_rnd:1102}
  \end{equation}
  where the homeomorphism is provided through a non-increasing rearrangement-invariant transform preserves the volume of the locally convex space $X$. With respect to the Hahn-Banach extension and separation theorems, $X^{\#}\left(t,\omega\right)$ is the diagonalization obtained from $\Delta_{\lambda_{f}\left(\alpha\right)}$, and $\widetilde{X}$ is the continuous spectral extension of $\Delta_{\lambda_{f}\left(\alpha\right)}$ operator \footnote{In this work, the bases of compressible topological vector space coincide with the orthonormal Hilbert-space basis. Diagonalization is followed, accordingly.}. Then, $\Delta_{\lambda_{f}\left(\alpha\right)}$ leads to reflexive homeomorphism
     \begin{equation}
        \begin{split}
        & \|X^{\Delta}\left(t\right)\|_{1,\infty} \longmapsto \|X^{L}\|_{1,\infty}\\
        & \|X^{L}\|_{1,\infty} \longmapsto \|X\|_{1,\infty}
        \end{split}
        \label{eq:mth:ta:h:1099}
    \end{equation}
Rephrasing \eqref{eq:mth:ta:h:1099} as 
    \begin{equation}
        \Delta^{-1}: \|X^{\Delta}\left(t\right)\|_{1,\infty} \longmapsto \|X\|_{1,\infty}
        \label{eq:mth:ta:h:1096b}
     \end{equation}
and
    \begin{equation}
        \Delta: \|X\|_{1,\infty} \longmapsto \|X^{\Delta}\left(t\right)\|_{1,\infty} 
        \label{eq:mth:ta:h:1096a}
     \end{equation}
Then, transition low provides 
    \begin{equation}
        \centering
        \|X\|_{1,\infty} \thicksim \|X^{\Delta}\left(t\right)\|_{1,\infty}
        \label{eq:mth:ta:h:1098}
    \end{equation}
From \eqref{eq:mth:ta:h:1098}, it follows that 
	\begin{equation}
        \centering
        \Delta_{\lambda_{f}\left(\alpha\right)}= \Delta^{-1}\Delta
        \label{eq:mth:ta:h:1098a}
    \end{equation}
Then, $\Delta_{\lambda_{f}\left(\alpha\right)}$ is linear functional that enables homeomorphism with respect to invariant volume in general. The relation in \eqref{eq:mth:ta:h:1098} is reflexive homeomorphism as the forward homeomorphism is provided through invariant $\Delta$ and inverse homeomorphism is due to invariant $\Delta^{-1}$.
    \label{thm:mth:ta:h:1099}
\end{thm}

\begin{IEEEproof}
In order to prove Theorem \ref{thm:mth:ta:h:1099}, we need to prove the forward and inverse homeomorphisms. Their proof are give in sections \ref{m:mth:fr}, \ref{m:mth:fr}, and consequently, reflexive homeomorphism is proved in section \ref{m:mth:vcb_fr} by demonstrating that all Cauchy nets defined on $X$ and $\left(X^{\ast}\right)^{\ast}$ converge to their limits in $X_{n}$ and $\left(X^{\ast}_{n}\right)^{\ast}$ simultaneously.
\end{IEEEproof}

\begin{rmk} The linear relation of the compressed sensing, $Y\left(t,\omega\right) = \Lambda\left(\omega\right) X\left(t\right)+e$, is neither homeomorphic nor isomorphic. However, Theorem \ref{thm:mth:ta:h:1099} defines the homeomorphic equivalence relation between $\Delta: X\rightarrow \left(X^{\ast}\right)^{\ast}$ through the Theorem \ref{thm:fdos:11010}. 
    \label{rmk:mth:ta:h:1100}
\end{rmk}

The continuity of linear functional is essential for the existence of homeomorphic and to establish the fact that the metric spaces defined on LHS and RHS of the Theorem \ref{thm:mth:ta:h:1099} preserve the structural properties of the mathematical object. 

\begin{thm}[Closure of Orthogonal Subspace of $X^{\ast}$] Let $X$ be a topological vector space with dual space $X^{\ast}$ that separates points on $X$. Also, let $V\in{X^{\ast}}$ and $U\in{X^{\ast}}$ be orthogonal and complement orthogonal. The closure of orthogonal subspace $V$ is
    \begin{equation}
        \begin{split}
        	& \overline{V} = \left(V^{\perp}\right)^{\perp}\\
			& \Rightarrow \overline{\Lambda X_{n}} = \mathcal{N}\left(\Lambda^{\ast}\right)^{\perp}	        	
        \end{split}
        \label{eq:mth:ta:ts:1102}
    \end{equation}
    \label{thm:mth:ta:ts:1102}
\end{thm}

\begin{rmk} The closure of orthogonal subspace of $X^{\ast}$ is analogous to the closed linear subspac of topological vector space. 
    \label{rmk:mth:ta:ts:1101}
\end{rmk}

\begin{rmk} Let $K^{-}_{1}\subset{K^{-}_{2}}\subset{K^{-}_{3}}\subset \cdots \subset{K^{-}_{n}}\subset{\overline{K^{-}}}\subset{E}\subset{X^{-}}$ be a compact pre-image for linear functional $\Lambda$. Then, there is a compact image $\left(\left(K^{\ast}_{1}\right)^{\ast}\right)^{-}\subset{\left(\left(K^{\ast}_{2}\right)^{\ast}\right)^{-}}\subset{\left(\left(K^{\ast}_{3}\right)^{\ast}\right)^{-}}\subset \cdots \subset{\left(\left(K^{\ast}_{n}\right)^{\ast}\right)^{-}}\subset{\overline{\left(\left(K^{\ast}\right)^{\ast}\right)^{-}}}\subset{U}\subset{\left(\left(X^{\ast}\right)^{\ast}\right)^{-}}$. We define pseudo-metric space $\left(X,d_{X}\right)$ such that $d_{X_{n}}$ on $K^{-}_{n}\subset{X^{-}}$ uniformly converges over compact subsets to $d_{X}$. Similarly, we define $\left(\left(X^{\ast}\right)^{\ast},d_{\left(X^{\ast}\right)^{\ast}}\right)$ such that $d_{\left(X^{\ast}_{n}\right)^{\ast}}$ converges uniformly over compact subnets to $d_{\left(X^{\ast}\right)^{\ast}}$. If weak-topology defined on $X$ separates points on $X^{\ast}$ and weak$^{\ast}$-topology defined on $X^{\ast}$ separates points on $X$, then, for invariant linear functional $\Lambda\in{X^{\ast}}$, we can conclude that the distances between coordinates of $\left(X^{\ast}\right)^{\ast}$ is the invariant scaled distance between coordinates of $X$ and vice-versa. Consequently, we have: (1) pseudo metric $d_{\left(X^{\ast}\right)^{\ast}}$ is compatible with weak-topology defined on $X$, and (2) pseudo metric $d_{X}$ is compatible with weak$^{\ast}$-topology defined on $\left(X^{\ast}\right)^{\ast}$.
	\label{rmk:mth:ta:h:1102}
\end{rmk}

\begin{thm} [Uniform Continuity of $\Delta_{\lambda_{f}\left(\alpha\right)}$] Let $X$ be a compact locally convex topology with double dual space $\left(X^{\ast}\right)^{\ast}$ where sequentially compact subsequences $K_{i}$ and $\left(K^{\ast}_{i}\right)^{\ast}$ are defined as in the Remark \ref{rmk:mth:ta:h:1102}. There is a family of convergence balls $\mathfrak{B}_{\delta}\left(x_{i}\right)$ of radius $\delta_{i}$ in the neighborhood of $x_{i}\in{X}$, for all $i\in{D}$, with a corresponding family of convergence balls $\mathfrak{B}_{\epsilon}\left(\left(x^{\ast}_{i}\right)^{\ast}\right)$ of radius $\epsilon$ at the neighborhood of $\left(x^{\ast}_{i}\right)^{\ast}\in{\left(X^{\ast}\right)^{\ast}}$ such that:
    \begin{equation}
        \centering
        \mathfrak{B}_{\delta}\left(x_{n}\right) = \{x_{n}\in{X}, \ | \ d_{X_{n}}\left(f_{n}-f\right)\leq\delta\}
        \label{eq:mthd_osl1100}
    \end{equation}
    \begin{equation}
        \centering
        \mathfrak{B}_{\epsilon}\left(\left(x^{\ast}_{n}\right)^{\ast}\right) = \{\left(x^{\ast}_{n}\right)^{\ast}\in{\left(X^{\ast}\right)^{\ast}} \ | \  d_{\left(X^{\ast}_{n}\right)^{\ast}}\left(\left(f^{\ast}_{n}\right)^{\ast}-\left(f^{\ast}\right)^{\ast}\right)\leq\epsilon\}
        \label{eq:mthd_osl1101}
    \end{equation}
    \label{thm:mth:ta:h:1103}
\end{thm}
where $f$ and $\left(f^{\ast}\right)^{\ast}$ are the supremum semi-norms defined on $X$ and $\left(X^{\ast}\right)^{\ast}$.

Theorem \ref{thm:mth:ta:h:1103} represents the collection of the convex balanced local basis for weak topology induced by $X_{n}$ (resp. $\left(X^{\ast}_{n}\right)^{\ast}$) on $X$ and $\left(X^{\ast}\right)^{\ast}$. Similarly, there is a family of directed semi-norms $f^{-}$ and $\left(\left(f^{\ast}\right)^{\ast}\right)^{-}$ corresponding to $f$ and $\left(f^{\ast}\right)^{\ast}$ that satisfy the uniform convergence conditions in Theorem \ref{thm:mth:ta:h:1103} as
\begin{equation}
	\centering
    \mathfrak{B}_{\delta}\left(x^{-}_{n}\right) = \{x^{-}_{n}\in{X^{-}_{n}}, \ | \ d_{X^{-}_{n}}\left(f^{-}_{n}-f^{-}\right)\leq\delta\}
    \label{eq:mthd:ta:l406}
\end{equation}
\begin{equation}
    \centering
    \mathfrak{B}_{\epsilon}\left(\left(\left(x^{\ast}_{n}\right)^{\ast}\right)^{-}\right) = \{\left(\left(x^{\ast}_{n}\right)^{\ast}\right)^{-}\in{\left(\left(X^{\ast}\right)^{\ast}\right)^{-}} \ | \  d_{\left(\left(X^{\ast}_{n}\right)^{\ast}\right)^{-}}\left(\left(\left(f^{\ast}_{n}\right)^{\ast}\right)^{-}-\left(\left(f^{\ast}\right)^{\ast}\right)^{-}\right)\leq\epsilon\}
    \label{eq:mthd:ta:l407}
\end{equation}
The convergence balls in \eqref{eq:mthd:ta:l406} and \eqref{eq:mthd:ta:l407} satisfy the condition (refer to Fig. \ref{fig:fs:1100} and the properties therein.)
\begin{equation}
	\centering
	\mathfrak{B}_{\delta}\left(x^{-}_{3}\right) \subset \mathfrak{B}_{\delta}\left(x^{-}_{1}\right) \bigcap \mathfrak{B}_{\delta}\left(x^{-}_{2}\right) 
	\label{eq:mthd:ta:l408}
\end{equation}
for $f^{-}_{1}\prec f^{-}_{3}$ and $f^{-}_{2} \prec f^{-}_{3}$. 

The generalization of \eqref{eq:mthd:ta:l408} can be formulated as: The intersection of all the bases in a ball of radii $\delta_{1}\prec \delta_{2} \prec \cdots \prec \delta_{n}$ provides a set of bases for $X^{-}_{n}$ corresponding to sequentially compact $K^{-}_{1}\subset{K^{-}_{2}}\subset \cdots \subset{K^{-}_{n}}\subset{X^{-}}$ as the weakest topology on $X$. In other words, a weak-compact topology on compressible topological vector space $X$ can be defined with respect to invariant volume of locally convex space $X$. In the same way, the intersection of all the bases of radii $\epsilon_{1}\prec \epsilon_{2} \prec \cdots \prec \epsilon_{n}$ provides a set of bases for $\left(X^{\ast}_{n}\right)^{\ast}$ corresponding to $\left(\left(K^{\ast}_{1}\right)^{\ast}\right)^{-}\subset{\left(\left(K^{\ast}_{2}\right)^{\ast}\right)^{-}}\subset \cdots \subset{\left(\left(K^{\ast}_{n}\right)^{\ast}\right)^{-}}\subset{\left(\left(X^{\ast}\right)^{\ast}\right)^{-}}$ as the weak$^{\ast}$-topology for $\left(\left(X^{\ast}\right)^{\ast}\right)^{-}$. Then, a set of locally convex bases in the neighborhood of $x^{-}_{i}$ and $\left(x^{\ast}_{i}\right)^{-}$ are
\begin{equation}
	\centering
	V^{-}_{i} = \{x^{-}_{i}\in{K_{i}} \vert \mathfrak{B}_{\delta}\left(x^{-}_{i}\right)<\delta_{i}\} \text{, } \forall{i\in{I}}
	\label{eq:mthd:ta:l409}
\end{equation}
\begin{equation}
	\centering
	\left(\left(V^{\ast}_{i}\right)^{\ast}\right)^{-} = \{\left(\left(x^{\ast}_{i}\right)^{\ast}\right)^{-}\in{\left(\left(K^{\ast}_{i}\right)^{\ast}\right)^{-}} \vert \mathfrak{B}_{\epsilon}\left(\left(\left(x^{\ast}_{i}\right)^{\ast}\right)^{-}\right)<\epsilon_{i}\} \text{, } \forall{i\in{I}}
	\label{eq:mthd:ta:l409}
\end{equation}
\subsection{Fundamental of Compressed Sensing Topological Analysis}
\begin{dfn}[Family of Separable Semi-norms] Let $X$ be an $n$-dimensional compressible topological vector space lying in $\mathds{R}^{d}$ with non-increasing rearrangement $X^{-}$. Let $\Omega\in{X^{-}}$ be an open set in a Euclidean space defined as the union of a finite number of nonempty compact subsets $\Omega = \{K^{-}_{1},K^{-}_{2},\cdots\}$ such that $K^{-}_{1}\subset{K^{-}_{2}}\cdots\subset{K^{-}_{i}}\cdots\subset{\overline{K}}$ for $i\in{\mathds{N}}$. Also let $f:\mathds{C}_{c}\left(\Omega\right) \rightarrow \mathds{R}^{+}\bigcup \{0\}$. $C_{c}\left(\Omega\right)$ can be approximated by $C_{c}\left(K_{i}\right)$ as the set of all continuous linear functional with compact support where $n$ is the dimension of compactness. The family of semi-norms defines a locally convex space for all $x^{-}\in{X^{-}}$ as 
\begin{equation}
    \centering
    \|f\|_{\left(K^{-}_{i},\infty\right)} = \sup_{\substack{i\in{\mathds{N}}\\
    x = \{K^{-}_{i}\}}} 
    \left|\mathbf{D}^{\alpha}f\left(x\right)\right|
    \label{eq:mth:ta:ts:1097}
\end{equation}
where $\mathbf{D}$ is a derivative operator and $\alpha = \left(\alpha_{1},\alpha_{2},\cdots,\alpha_{n}\right)$ such that $\left|\alpha\right|=\sum_{i}\alpha_{i}$.

In order to measure the coordinates of $X^{-}$ for which the total variation vanishes, we generate a family of direct semi-norms as $\left(P^{-},\prec\right)$
\begin{equation}
	\begin{split}
	P^{-} & = \sum^{d}_{i=1} \|f\|_{K_{i},\infty}\\
	& = \Bigg\{\|f\|_{K^{-}_{1},\infty},\|f\|_{K^{-}_{1},\infty}+\|f\|_{K^{-}_{2},\infty},\|f\|_{K^{-}_{1},\infty}+\|f\|_{K^{-}_{2},\infty}+\|f\|_{K^{-}_{3},\infty},\cdots, \\
	& \|f\|_{K^{-}_{1},\infty}+\|f\|_{K^{-}_{2},\infty}+\cdots+\|f\|_{K^{-}_{n},\infty}\Bigg\}
	\end{split}
	\label{eq:mth:ta:ts:1401}
\end{equation}
$P^{-}$ generates a smoothly and monotonically increasing series that saturates at some points. In fact, function $f\left(x\right)$ should be selected such that $P^{-}$ generates a smoothly increasing and saturating graph. In this work, we fix $f\left(x\right)\subset{U}$ with $\alpha = 2$. The larger $\alpha$ will suppress the total variation even further.

Considering the sequentially compact structure of the $\{K^{-}_{1},K^{-}_{2},\cdots,K^{-}_{i},\cdots,K^{-}_{n}\}$ for $i\in{\mathds{N}}$, the uniform convergence can be generalized as uniform convergence over a compact subset or equivalently as 
\begin{equation}
    \centering
    \mathfrak{B}_{n,\epsilon}\left(x\right) = \{\sup_{i\in{\left[1,n\right]}} 
    \{\left|\mathbf{D}^{\alpha}f\left(K_{i+1}\right)-\mathbf{D}^{\alpha}f\left(K_{i}\right)\right|<\epsilon\}
     \label{eq:mth_osl_1209}
\end{equation}
The argument of $\sup$ function measures where the the supremum of the derivative of total variation vanishes. \textit{That is, the optimum subspace dimension does not change anymore.}
    \label{dfn:mth:ta:ts:1098}
\end{dfn}

The main question we aim to answer is what $n$ is? More important, how can we measure the inner regularity to compute optimum total variation vanishing point when there are many candidate points that can potentially satisfy the uniform convergence condition in \eqref{eq:mth_osl_1209}? Knowing $n$, one can determine the $dim\left(\mathcal{N}\left(\Lambda\right)^{\perp}\right) = dim\left(\overline{\Lambda X}\right)$.

The orthogonal subspace $\overline{\Lambda X} = X^{\ast}\slash\mathcal{N}\left(\Lambda^{\ast}\right)_{\perp}$ can be obtained using the $\ell_{1}$ optimization or greedy algorithms. The point we are interested in here is how $\Lambda X$ coincides with the Radon measure discussed in section \ref{m:mth:rg}. In particular, the orthogonal subspace $\Lambda^{\ast}\in{\left(X^{\ast}\right)^{\ast}}$ that satisfies $X^{\ast}\slash\mathcal{N}\left(\Lambda^{\ast}\right)_{\perp}$ represents the inner regularity over the compact $K$ subset of $\sigma$-algebra set $\mathcal{B}$. Accordingly, the orthogonal complement subspace $\mathcal{N}\left(\Lambda^{\ast}\right)\in{X^{\ast}}$ represents the outer regularity measured on $\mathcal{B}$.

\begin{dfn}[Compact Continuous Functions $C_{c}\left(K\right)$]Let $X$ be a topological vector space. Also, let $\Omega\in{X}$ be an open set in a Euclidean space defined as the union of a finite number of countable and nonempty compact subsets such that $K_{1}\subset{K_{2}}\cdots\subset{K_{n}}\subset{\Omega}$ for $i\in\left[N\right]$ and $n \ll N$. That is every $K_{i}$ is the interior of the $K_{i+1}$. Also, let $\Lambda:\mathds{C}\left(\Omega\right) \rightarrow Y$. Then, $C\left(\Omega\right)$ can be replaced by $C_{c}\left(K_{n}\right)$ where $C_{c}\left(K_{n}\right)$ is the set of all continuous linear functional with compact support. In the remaining of the paper, we drop the index $n$ denoting the family of all compact continuous functions in Fr\'echet space as $C_{c}\left(K\right)$.
    \label{dfn:mth:ta:fr:1102}
\end{dfn}
 
\begin{dfn}[Equicontinuity] Let $d$ be a distance function, and $f:X \rightarrow \mathds{R}$ be a metric defined on $X$. Also, assume that $K^{\ast}\subset{C\left(X\right)}$ and $f^{\ast}\in{K^{\ast}}$. For all $x,y\in{X}$, the equicontinuity is defined as
    \begin{equation}
        \centering
        d\left(x,y\right)<\delta \Longrightarrow \left|f^{\ast}\left(x\right)-f^{\ast}\left(y\right)\right|<\epsilon
        \label{eq:mth:ta:ts:1106}
    \end{equation}
    where $\delta$ is \textit{uniformly the same} for all $f^{\ast}\in{K^{\ast}}$.
    \label{dfn:mth:ta:ts:1101}
\end{dfn}

\begin{thm}[Arzel\'a-Ascoli Theorem for Compact Subset of Linear Functional] \cite[Theorem 2.38]{einsiedler2017functional} Let $\left(X,d\right)$ be a metric space. Then $K^{\ast}\subset{C\left(X\right)}$ is compact if and only if $K^{\ast}$ is closed, bounded, and equicontinuous. For all linear functional $\Lambda_{1},\Lambda_{2},\cdots, \Lambda_{n}\in{K^{\ast}}$
    \begin{equation}
        \centering
        K^{\ast}\subseteq{\bigcup^{N}_{i=1}\mathfrak{B}_{\epsilon}\left(\Lambda_{i}\right)}
         \label{eq:mth_osl_1203a}
    \end{equation}
    where $\mathfrak{B}_{\epsilon}\left(\Lambda_{i}\right)$ is an open ball of fixed $\Lambda\in{K}$. Then, for all $x_{1},x_{2}\in{X}$ 
    \begin{equation}
        \centering
            \mathfrak{B}_{\epsilon}\left(\Lambda_{i}\right) = \{\Lambda_{i}\left(x\right)\in{K^{\ast}} \vert \|\Lambda\left(x_{2}\right)-\Lambda_{i}\left(x_{1}\right)\|_{\infty}\leq \epsilon \}
        \label{eq:mth_osl_1203}
    \end{equation}
	for $\epsilon\in{\mathds{R}^{+}\cup \{0\}}$.
    \label{thm:mth:ta:ts:1103}
\end{thm}

\begin{thm}[Arzel\'a-Ascoli on Metric Space]\cite[Corallary 1.8.25]{tao2010epsilon} Let $f:X \rightarrow Y$ be a separating semi-norm defined on compact metric space $X$. Also, let $f_{j}:X \rightarrow Y$ be a countably finite sequence of semi-norms in $f$, where $j\in{\mathds{N}}$. Then, there is an $n\in{\mathds{N}}$ and $j \preccurlyeq n$ and corresponding subsequence $f_{j}$ of $f_{n}$ that compactly converges to a $f$. That is, for certain $n$, then, $\left|f-f_{n}\right|<\epsilon$.
\label{thm:mth:ta:ts:1104}
\end{thm}

\begin{dfn}[Equivalent Topologies] The topologies $\mathscr{T}_{1}$ and $\mathscr{T}_{2}$ induced by linear functional $\Lambda$ and semi-norm $f$ are equivalent if the following property is satisfied.
	\begin{enumerate}
		\item For all $x\in{X}$, either $\Lambda x \geq \mu\left(x\right)$ or $\Lambda x \leq \mu\left(x\right)$.
		\item Let $X^{-}$ be a non-increasing rearrangement-invariant transform of $X$, where $X$ can be any sequence of numbers. Also, assume that $\Lambda\left(x^{-}\right)$ and $\mu\left(x^{-}\right)$, for all $x^{-}\in{X^{-}}$, take their entries from $X^{-}$ with respect to directed sets $D_{\Lambda}$ and $D_{\mu}$. Then, $D_{1} = D_{2}$ point-wise with the same order.
		\item Two toplologies uniformly converge over the equivalent directed nets.
		\item Every non-empty open set of $\mathscr{T}_{1}$ contains a non-empty set of $\mathscr{T}_{2}$, and every non-empty open set of $\mathscr{T}_{2}$ contains a non-empty set of $\mathscr{T}_{1}$.
		\item $\mathscr{T}_{1}$ and $\mathscr{T}_{2}$ are equidecomposible, that is, given partitions of open sets $A = \bigcup^{n}_{i=1} A_{i}$ and $B = \bigcup^{n}_{i=1} B_{i}$ on $\mathscr{T}_{1}$ and $\mathscr{T}_{2}$, then operator norm equivalence is satisfied as
		\begin{equation}
			\centering
			\|\Lambda\|_{Y} = C\|\mu\|_{Y}
			\label{eq:mth:ta:ts:1404}
		\end{equation}
		where $C\in{\mathds{R}^{+}}$ is a constant. Given distribution functions $\lambda_{\Lambda}\left(\alpha\right)$ and $\lambda_{\mu}\left(\alpha\right)$, \eqref{eq:mth:ta:ts:1404} implies 
		\begin{equation}
			\centering
			\lambda_{\Lambda}\left(\alpha^{\prime}\right) = \lambda_{\mu}\left(\alpha\right) \text{, for some } \alpha,\alpha^{\prime}\in{\mathds{R^{+}}}
			\label{eq:mth:ta:ts:1405}
		\end{equation}	
	\end{enumerate}
	\label{dfn:mth:ta:ts:1400}
\end{dfn}
\subsection{Froward Relation of Homeomorphism (Compressed Representation)}\label{m:mth:fr}
\begin{figure}[t!]
	\centering
	\includegraphics[width = 3.0in]{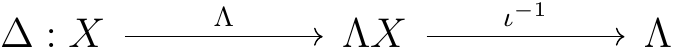}
	\caption{Forward relation of the reflexive homeomorpshism.}
	\label{fig:mth:ta:fr:1101}
\end{figure}

In order to prove the homeomorphic relation \eqref{eq:mth:ta:h:1098} proposed in Theorem \ref{thm:mth:ta:h:1099}, we need to show the bidirectional topological relation between RHS and LHS. In the following, we define metrizability on the LHS sides of the \eqref{eq:mth:ta:h:1096a} using idea of dual space, and will prove the homeomorphism in the forward direction. Fig. \ref{fig:mth:ta:fr:1101} shows the forward relation of the reflexive homeomorphism with respect to linear functional $\Lambda: X\rightarrow Y$.

\begin{thm} [Forward Relation] Let $X$ be a compressible locally convex space with sequentially compact subsets $K\subset{X}$ and corresponding dual space $K^{\ast}\subset{X^{\ast}}$ that separates point on $X$. Then, the RHS of the forward relation in \eqref{eq:mth:ta:h:1098} has the same internal structure as LHS. The relation that preserves the structure is called forward homeomorphic relation between $X$ and $X^{\ast}$
\begin{equation}
    \centering
    \|X\|_{1,\infty} \longmapsto \|X^{\Delta}\left(t\right)\|_{1,\infty}
    \label{eq:mth:ta:fr:1099}
\end{equation}
where $\Delta$ is the approximating algorithm operator acting on $X$ with respect to random process path $\lambda_{f}\left(\alpha\right)$. 
    \label{thm:mth:ta:fr:1099}
\end{thm}
The proof is given at the end of this section.

\begin{dfn}[Weak Topology on $X$] Let $X$ be topological vector space with dual space $X^{\ast}$ whose separates points on $X$. The weak topology $\mathscr{T}_{w}$ defined on $X$ is the coarsest topology on $X$ that makes $X^{\ast}$ continuous. That is, at neighborhood of $x_{0}\in{X}$ there is $\epsilon\in{\mathds{R}^{+} \cup \{0\}}$ and $n\in{\mathds{N}}$ such that:
    \begin{equation}
        \centering
        N_{\Lambda_{1},\cdots\Lambda_{n} \ | \epsilon}\left(x_{0}\right)= \bigcap^{n}_{i=1}\{x\in{X}\ | \ \left|\Lambda_{i}x-\Lambda_{i}x_{0}\right|<\epsilon\}
        \label{eq:mth:ta:fr:1100}
    \end{equation}
    \label{dfn:mth:ta:fr:1100}
\end{dfn}
Note that $N_{\Lambda_{1},\cdots\Lambda_{n} \ | \epsilon}\left(x_{0}\right)$ describes decaying behavior of the components at the neighborhood, mainly at the tail, of $x_{0}$.

It is obvious that for finite-dimensional vector space, the weak-topology defined on $X$ with respect to $\Lambda_{i}$, $i\in{\mathds{N}}$, and norm topology are the same. However, \textit{one still needs weak-topology on finite-dimensional compressible signal to (1) emphasize on sequentially compactness of the $X$, and (2) to define projective topology that approximates the compressible topological vector space $X$ over $n$ sequentially compact subsets}. On the other hand, semi-infinite and infinite-dimensional compressible signals $X$ (resp. $X^{\ast}$) only satisfies weak topology (resp. weak$^{\ast}$ topology). The compactness is quantified by $n$, which is directly proportional to the compressibility of vector space $X$. Consequently, we define the weak topology on compressible signal $X$ in the following definition.

\begin{dfn}[Weak Topology on Compressible Vector Space] Let $X\in{\mathds{K}^{n}}$ be a compressible topological vector space lies in $\mathds{K}^{d}$, $d \gg n$, and $X^{\ast}$ be a dual space that separates points on $X$. The weakest topology $\mathscr{T}_{w}$ defined on $X$ is the topology that satisfies one of the following properties:
    \begin{enumerate}[(a)]
        \item Makes the $X^{\ast}$ continuous as defined in \eqref{eq:mth:ta:fr:1100}.
        \item Satisfy the RIP condition.
        \item Minimizes the compressive $m$-width considering sequentially compact subsets $K^{\ast}_{1}\subset{K^{\ast}_{2}}\subset\cdots\subset{K^{\ast}_{n}}\subset{X^{\ast}}$ (resp. $K_{1}\subset{K_{2}}\subset\cdots\subset{K_{n}}\subset{X}$) by uniformly converging over compact subsets to $X^{\ast}$ (resp. $X$).
    \end{enumerate}
    \label{dfn:mth:ta:fr:1101}
\end{dfn}
Definition \ref{dfn:mth:ta:fr:1101} brings us to the following theorem on the metrizability of the compressible vector space.

\begin{thm}[Metrizibile Compressible Topological Vector Space] Let $X$ be a compact topological vector space with dual space $X^{\ast}$ that separates point $X$. Then, $X$ is metrizable as the union of countable finite number of Borel sets.
    \label{thm:mth:ta:fr:1100}
\end{thm}

\begin{figure}[t!]
	\centering
	\includegraphics[width = 4.0in]{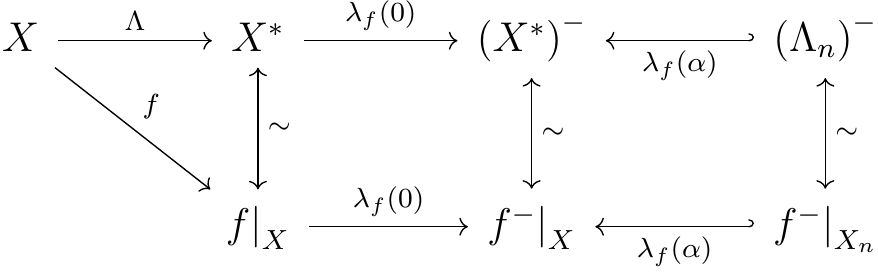}
	\caption{The first and second rows summarize the implication of dual space $X^{\ast}$ and $\widehat{X}^{\ast}$ through weak-compact topology $\mathscr{T}_{w}$ and measure space $\left(X,f\left(X\right)\right)$. The equivalent relation denoted by $\thicksim$ will be proved later in the section \ref{m:mth:vcb_fr}.}
	\label{fig:mth:ta:fr:1100}
\end{figure}

Compact continuous functions $C_{c}\left(K\right)$ in Definition \ref{dfn:mth:ta:fr:1102} provides several favorable properties, here. Two of the most important ones are formulated in the next two theorems. The Arzel\'a-Ascoli theorem first is applied to the a family of semi-norms $f$ defined on $X$ to prove the compactly convergence of $X$ with respect to $f$. Then, it is applied to the linear functional $\Lambda\in{X^{\ast}}$ to prove the compactly convergence of $\Lambda$ over $X^{\ast}$. Finally, the uniqueness of the Riesz representation theorem will be used to show that compactly convergence of $f$ implies uniquely convergence of $\Lambda_{n}$ to $\Lambda$. Diagram in Fig. \ref{fig:mth:ta:fr:1100} summarizes the Lemmas \ref{lm:mth:ta:fr:1101} and \ref{lm:mth:ta:fr:1102}.

\begin{lm}[Applying Arzel\'a-Ascoli on Measure Space $\left(X,f\left(X\right)\right)$] Let $X$ be a compressible topological vector space induced with weak-compact topology $\mathscr{T}_{w}$ and compatible with separable function $f:X \rightarrow \widehat{X}^{\ast}\left(X\right)$. Let $K_{n}$ be a sequentially compact, bounded, and equicontinuous subsets of $X$ such that $K_{1}\cdots\subset{K_{n}}\subset{\overline{\mathfrak{B}_{1}\left(X\right)}}$, that is, $K_{i}$ lies in the interior of $K_{i+1}$. Then, there is $\widehat{K}^{\ast}_{1}\subset{\widehat{K}^{\ast}_{2}}\subset\cdots\subset{\widehat{K}^{\ast}_{n}}\subset{\overline{\mathfrak{B}_{\epsilon}\left(\widehat{X}^{\ast}\right)}}$ such that as $K_{n}$ uniformly converges over compact subsets to $X$, then $f_{n}\coloneqq {f\vert}_{K_{n}}$ uniformly converges over compact subsets to $f\left(X\right)$. The sequentially compact subsets $\{\widehat{K}^{\ast}_{1},\widehat{K}^{\ast}_{2},\cdots,\widehat{K}^{\ast}_{n}\} = \{\{f_{1}\},\{\{f_{1},f_{2}\},\{f_{1},f_{2},\cdots,f_{n}\}\}$ all are in $\widehat{X}^{\ast}$.
     \label{lm:mth:ta:fr:1101}
\end{lm}
\begin{IEEEproof}
Let $X^{-}$ be a non-increasing rearrangement-invariant transform of $X$. We can find a sequence of $f^{-}_{i}\left(x\right)$, $i\in{D}$. Then, there is $K^{-}_{1}\subset{K^{-}_{2}}\subset\cdots\subset{K^{-}_{n}}\subset{\overline{\mathfrak{B}_{\epsilon}\left(X^{-}\right)}}$.  By Corollary \ref{crl:bck_rit_1100}, $\overline{\mathfrak{B}_{\epsilon}\left(X^{-}\right)} \equiv \overline{\mathfrak{B}_{\epsilon}\left(X\right)}$. Finally, the proof follows from sequentially completeness and \eqref{eq:lcs:1406}.
\end{IEEEproof}

\begin{lm}[Applying Arzel\'a-Ascoli on Weak-Compact Topology $\mathscr{T}_{w}$] Let $X$ be a compact topology with dual space $X^{\ast}$ that separates points on $X$. Also, let $\Lambda:X \rightarrow X^{\ast}\left(X\right)$ be a linear functional acting on $X$.
Let $K_{n}$ be a sequentially compact, bounded, and equicontinuous subset of $X$ such that $K_{1}\subset{K_{2}}\cdots\subset{K_{n}}\subset{\overline{\mathfrak{B}_{\epsilon}\left(X\right)}}$, that is $K_{i}$ lies in the interior of $K_{i+1}$. Then, there is $K^{\ast}_{1}\subset{K^{\ast}_{2}}\cdots\subset{K^{\ast}_{n}}\subset{\overline{\mathfrak{B}_{\epsilon}\left(X^{\ast}\right)}}$ such that as $K_{n}$ uniformly converges over compact subsets to $X$, then $\Lambda_{n} \coloneqq {\Lambda\vert}_{K_{n}}$ uniformly converges over compact subsets to $\Lambda\left(X\right)$. The sequentially compact subsets $\{K^{\ast}_{1},K^{\ast}_{2},\cdots,K^{\ast}_{n}\} = \{\{\Lambda_{1}\},\{\{\Lambda_{1},\Lambda_{2}\},\{\Lambda_{1},\Lambda_{2},\cdots,\Lambda_{n}\}\}$ all are in $X^{\ast}$.
    \label{lm:mth:ta:fr:1102}
\end{lm}
\begin{IEEEproof}
The proof is similar to proof of Lemma \ref{lm:mth:ta:fr:1101}. 
\end{IEEEproof}

For locally convex space $X$ with dual space $X^{\ast}$ that separates points on $X$, Lemma \ref{lm:mth:ta:fr:1102} states that there is convex combination of finite number of $\Lambda_{i}\in{X^{\ast}}$ for $i\in{[1,n]}$ that converges uniformly to $X^{\ast}$. 

Proof of Theorem \ref{thm:mth:ta:fr:1099} is given in the following.
\begin{IEEEproof}

In order to prove \ref{thm:mth:ta:fr:1099}, (a) we need to show that Lemma \ref{lm:mth:ta:fr:1101} and Lemma \ref{lm:mth:ta:fr:1102} are equivalent. (b), we need to show that there is an open ball at the neighborhood of $\Lambda_{i}\in{X^{\ast}}$ as the intersection of local bases in $X^{\ast}$ with respect to $\|\cdot\|_{\infty}$ \eqref{eq:mth_osl_1203}. And (c), we need to prove LHS and RHS of \eqref{eq:mth:ta:h:1098} are equivalent.

(a) Considering the uniqueness of the Riesz-representation theorem, from \ref{lm:mth:ta:fr:1101} and \ref{lm:mth:ta:fr:1102}, we can conclude that for locally convex space $X$ with compact topology, both $X$ and dual space $X^{\ast}$ share the same Euclidean structure measured by $f$ and are compatible with continuous linear functional $\Lambda$. 
The Riesz representation theorem indicates that the weak topology defined on $X$ by $K^{\ast}\in{C_{c}\left(K\right)}$ can be uniquely determined by a locally defined $f_{K}\in{C_{c}\left(K\right)}$ generated by Borel set $\{f_{1} = \mathcal{B}_{1},f_{2} = \mathcal{B}_{2},\cdots,f_{n} = \mathcal{B}_{n}\}$.

(b) From Arzel\'a-Ascoli theorem and Lemma \ref{lm:mth:ta:fr:1102}, sequentially compact subset $K^{\ast}_{n}\subset{X^{\ast}}$ is the closed inside the intersection of local bases at the neighborhood $\Lambda_{i}$ of $X^{\ast}$ for all $i\in{[1,n]}$ defined in \eqref{eq:mth_osl_1203a}.

(c) The compactness of $X^{\ast}$ indicates that the compressible topological vector space $X$ and $X^{\ast}$ are in the $L_{1}$-space. Also, the Arzel\'a-Ascoli theorem indicates that the $X$ and $X^{\ast}$ are in the $L_{\infty}$-space. Considering these, we can conclude that the equivalence relation between LHS and RHS lies in the $L_{1,\infty}$ space. Finally, the Kolmogorov extension theorem indicates the existence of a distribution corresponding to algorithm $\Delta$ for which the LHS and RHS in \eqref{eq:mth:ta:fr:1099} are equidistibution. And, this proves Theorem \ref{thm:mth:ta:fr:1099}.
    \label{prf:mth:ta:fr:1100}
\end{IEEEproof}
\subsection{Inverse Relation of Homeomorphism (Compressed Recovery)}\label{m:mth:rr}
\begin{figure}[t!]
	\centering
	\includegraphics[width = 3.25in]{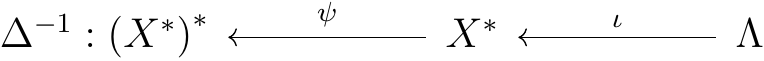}
	\caption{Inverse relation of the reflexive homeomorpshism.}
	\label{fig:mth:ta:rr:1101}
\end{figure}

In this section, we prove the inverse homeomorphic relation in Theorem \ref{thm:mth:ta:h:1099}. In order to homeomorphism to be satisfied, both compact subset of $K_{n}\subset{X}$ and $K^{\ast}_{n}\subset{X^{\ast}}$ have to be continuous. The continuity of the compact pre-image is satisfied through weak topology induced on $X$ by $X_{n}$, and then by Riesz extension theorem, as we have studied in section \ref{m:mth:fr}. The continuity of the image will be studied in the following. Fig. \ref{fig:mth:ta:rr:1101} shows the inverse relation the reflexive homeomorpshism with respect to linear function $\Lambda: X\rightarrow Y$.

\begin{thm}[Inverse Homoemorphism] Let $X$ be a compressible topological vector space with weak topology with $X^{\ast}$ that is equipped with weak$\ast$-compact topology. The relation that reflects the structure of RHS onto the LHS in \eqref{eq:mth:ta:rr:1099} is called inverse homeomorphic relation from $X^{\ast}$ to $X$.
\begin{equation}
    \centering
    \|X\left(t\right)\|_{1,\infty} 
    \underset{c}{\reflectbox{$\longmapsto$}}
    \|X^{\Delta}\left(t\right)\|_{1,\infty}
    \label{eq:mth:ta:rr:1099}
\end{equation}
where $\underset{c}{\reflectbox{$\longmapsto$}}$ refers to inverse homeomrphism over compact subset.
    \label{thm:mth:ta:rr:1099}
\end{thm}
The proof is given at the end of this section.

\begin{dfn}[Weak$^{\ast}$ Topology] \cite[Definition 8.4]{einsiedler2017functional} Let $X$ be a topological vector space with dual space $X^{\ast}$ that separates\footnote{The fact that $X^{\ast}$ separates points on $X$ is essential to make $X$ continuous.} points on $X$. Also, let $\Lambda \rightarrow \Lambda x$ be an evaluation map that evaluates every $x\in{X}$ on $X^{\ast}$. Then, the weak$^{\ast}$ topology is the weakest topology on $X^{\ast}$ that makes $\Lambda \rightarrow \Lambda x$ for all $x\in{X}$. This implies the continuity of $X^{\ast}$ at the neighborhood of $\Lambda_{0}$ for some $\epsilon\in{\mathds{R}^{+} \cup \{0\}}$ and for all $x\in{X}$ such that:
    \begin{equation}
        \centering
        N_{x_{1},\cdots,x_{n}|\epsilon}\left(\Lambda_{0}\right) = \bigcap^{n}_{i=1}\{\forall{\Lambda}\in{X^{\ast}} \ | \ \left|\Lambda x-\Lambda_{0} x\right|<\epsilon\}
        \label{eq:mth:ta:rr:1100}
    \end{equation}
    \label{dfn:mth:ta:rr:1100}
\end{dfn}

\begin{rmk} Weak$^{\ast}$ topology defined on $X^{\ast}$ separates points on $X^{\ast}$, for all $x\in{X}$ \cite[Section 3.14]{rudin1991functional}. In addition, it guarantees continuity of $X^{\ast}$ space at the neighborhood of $\Lambda_{0}\in{X^{\ast}}$ as formulated in \eqref{eq:mth:ta:rr:1100}.
    \label{rmk:mth:ta:rr:1100}
\end{rmk}

In the following theorem, the compactness property of the weak$^{\ast}$ topology has been explained. The Banach-Agao\u{g}lu theorem is essential to formulate compressible topological vector space.

\begin{thm}[Banach-Agao\u{g}lu Theorem] Let $X^{\ast}$ be a dual space of separable topological vector space $X$. Also, let $f:X \rightarrow \mathds{R}$ be a separable metric defined on $X$. Then,
    \begin{enumerate}[(a)]
        \item Closed unit ball
            \begin{equation}
                \centering
                \overline{\mathfrak{B}^{X^{\ast}}_{1}} = \{\Lambda\in{X^{\ast}} \ | \ \|\Lambda x\|\leq 1\}
                \label{eq:mth:ta:rr:1101}
            \end{equation}
        is weak$^{\ast}$-compact in $X^{\ast}$ \cite[Section 3.14]{rudin1991functional} \cite[Theorem 8.10]{einsiedler2017functional}.
        \item If $\overline{\mathfrak{B}^{X^{\ast}}_{1}}\in{X^{\ast}}$ and $\overline{\mathfrak{B}^{X^{\ast}}_{1}}$ is a weak$^{\ast}$-compact topology, then $\overline{\mathfrak{B}^{X^{\ast}}_{1}}$ is metrizable in weak$^{\ast}$ topology \cite[Section 3.15]{rudin1991functional}.
\end{enumerate}
    \label{thm:mth:ta:rr:1100}
\end{thm}

\begin{rmk}[$X$ Separates Subspaces in $X^{\ast}$] From the Banach-Agao\u{g}lu theorem, we conclude that $X$ separates subspaces in $X^{\ast}$. The compactness is an important specification because without separability, we can not decompose\footnote{It does not need to be necessary optimal decomposition.} $X$ to subspaces with respect to a countable number of linear functionals $\Lambda_{n}\in{X^{\ast}}$.
	\label{prf:mth:ta:rr:1401}
\end{rmk}

\begin{figure}[t!]
	\centering
	\includegraphics[width = 4.0in]{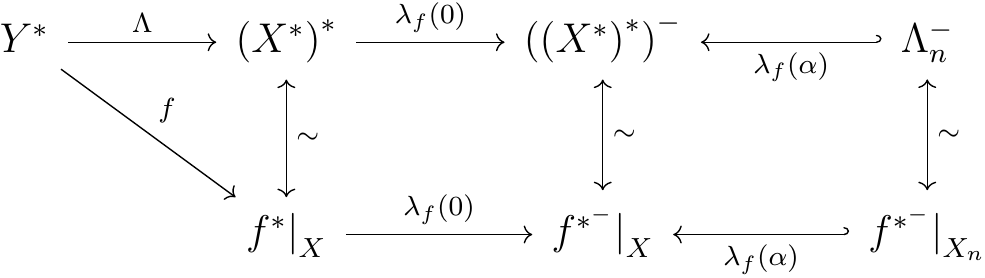}
	\caption{The first and second rows summarize the implication of double dual space $\left(X^{\ast}\right)^{\ast}$ through weak$^{\ast}$-compact topology $\mathscr{T}^{\ast}_{w}$ and measure space $\left(Y^{\ast},f^{\ast}\left(Y^{\ast}\right)\right)$. The equivalent relation denoted by $\thicksim$ will be proved later in section \ref{m:mth:vcb_fr}.}
	\label{fig:mth:ta:rr:1100}
\end{figure}

Now that we have the ingredients to apply the Arzel\'a-Ascoli Theorem to the weak$^{\ast}$-compact topology on $X^{\ast}$ and prove Theorem \ref{thm:mth:ta:rr:1099}. In the following, $\Omega^{\ast}$ is the image of the $\Omega\in{X}$. Under continuous linear functional $\Omega$ is the pre-image of $\Omega^{\ast}$. Diagram in Fig. \ref{fig:mth:ta:rr:1100} summarizes the Lemmas \ref{lm:mth:ta:rr:1103} and \ref{lm:mth:ta:rr:1102}.

\begin{lm}[Applying Arzel\'a-Ascloi Theorem on Measure Space $\left(Y^{\ast},f^{\ast}_{n}\left(Y^{\ast}\right)\right)$] Let $X$ be a compressible topological space equipped with a weak-compact topology. Assume that $\widehat{X}^{\ast}$ and $\left(\widehat{X}^{\ast}\right)^{\ast}$ are dual space and double dual space of $X$, both induced with a weak$^{\ast}$-compact topology. Also, let $f^{\ast}: Y^{\ast}\rightarrow \left(\left(\widehat{X}^{\ast}\right)^\ast\right)\left(Y^{\ast}\right)$. Assume that there is a sequentially compact, bounded, and equicontinuous subsets of $\widehat{X}^{\ast}$ such that $\widehat{K}^{\ast}_{1}\subset{\widehat{K}^{\ast}_{2}}\subset\cdots\subset{\widehat{K}^{\ast}_{n}}\subset{\overline{\mathfrak{B}_{\epsilon}\left(Y^{\ast}\right)}}$, that is, $\widehat{K}^{\ast}_{i}$ lies in the interior of $\widehat{K}^{\ast}_{i+1}$. Then, there is $\left(\widehat{K}^{\ast}_{1}\right)^{\ast}\subset{\left(\widehat{K}^{\ast}_{2}\right)^{\ast}}\subset\cdots\subset{\left(\widehat{K}^{\ast}_{n}\right)^{\ast}}\subset{\overline{\mathfrak{B}_{\epsilon}\left(\left(\widehat{X}^{\ast}\right)^{\ast}\right)}}$ such that as $\widehat{K}^{\ast}_{n}$ uniformly converges over compact subsets to $\widehat{X}^{\ast}$, then $f^{\ast}_{n} \coloneqq f^{\ast}\vert_{\widehat{K}^{\ast}_{n}}$ uniformly converges over compact subsets to $f^{\ast}\left(Y^{\ast}\right)$. The sequentially compact subsets $\{\left(\widehat{K}_{1}^{\ast}\right)^{\ast},\left(\widehat{K}_{2}^{\ast}\right)^{\ast},\cdots,\left(\widehat{K}_{n}^{\ast}\right)^{\ast}\} = \{\{f^{\ast}_{1}\},\{\{f^{\ast}_{1},f^{\ast}_{2}\},\{f^{\ast}_{1},f^{\ast}_{2},\cdots,f^{\ast}_{n}\}\}$ all are in $\left(\widehat{X}^{\ast}\right)^{\ast}$.
    \label{lm:mth:ta:rr:1103}
\end{lm}

\begin{IEEEproof}
The proof is similar to the proof of the Lemma \ref{lm:mth:ta:fr:1101}.
\end{IEEEproof}

Distinct $\left(\widehat{K}^{\ast}\right)^{\ast}$ and $\left(K^{\ast}\right)^{\ast}$ have been used to distinguish the range spaces of the linear functional space $\Lambda$ and function $f^{\ast}$.

\begin{lm}[Applying Arzel\'a-Ascloi Theorem on Weak$^{\ast}$-Compact $\mathscr{T}^{\ast}_{w}$] Let $X$ be a compressible topological space equipped with a weak-compact topology. Assume that $X^{\ast}$ and $\left(X^{\ast}\right)^{\ast}$ are dual space and double dual space of $X$, both induced with a weak$^{\ast}$-compact topology. Also, let $\Lambda^{\ast}: Y^{\ast}\rightarrow \left(\left(X^{\ast}\right)^\ast\right)\left(Y^{\ast}\right)$ be a set of linear functional acting on $Y^{\ast}$. Assume that there is a sequentially compact, bounded, and equicontinuous subsets of $X^{\ast}$ such that $K^{\ast}_{1}\subset{K^{\ast}_{2}}\subset\cdots\subset{K^{\ast}_{n}}\subset{\overline{\mathfrak{B}_{\epsilon}\left(Y^{\ast}\right)}}$, that is, $K^{\ast}_{i}$ lies in the interior of $K^{\ast}_{i+1}$. Then, there is $\left(K^{\ast}_{1}\right)^{\ast}\subset{\left(K^{\ast}_{2}\right)^{\ast}}\subset\cdots\subset{\left(K^{\ast}_{n}\right)^{\ast}}\subset{\overline{\mathfrak{B}_{\epsilon}\left(\left(X^{\ast}\right)^{\ast}\right)}}$ such that as $K^{\ast}_{n}$ uniformly converges over compact subsets to $Y^{\ast}$, then $\Lambda^{\ast}_{n} \coloneqq \Lambda^{\ast}\vert_{K^{\ast}_{n}}$ uniformly converges over compact subsets to $\Lambda^{\ast}\left(Y^{\ast}\right)$. The sequentially compact subsets $\{\left(K_{1}^{\ast}\right)^{\ast},\left(K_{2}^{\ast}\right)^{\ast},\cdots,\left(K_{n}^{\ast}\right)^{\ast}\} = \{\{\Lambda^{\ast}_{1}\},\{\{\Lambda^{\ast}_{1},\Lambda^{\ast}_{2}\},\cdots,\{\Lambda^{\ast}_{1},\Lambda^{\ast}_{2},\cdots,\Lambda^{\ast}_{n}\}\}$ all are in $\left(X^{\ast}\right)^{\ast}$.
    \label{lm:mth:ta:rr:1102}
\end{lm}

Proof of Theorem \ref{thm:mth:ta:rr:1099} is given in the following.
\begin{IEEEproof} Theorem \ref{thm:mth:ta:rr:1099} can be concluded by applying Riesz representation to Lemma \ref{lm:mth:ta:rr:1102} and Lemma \ref{lm:mth:ta:rr:1103}.
    \label{prf:mth:ta:rr:1102}
\end{IEEEproof}
\subsection{Reflexive Homeomorphism}\label{m:mth:vcb_fr}
In this section, first, we provide some basic updates on the convexity of the locally convex space focusing on Fr\'echet space. First, we discuss the closure of the convex hull and will explain the Krein-Milman Theorem applied to the compressible topological vector space. We make clear some of the concepts and theorems that we have not provided in previous sections. The definitions are included in the context they are used. Second, we conclude the reflexive homeomorphism by proving that the forward and inverse homeomorphisms converge over the equivalent Cauchy nets. Third, we discuss the metric to measure the optimum compactness. We will discuss that since the absorbing null space is the connected space that provides negligible Lebesgue measure, then Gelfand's constraint $K \bigcap V^{\perp}$ can be replaced by more proper constraint measured on the compressible signal.

We have emphasized widely in the previous sections that the non-increasing rearrangement-invariant space, in particular, weak $L_{1,\infty}\left(X\right)$ Lorentz space is crucial to define compressible topological vector space. In order to obtain direct nets of semi-norms, $f\left(x\right)$ acting on $X$ is reordered ($\preccurlyeq$) as non-increasing rearrangement-invariant. In Lemma \ref{lm:bck_rit_1100}, it has been proved that the volume of the locally convex body will be preserved by rearrangement-invariant transform. The resultant rearranged compressible topological vector space $X^{-}$ and semi-norms $f^{-}\left(x\right)$ (resp. $\left(X^{\ast}\right){-}$ and $\left(f^{-}\right)^{\ast}\left(x\right)$) have Lizorkin distribution. Since $f\left(x\right)$ converges locally and monotonically to its limit, $\mathbf{D}^{\alpha}f^{-}\left(x\right)$ (resp. $\left(f^{-}\right)^{\ast}\left(x\right)$) has Lizorkin distribution.

We need to show that the compact convex subsets $K^{\ast}\subset{X^{\ast}}$ can be used for inner measurement of locally convex space $X^{\ast}$. For this purpose, \textit{we need to prove that $K^{\ast}$ is closed convex hull of its extreme points}. This formulates the problem upward monotone convergence that can be measured using inner regularity according to sequentially compact subsets of given topological vector space $X^{\ast}$. \footnote{From an algorithm point of view, the reconstruction error decreases monotonically over the sequentially compact subsets $K^{\ast}_{i}\subset{X^{\ast}}$ up to certain coordinates $i=n$.} This is equivalent to an optimum measurement over sequentially compact subset $K$ in compressible topological vector space $X$ for some compactness quantified by $n$. In fact, $K^{\ast}_{n}$ coincides with Gelfand's $n$-width $ \dim\left(K\bigcap V^{\perp}\right)$, $K\subset{X}$. \textit{We will show that $n$ obtained from the supremum of Fr\'echet metric is the optimum width for the inner measure over $X^{\ast}$ (resp. $X$)}. Since the compact subset $K_{n}$ is assumed to have closure with respect to $n$, the transition point is expected at $n$ for which the performance metric will start to decrease non-monotonically. The transition point will generate a skewed bell curve that increases rapidly for $K^{\ast}_{1}$ to $K^{\ast}_{n}$ and decreases for $i\succ n$. The coordinate $i=n$ that the transition point takes place gives the optimum compact subset $K^{-}_{n}$ that uniformly converges to $X^{-}$ (resp. $X$). Finally, it will be shown that given pseudo-metric \eqref{eq:mth_osl_1210}, the Fr\'echet distance metric generates a sequence space, in particular, K\"othe sequence that induces equivalent weak$^{\ast}$-compact topology for $\left(X^{\ast}\right)^{-}$. In section \ref{m:mth:fd}, we examine the Fr\'echet distance metric to measure the optimum subspace given the compressed sensing measurement vector.

\begin{dfn}[Extreme Point of Convex Subset $K$] \cite[Proposition 7.3]{conway1997course} Let $K$ be a convex subset of topological vector space $X$. Then, a point $a\in{K}$ is an extreme point of $K$ if it can not be represented as a convex combination of a points $x,y\in{K}$, i.e., $a\neq \theta x + \left(1-\theta\right)y$ for all $\theta\in{[0,1]}$. The extreme point $a\in{K}$ satisfies the following properties:
	\begin{enumerate}
		\item If $a=\frac{1}{2}\left(x+y\right)$, for all $x,y\in{X}$, then at least one of the points is not in $K$, i.e., $x\notin{K}$, or $y\notin{K}$, or $x = y = a$.
		\item If $a=\theta x +\left(1-\theta\right)y$, for all $x,y\in{X}$, then at least one of the points is not in $K$, i.e., $x\notin{K}$, or $y\notin{K}$, or $x = y = a$. For $\theta = \frac{1}{2}$, if $a = \frac{1}{2}\left(x+y\right)$, and if both $x,y\in{K}$, then $x = y = a$
		\item If $\{x_{1}, x_{2}, \cdots, x_{n}\}\in{K}$ and $a\in{\conv\{x_{1}, x_{2}, \cdots, x_{n}\}}$, then $a = x_{i}$ for $i\in{[0,1]}$.
		\item Let $\{a\}$ be a set of extreme points of $K$, then $X\slash\{a\}$ is an open convex set of $X$.
	\end{enumerate}   
\end{dfn}

\begin{thm}[Compact Convex Hull in Fr\'echet Space] \cite[Theorem 3.20]{rudin1991functional} If $X$ is Fr\'echet Space, i.e., Hausdorff locally convex space, and $K\subset{X}$ is compact, then $\overline{conv}\{K\}$ is compact.
    \label{thm:mth_dist_intro_1100}
\end{thm}
The next classical theorem establishes the relation between Fr\'echet space and closure of a compact convex subset $K\subset{X}$ with respect to the extreme points in $K$.

\begin{thm}[Krein-Milman Theorem] \cite[Theorem 3.20]{rudin1991functional} Let $X$ be a topological space with dual space $X^{\ast}$ that separates points on $X$. Then, non-empty compact convex set $K\subset{X}$ is the closed convex hull of the sets of extreme points of $K$, or equivalently $K=\overline{\conv}\left(E\left(K\right)\right)$, where $E\left(K\right)$ is the set of extreme points of $K$. 
    \label{thm:mth_dist_intro_1101}
\end{thm}

By Krein-Milman Theorem and the fact that $X^{\ast}$ separates points on $X$, we can find a set of linear functional $\Lambda_{n}\in{X^{\ast}}$ corresponding to $K\subset{X}$ that contains set of extreme points $E\left(K\right)$. In other words, $\Lambda_{n}$ generates closed convex hull for $X$. This is stated in the following corollary.

\begin{crl}[Optimum Subspace and Extreme Points] Let $X$ be a locally convex space equipped with weak topology. $X^{\ast}$ is the dual space of $X$ that separates points on $X$. Assume that $X^{\ast}$ is equipped with weak$^{\ast}$-compact topology that separates points $X$. Also, let $\overline{\mathfrak{B}_{1}\left(X^{\ast}\right)}$ be a closed unit ball with respect to Banach-Agao\u{g}lu Theorem \ref{thm:mth:ta:rr:1100}. Then, there is an optimum family of linear functional $\Lambda_{n}\in{\overline{\mathfrak{B}_{1}\left(X^{\ast}\right)}}$ such that 
    \begin{equation}
        \Lambda_{n} = \{\Lambda x = \Lambda E\left(K_{n}\right) \vert \Lambda\in{\overline{B_{1}\left(X^{\ast}\right)}} \text{, } x\in{\overline{K}} \text{, } K_{n}\subset{\overline{K}}\subset{X} \text{, } \overline{\conv}\left(K\right) = \overline{\conv}\left(E\left(K\right)\right) \}
        \label{eq:mth:ta:vcb_fr:1100}
    \end{equation}
for $i\in{\mathds{N}}$. Then, we can define the closure of $\overline{K^{\ast}}$ is defined as
	\begin{equation}
		\overline{\Lambda_{n}} = \overline{\conv}\left(E\left(K^{\ast}\right)\right)
		\label{eq:mth:ta:vcb_fr:1100a}
	\end{equation}
    \label{lm:mth_dist_intro_1102}
\end{crl}

\begin{rmk} From \eqref{eq:mth:ta:vcb_fr:1100a}, we can conclude that $\overline{\conv}\left(E\left(K^{\ast}_{n}\right)\right) = \mathcal{N}\left(\Lambda^{\ast}\right)_{\perp}$. In other words, the closure of the optimum subspace $K_{n}\in{X}$ is
	\begin{equation}
		\centering
			\overline{\conv}\left(K_{n}\right) = \left(\mathcal{N}\left(\Lambda^{\ast}\right)_{\perp}\right)^{\perp}
		\label{eq:mth:ta:vcb_fr:1100b}
	\end{equation}

	This is equivalent to say $\Lambda^{\ast}\left(y^{\ast}\right) = X_{n}$. \footnote{Remark \ref{rmk:mth_dist_intro_1400} perfectly fits the definition of the double dual space $\left(X^{\ast}\right)^{\ast}$ with an evaluation map $\phi:X\rightarrow \left(X{^\ast}\right)^{\ast}$. According to discussion in \cite[Section 4.5]{rudin1991functional}, $x^{\ast}_{i}\in{X^{\ast}}$, for all $x\i\in{\mathds{N}}$ is the collection of linear functional on $X$ that separate points on $X^{\ast}$. Then, $\Lambda^{\ast}$ satisfies \ref{eq:mth:ta:vcb_fr:1100b} refers to linear functional such that
	\begin{equation}
		\centering
			\langle x,x^{\ast} \rangle = \langle x^{\ast},\phi\left(x\right) \rangle
		\label{eq:mth:ta:vcb_fr:1100d}
	\end{equation}
	where $\phi$ is isometric isomorphic as
	\begin{equation}
		\centering
			\phi: X \rightarrow \left(X^{\ast}\right)^{\ast}
		\label{eq:mth:ta:vcb_fr:1100e}
	\end{equation}
	In other words, $\Lambda^{\ast}\left(x\right) \subset X_{n}$, where $\Lambda^{\ast}: Y^{\ast} \rightarrow \left(X^{\ast}\right)^{\ast}$}	
	\label{rmk:mth_dist_intro_1400}
\end{rmk}

\begin{thm}[Extreme Points and Closure of Convex Hull] \cite[Theorems 3.24 and 3.25]{rudin1991functional} Let $X$ be a locally convex space, and $K\subset{X}$ is compact. Then the following two state are equivalent:
    \begin{enumerate}[(a)]
        \item $K\subset{\overline{conv}\left(E\left(K\right)\right)}$
        \item $\overline{conv}\left(K\right)\subset{\overline{conv}\left(E\left(K\right)\right)}$
        \item If $x\in{\overline{\conv}\left(K\right)}$ then $x\in{\overline{K}}$ 
    \end{enumerate}
    \label{thm:mth_dist_intro_1101a}
\end{thm}

The $\overline{\conv}\left(K\right)$ agrees the $k$-face neighborhood of polytopes \cite{donoho2006high} where the extreme points of the convex sets construct the vertices of the $k$-face neighborhood surfaces of $d$-polytope.

\begin{thm}[Reflexive Homeomorphism in Fr\'echet Space - Unified Measure Spaces, Weak-Compact and Weak$^{\ast}$-Compact Topologies] Let $X$ be a locally convex space with a weak topology $\mathscr{T}_{w}$ that is compatible with measure space $\left(X,f\left(K\right)\right)$. Also, let $X^{\ast}$ be a dual space of $X$ with weak$^{\ast}$-compact topology $\mathscr{T}_{w^{\ast}}$ compatible with measure space $\left(X^{\ast},f^{\ast}\left(K\right)\right)$. Then, 
    \begin{equation}
        \centering
        \|X\|_{1,\infty}
        \thicksim 
        \|X^{\Delta}\|_{1,\infty}
        \label{eq:mth:ta:vcb_fr:1099}
    \end{equation}
\eqref{eq:mth:ta:vcb_fr:1099} is reflexive homeomorphism in Fr\'echet space, if one of the following three statements is satisfies
    \begin{enumerate}
        \item Weak- and weak$^{\ast}$-compact typologies induced by measure spaces and linear functionals on $X$ and $X^{\ast}$ are equivalent such that
        \begin{equation}
            \centering
            \mathscr{T}_{w} \thicksim \left(X,f\left(K\right)\right) \thicksim \mathscr{T}^{\ast}_{w} \thicksim \left(X^{\ast},f^{\ast}\left(K^{\ast}\right)\right)
            \label{eq:mth:ta:vcb_fr:1101}
        \end{equation}
    \item There exist $\left(X^{\ast}\right)^{\ast}$ such that $\iota: X \hookrightarrow \left(X^{\ast}\right)^{\ast}$ is evaluation map. Equivalently, one can check for $\langle x,x^{\ast}\rangle = \langle x^{\ast}, \left(x^{\ast}\right)^{\ast}\rangle$.
    \item There is an evaluation map $\iota: \Lambda \rightarrow \Lambda x$ that evaluates all points $x\in{X}$ using the linear functional $\Lambda$.
    \end{enumerate}
    \label{thm:mth_dist_intro_1103}
\end{thm}

Proof of Theorem \ref{thm:mth_dist_intro_1103} has three steps. First, using inner regularity, we show that 
    \begin{equation}
            \centering
            \mathscr{T}_{w} \thicksim \left(X,f\left(K\right)\right)
            \label{eq:mth:ta:vcb_fr:1101a}
    \end{equation}
Then, in the second step, we use the inner regularity to prove
    \begin{equation}
            \centering
            \mathscr{T}^{\ast}_{w} \thicksim \left(X^{\ast},f^{\ast}\left(K^{\ast}\right)\right)
            \label{eq:mth:ta:vcb_fr:1101a}
    \end{equation}
and finally, we prove $K_{n} \thicksim K^{\ast}_{n}$. The equivalence of $K$ and $K^{\ast}$ indicates that the weak and weak$^{\ast}$ topology defined on $X$ and $X^{\ast}$ are inducing the same topological space.

We formulate Lemma \ref{lm:mth:ta:cvb_fr:1100} in the following to establish the relation between the weak topology defined on $X$ and $\|f\left(x\right)\|_{1,\infty}$ over compact subsets of $K\subset{\Omega}\subset{X}$. Directed nets and non-increasing rearrangement-invariant semi-norms will be represented in slightly different forms but they are completely consistent with the previous formulation.

\begin{lm}[Inner Regularity Compatible with Weak-Compact topology] Let $X$ be a compact locally convex space with dual space $X^{\ast}$ separates points on $X$. If measure space $\left(X,f_{n}\left(K_{n}\right)\right)$ uniformly converges over compact subspace to measure space $\left(X,f\left(\Omega\right)\right)$ where $f\left(X\right)$ is identifiable using inner regularity, i.e., measure space $\left(X,f_{n}\left(K_{n}\right)\right)$ is compact with respect to Tychonoff's theorem. Then, there is a compact $K_{n}\in{X}$ identifiable with a weak-compact topology on $X$. The Reisz representation theorem guarantees the existence of weak$^{\ast}$-topology defined on linear functional in the closed convex hull of $\overline{\conv}\left(K_{n}\right)\in{X}$ closed in unit ball $\Lambda_{n}\subset{\overline{\mathfrak{B}_{1}\left(X^{\ast}\right)}}$ compatible with measure space $\left(X,f_{n}\left(K_{n}\right)\right)$.
    \label{lm:mth:ta:cvb_fr:1100}
\end{lm}

Lemma \ref{lm:mth:ta:cvb_fr:1100} establishes the bridge between the Arzel\'a-Ascoli Theorem applied to measure space $\left(X,f\right)$ in Lemma \ref{lm:mth:ta:fr:1101} and to weak-compact topology on $X$ in Lemma \ref{lm:mth:ta:fr:1102} by showing that both Lemmas measure the inner regularity of compressible topological vector space. 

We formulate Lemma \ref{lm:mth:ta:cvb_fr:1101} in the following to establish the relation between the weak$^{\ast}$-compact defined on $X^{\ast}$ and semi-norms $\|f^{\ast}\left(x^{\ast}\right)\|_{1,\infty}$ over compact subsets of $K^{\ast}_{n}\subset{X^{\ast}}$ with respect to the Riesz representation theorem.

\begin{lm} [Inner Regularity Compatible with Weak$^{\ast}$-Compact topology] Let $X$ be a compact locally convex space with dual space $X^{\ast}$ separate points in $X$. Assume that $X$ and $X^{\ast}$ are equipped with a weak-compact and a weak$^{\ast}$-compact topologies, respectively. If measure space $\left(X^{\ast},f^{\ast}_{n}\left(K^{\ast}_{n}\right)\right)$ uniformly converges over compact subspace to measure space $\left(X^{\ast},f^{\ast}\left(X^{\ast}\right)\right)$ where $f^{\ast}\left(X^{\ast}\right)$ is identifiable using inner regularity, i.e., measure space $\left(X^{\ast},f_{n}\left(K^{\ast}_{n}\right)\right)$ is complete with respect to Tychonoff's theorem, then, there is compact $K^{\ast}_{n}\in{X^{\ast}}$ identifiable with weak$^{\ast}$-compact topology on $X^{\ast}$. The Reisz representation theorem guarantees the existence of weak$^{\ast}$-topology defined on linear functional in closed convex hull of $K^{\ast}_{n}\in{X^{\ast}}$ closed in unit ball $\Lambda_{n}\subset{\overline{\mathfrak{B}_{1}\left(X^{\ast}\right)}}$ compatible with measure space $\left(X^{\ast},f_{n}^{\ast}\left(K^{\ast}_{n}\right)\right)$.
    \label{lm:mth:ta:cvb_fr:1101}
\end{lm}

We finalize the proof of the Theorem \ref{thm:mth_dist_intro_1103} in the following by showing the inner regularity Lemmas in \ref{lm:mth:ta:cvb_fr:1100} and \ref{lm:mth:ta:cvb_fr:1101} are equivalent.
 
\begin{IEEEproof}
    \label{prf:mth:ta:rr:1102}
In the following, we use the notation $X_{n}\in{X}$ and $X^{\ast}_{n^{\prime}}\in{X^{\ast}}$. We aim to prove $n = n^{\prime}$.
In order to prove Theorem \ref{thm:mth_dist_intro_1103}, we need to show that for a sequentially compact $K_{n}$ and $K^{\ast}_{n^{\prime}}$, then  $K^{\ast}_{n^{\prime}}\thicksim K_{n}$. We can show that in fact $X^{\ast} \rightarrow \left(X^{\ast}\right)^{\ast} \hookrightarrow X$. This leads to evaluation map $X \hookrightarrow \left(X^{\ast}\right)^{\ast}$ and establishes the reflexive homeomorphism between $X$ and $\left(X^{\ast}\right)^{\ast}$. For this purpose, we are employing directed nets where $X$ and $X^{\ast}$ are ordered in the form of non-increasing rearrangement. We need to show the followings:
\begin{quote}
    Let $\left(D_{X},\preccurlyeq\right)$ and $\left(D_{X^{\ast}},\preccurlyeq\right)$ be directed nets on topological vector space $X$ and $X^{\ast}$. Then, Theorem \ref{thm:mth:reg:reg:1100} implies ${\left(D_{X},\preccurlyeq\right) \thicksim \left(D_{X^{\ast}},\precsim\right)}$.
\end{quote}

Let $X$ be a topological vector space with weak$^{\ast}$-compact topology $\mathscr{T}_{w}$ defined on $X$. Also let $X^{\ast}$ be dual space of $X$ with weak$^{\ast}$-compact topology $\mathscr{T}^{\ast}_{w}$. It is obvious that $X$ and $X^{\ast}$ can be represented as direct sets $\|X\|_{1,\infty}$ and $\|X^{\ast}\|_{1,\infty}$\footnote{$\|X^{\ast}\|_{1,\infty} = \sup_{x^{\ast}\in{X^{\ast}}}\|x^{\ast}\left(x\right)\|$ is the operator norm.}. Note that the non-increasing rearrangement-invariant transform coincides perfectly with the well-ordered $X$ (resp. $X^{\ast}$) as sequentially compact subsets. 
The directed net on a typical directed set $\left(D_{X},\precsim\right)$ is defined as 
\begin{equation}
	f:D \rightarrow Z
	\label{eq:mth:ta:vcb_fr:1500}
\end{equation}
In our formulation, $D$ is a domain that takes its value from topological space. In \eqref{eq:mth:ta:vcb_fr:1500}, $f$ (1) can refer to linear function in $\left(X^{-}\right)^{\ast}$ acting on $X^{-}$ as $f_{\Lambda}:X^{-} \rightarrow a^{-}_{\Lambda}$ where $a^{-}_{\Lambda} = \{\{\Lambda^{-}_{1}\},\{\Lambda^{-}_{1},\Lambda^{-}_{2}\},\cdots,,\{\Lambda^{-}_{1},\Lambda^{-}_{2},\cdots,\Lambda^{-}_{N}\}\}$, (2) refer to the Lebesgue measure defined on $f: X^{-} \rightarrow a^{-}_{X}$ where $a^{-}_{X} = \{\mathds{R}^{+} \cup \{0\}\}$ is a Cauchy net with members from Lebesgue measure on $X^{-}$, or (3) be the Lebesgue measure defined on $f:\left(X^{-}\right)^{\ast} \rightarrow a^{-}_{X^{\ast}}$ where $a^{-}_{X^{\ast}} = \{\mathds{R}^{+} \cup \{0\}\}$ is a Cauchy net with members from Lebesgue measure on $X^{\ast}$. \textit{We aim to prove that ${\left(D_{X_{n}},\preccurlyeq\right) \thicksim \left(D_{X^{\ast}_{n}},\precsim\right)}$ provides a reflexive homeomorphism $\langle x,x^{\ast}\rangle = \langle x^{\ast},\left(x^{\ast}\right)^{\ast}\rangle$ satisfied between $X$ and $X^{\ast}$ over the compact Cauchy nets. Consequently, the result of transitivity will prove that $\Lambda_{n} \thicksim \Lambda_{n^{\prime}}$. Finally, we prove that $n=n^{\prime}$.}

Let $g_{X}:D_{X} \rightarrow a^{-}_{X}$ and $g_{X^{\ast}}:D_{X^{\ast}} \rightarrow a^{-}_{X^{\ast}}$ be directed nets defined on $D_{X}$ and $D_{X^{\ast}}$, respectively. This is direct application of a sequentially compact non-increasing rearrangement-invariant transform of $X$ that generates convergent Cauchy nets in $X$ and $X^{\ast}$. Then, there is a neighborhood of $n\in{V}$ $p\in{K_{i}},q\in{K_{i+1}}$ such that
\begin{equation}
    \centering
     p \preccurlyeq q \Longrightarrow g\left(q\right)\in{V}
    \label{eq:mth:ta:vcb_fr:1102}
\end{equation}
where $V\subset{g}$. \textit{Equivalently one can say that $g: D_{X} \rightarrow a^{-}_{X}$ uniformly converges to its limit at $n$.} Note that $p\in{K_{i}},q\in{K_{i+1}}$ imply $K_{i}\subset{K_{i+1}}\subset{D_{X}}$. Finally, we should emphasize that $U\subset{\overline{\mathfrak{B}_{\epsilon}\left(X\right)}}$.

By Lemma \ref{lm:mth:ta:cvb_fr:1100}, we know that there is a measurement space $\left(X,f_{n}\right)$ with a function defined as $f_{n}: \left(X_{n}\subset{X}\right) \rightarrow \mathds{R}^{+} \cup \{0\}$ compactly converges to measure space $\left(X,f\right)$ with $f:X \rightarrow \mathds{R}^{+} \cup \{0\}$ acting on $X$ such that $X_{n}$ compactly converges to $X$. By \eqref{eq:mth:ta:vcb_fr:1102} and considering $\|f_{1}\|_{\infty}\leq\|f_{2}\|_{\infty}\cdots \leq \|f_{n}\|_{\infty} \leq \cdots $, and Zorn's lemma, we conclude that
\begin{equation}
    \centering
    g_{n} \thicksim f_{n} \Longrightarrow g \thicksim f  
    \label{eq:mth:ta:vcb_fr:1103}
\end{equation}
From Lemma \ref{lm:mth:ta:fr:1101}, there is a Cauchy sequence $D_{X_{n}}\subset{D_{X}}$ converges to ${D_{X}}$ (resp. $X_{n}$ converges to $X$). Finally, by Lemma \ref{lm:mth:ta:cvb_fr:1100}, we can find $\Lambda_{n}\in{\overline{\mathfrak{B}^{X^{\ast}}_{1}}}$ corresponding to $D_{X_{n}}$ that compactly converges to $\Lambda\in{X^{\ast}}$. This is equivalent to say
\begin{equation}
	\begin{split}
		&\Lambda_{n} \thicksim f_{n} \\
		& f_{n} \thicksim f_{n}\\
		& \underset{a}{\Rightarrow} \Lambda_{n} \thicksim g_{n}
	\end{split}
	\label{eq:mth:ta:vcb_fr:1400}
\end{equation}
where (a) is due to the transition law. 

Similar to \eqref{eq:mth:ta:vcb_fr:1102}, for neighborhood $n^{\ast}\in{V^{\ast}}$ in $X^{\ast}$, there is $p^{\ast}\in{K^{\ast}_{i}},q^{\ast}\in{K^{\ast}_{i+1}}$ such that
\begin{equation}
    \centering
     p^{\ast} \preccurlyeq q^{\ast} \longrightarrow g^{\ast}\left(q^{\ast}\right)\in{V^{\ast}}
    \label{eq:mth:ta:vcb_fr:1104}
\end{equation}
where $V^{\ast}\subset{g}^{\ast}$. \textit{Equivalent, one can say that $g^{\ast}: D_{X^{\ast}} \rightarrow a^{-}_{X^{\ast}}$ uniformly converges to its limit at $n^{\prime}$.} Note that $p^{\ast}\in{K^{\ast}_{i}},q^{\ast}\in{K^{\ast}_{i+1}}$ also imply $K^{\ast}_{i}\subset{K^{\ast}_{i+1}}\subset{D_{X^{\ast}}}$. Finally, we should emphasize that $V^{\ast}\subset{\overline{\mathfrak{B}_{\epsilon}\left(X^{\ast}\right)}}$.

By Lemma \ref{lm:mth:ta:cvb_fr:1101}, we know that there is a measurement space $\left(X^{\ast},f^{\ast}_{n^{\prime}}\right)$ with a function defined as $f^{\ast}_{n}: \left(X^{\ast}_{n}\subset{X^{\ast}}\right) \rightarrow \mathds{R}^{+} \cup \{0\}$ compactly converges to measure space $\left(X^{\ast},f^{\ast}\right)$ with $f^{\ast}:X^{\ast} \rightarrow \mathds{R}^{+} \cup \{0\}$ acting on $X^{\ast}$ such that $X^{\ast}_{n^{\prime}}$ compactly converges to $X^{\ast}$. By \eqref{eq:mth:ta:vcb_fr:1104} and considering $\|f^{\ast}_{1}\|_{\infty}\leq\|f^{\ast}_{2}\|_{\infty}\cdots\leq\|f^{\ast}_{n^{\ast}}\|_{\infty}$, Zorn's lemma implies that 
\begin{equation}
    \centering
    g^{\ast}_{n^{\prime}} \thicksim f^{\ast}_{n^{\prime}} \Longrightarrow g^{\ast} \thicksim f^{\ast}  
    \label{eq:mth:ta:vcb_fr:1103}
\end{equation}

Accordingly, there is a Cauchy sequence $D^{\ast}_{X^{\ast}_{n^{\prime}}}\subset{D_{X^{\ast}}}$ converges to ${D_{X^{\ast}}}$ (resp. $X^{\ast}_{n^{\prime}}$ converges to $X^{\ast}$). By Lemma \ref{lm:mth:ta:cvb_fr:1101}, we can find $\Lambda_{n^{\prime}}\in{\overline{\mathfrak{B}_{1}\left(X^{\ast}\right)}}$. This is equivalent to say that
\begin{equation}
	\begin{split}
		& \Lambda_{n^{\prime}} \thicksim f^{\ast}_{n^{\prime}}\\
		& f^{\ast}_{n^{\prime}} \thicksim g^{\ast}_{n^{\prime}}\\
		& \Rightarrow \Lambda_{n^{\prime}} \thicksim g^{\ast}_{n^{\prime}}
	\end{split}
	\label{eq:mth:ta:vcb_fr:1401}
\end{equation}

Assuming that $X^{\ast}$ is a unitary dual space of the $X$, then we can conclude that both $\Lambda_{n}\in{\overline{\mathfrak{B}_{1}\left(X^{\ast}\right)}}$ and $\Lambda_{n^{\prime}}\in{\overline{\mathfrak{B}_{1}\left(X^{\ast}\right)}}$. We can conclude that 
\begin{equation}
	\centering
	\Lambda_{n^{\prime}} \thicksim \Lambda_{n} 
	\label{eq:mth:ta:vcb_fr:1401}
\end{equation}
that is $n = n^{\ast}$. And, this ends the proof of Theorem \ref{thm:mth_dist_intro_1103}.
    \label{prf:mth:ta:vcb_fr:1100}
\end{IEEEproof}

By proof of Theorem \ref{thm:mth_dist_intro_1103}, the question is how to determine optimum $n$ that establishes uniform convergence of $\Lambda_{n}$ to $\Lambda$.
In the remaining of this section, we answer this question. For this, we let the $X$ and $\left(X^{\ast}\right)^{\ast}$ be in Fr\'chet space by defining weak-topology and weak$^{\ast}$-compact topology, respectively. Then, we define supremum semi-norm as in the following. 

\begin{equation}
    \centering
    d\left(g,h\right) = \sum^{n}_{i=1}\frac{c_{i}\|D^{\alpha}g_{i}-D^{\alpha}h_{i}\|_{\left(K_{i},\infty\right)}}{1+\|D^{\alpha}g_{i}-D^{\alpha}h_{i}\|_{\left(K_{i},\infty\right)}}
    \label{eq:mth_osl_1210}
\end{equation}
where $c_{i}$, and $g_{i} = g^{\ast}_{i:N}$, and $h_{i}=g^{\ast}_{i+1:N}$.

Considering the Lizorkin distribution of compressible topological vector space and the fact that $c_{i}<c_{i+1}$, we can determine $n$ as
\begin{equation}
	\centering
	\lim_{i \rightarrow n} \left|d\left(g_{i},h_{i}\right)-d\left(g_{i+1},h_{i+1}\right)\right| \rightarrow 0
	\label{eq:mth_osl_1211}
\end{equation}
\section{Candidate Algorithms and Relation to Distribution Theory} \label{m:mth:ca}
\begin{figure}[t!]
	\centering
	\includegraphics[width = 6in]{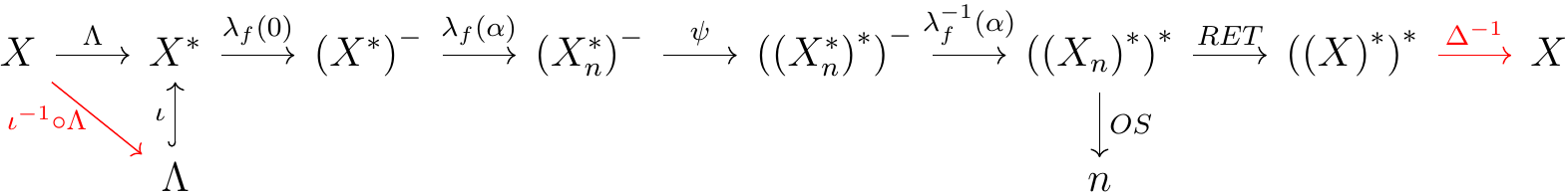}
	\caption{Reflexive homeomorpshism and optimum subspace estimation using the model in Theorem \ref{thm:mth:ta:h:1099}. RET and OS stand for the Riesz Extension Theorem and Optimum Subspace.}
	\label{fig:mth:ca:11001}
\end{figure}

Hitherto, we have studied uniform convergence of the compressible topological vector space without giving any explicit algorithm. In the following, first, we study the structure of the $\Delta_{\lambda_{f}\left(a\right)}=\Delta^{-1}\Delta$ in Theorem \ref{thm:mth:ta:h:1099}. We give two examples for Theorem \ref{thm:mth:ta:h:1099}. As a special case, we show that Theorem \ref{thm:mth:ta:h:1099} decays to greedy projective algorithms, in particular, Orthogonal Matching Pursuit. Then, a more general form is derived for $\Delta_{\lambda_{f}\left(a\right)}=\Delta^{-1}\Delta$ with respect to diagrams in Fig. \ref{fig:mth:ca:11001}. We prove that CS-KLE \cite{robaeiP7CSKLE} is an integral operator that satisfies reflexive homeomorphism in Fr\'echet space. We review the test functions in Schwartz space. We prove that the double dual space formulation in the Schwartz space of Theorem \ref{thm:mth:ta:h:1099} leads to an integral operator that extends to comobinatorial search region in the $\ell_{1}$ theoretical transition phase diagram. Finally, we propose a novel approach to measure optimum subspace dimension of finite- and infinite-dimensional compressible topological vector spaces in the Fr\'echet space.

\subsection{Specialized and Generalized Candidate Algorithms} \label{m:mth:sgca}
Theorem \ref{thm:mth:ta:h:1099} is rephrased with respect to diagrams in Fig. \ref{fig:mth:ca:11001} as
\begin{equation}
	\begin{split}
		\Delta_{\lambda_{f}\left(\alpha\right)} & = \Delta^{-1}\iota \iota^{-1} \Delta \\ 
		& = \Delta^{-1}\kappa
	\end{split}	
	\label{eq:part2:example_algo_1:gpa:1100}
\end{equation}
where $\kappa = \iota\iota^{-1}\Delta$. Since $\Delta^{-1}$ is mainly determined by dual space $X^{\ast}$ of $X$, the question is what are the candidates for $\iota\iota^{-1}$. In the following, first, we give one special case for greedy projective algorithms. Then, a general form of \eqref{eq:part2:example_algo_1:gpa:1100} are derived.

\textbf{\textit{Greedy Projective Algorithms -}} Let $\iota$ be an unitary operator that estimates the linear functional $\Lambda\in{X^{\ast}}$. Obviously, $\iota \iota^{-1} = \mathbf{I}$, where $\mathbf{I}$ is an identity operator. Consequently, \eqref{eq:part2:example_algo_1:gpa:1100} can be written as
\begin{equation}
	\centering
	\Delta_{\lambda_{f}\left(\alpha\right)} = \Delta^{-1}\Delta 
	\label{eq:part2:example_algo_1:gpa:1101}
\end{equation}

Letting $\Delta$ to be chosen from $X^{\ast}$ randomly, $\Delta$ is a projective operator estimates $\Lambda$. The well-known examples is the OMP algorithm, which estimates the supports of the sparse signal iteratively by projecting $\Gamma$ onto the compressed sensing measurement vector $\overline{y}$. It is well known, the estimated support subset satisfies the RIP condition. Since we aim to find the optimum subspace $X_{n}$ of $X$, estimating only small subset of $\Lambda$ is not sufficient for this purpose. Due to estimation of a few dominant subspace of $X$, greedy projective algorithms are not sufficient for the continuous approximation of the spectrum of the compressible signal. As a result, the class of greedy projective algorithms does not look sufficient for our scope in this work.

\textbf{\textit{Generalized Homeomorphic Algorithm -}} Let $\iota \iota^{-1} = \psi$, then we can write
\begin{equation}
	\centering
	\psi\Delta = \kappa 
	\label{eq:part2:example_algo_1:gpa:1102}
\end{equation}

Assuming that dual space $X^{\ast}$ is known, we need to find a candidate for $\psi$. Then, the generalized homeomorphic relation in Theorem \ref{thm:mth:ta:h:1099} can be established as
\begin{equation}
	\centering
	\Delta_{\lambda_{f}\left(\alpha\right)} = \Delta^{-1}\psi\Delta
	\label{eq:part2:example_algo_1:gpa:1103}
\end{equation}

From functional analysis and optimization theory, we know that $\psi$ needs to be a positive semi-definitive operator. Alternatively, there must be positive measurement $\mu$ on $X$ corresponding to operator $\psi$. In section \ref{m:mth:dt}, we prove that one natural candidate for $\psi$ is the eigenfunction of the compressible topological vector space, or equivalently eigenfunction of compressed sensing measurement vector $\overline{y}$.
\subsection{Relation to Distribution Theory}\label{m:mth:dt}
\begin{dfn}[Dirac Delta function and Sifting Property]
Dirac Delta function is a useful tool in physics and signal processing. In signal processing, Dirac Delta function is fundamentally important to find impulse response of the Linear Time-Invariant systems. Such a relation is an essential tool to characterize the frequency response of a LTI system for any given input. For this purpose, Dirac Delta function is defined as the conceptual mathematical function as
\begin{equation}
    \centering
    \delta\left(x-x_{0}\right) = 
    \begin{cases}
        0 & ,x\in{\mathds{R}}, x\neq x_{0}\\
        \infty & ,x=x_{0}
    \end{cases}
    \label{eq:mth:ta:dt:1099a} 
\end{equation}

Sifting property is an important property of Dirac Delta function, which can be used to find a point value of the $f\left(x\right)$ as
\begin{equation}
    \centering
    \int_{\mathds{R}}f\left(x\right)\delta\left(x-x_{0}\right)dx = f\left(x_{0}\right)
    \label{eq:mth:ta:dt:1098a} 
\end{equation}

Definition of the Delta function in \eqref{eq:mth:ta:dt:1099a} is not integrable, so the sifting property in \eqref{eq:mth:ta:dt:1098a}. The formulation in \eqref{eq:mth:ta:dt:1098a} with respect to the Definition of the Delta function \eqref{eq:mth:ta:dt:1099a} has a zero Lebesgue integral\footnote{In $X-Y$ Cartesian product space, the integral in \eqref{eq:mth:ta:dt:1098a} can be computed as the sum of area of the countably infinite number of horizontal boxes with a width of zero, leading to zero Lebesgue integral.}, and this contradicts \eqref{eq:mth:ta:dt:1099a} for which $\delta\left(x-x_{0}\right) = \infty$ at $x = x_{0}$. The arising contradiction can be avoided if \eqref{eq:mth:ta:dt:1099a} is relaxed by defining a smoother function instead of \eqref{eq:mth:ta:dt:1099a}. For example, Delta function defined on an open subset of real line 
\begin{equation}
    \centering
    \psi\left(x\right)=
    \begin{cases}
        \frac{1}{2\Delta} & ,x_{0}-\Delta<x<x_{0}+\Delta \\
        0 & ,Otherwise
    \end{cases}
    \label{eq:mth:ta:dt:1097} 
\end{equation}
then the sifting property can be defined as the limit of $\psi\left(x\right)$ as
\begin{equation}
    \begin{split}
        I_{\psi}\left(f\right) & =  \int^{x_{0}+\Delta}_{x_{0}-\Delta}g\left(x\right)f\left(x\right)dx\\
        & = \lim_{\Delta \rightarrow 0}\int^{x_{0}+\Delta}_{x_{0}-\Delta}\frac{1}{2\Delta}f\left(x\right)dx\\
        & = \lim_{\Delta \rightarrow 0}\frac{1}{2\Delta}\int^{x_{0}+\Delta}_{x_{0}-\Delta}f\left(x\right)dx\\
        & = \lim_{\Delta\rightarrow 0}\frac{ \left(F\left(x_{0}+\Delta\right)-F\left(x_{0}-\Delta\right)\right)}{2\Delta}\\
        & = f\left(x_{0}\right)
        \end{split}
    \label{eq:mth:ta:dt:1096} 
\end{equation}
Another example of smooth function that its limit can be used as Delta function is Gaussian function with dilation limit approaches zero $\left(\psi\left(x\right)=\lim_{\Delta \rightarrow 0}\frac{1}{\sqrt{\pi}\Delta}e^{-x^{2}/\Delta^{2}}\right)$.
    \label{dfn:mth:ta:dt:1100}
\end{dfn}

\begin{dfn}[Distribution Function]
Smooth Delta function can be used to find point value of the compressible topological vector space over its supports (refer to \eqref{eq:bck:cs:cv:1099}). Then, one can define an integral operator $\left(\mathbf{I}_{\psi}\right)\left(x\right)$ acting on compressible topological vector space $g\left(x\right)$ as
\begin{equation}
    \centering
    \left(\mathbf{I}\psi\right)\left(x\right) = \int_{\mathds{R}^{d}} g\left(x\right)\psi\left(x\right)dx
    \label{eq:mth:ta:dt:1095} 
\end{equation}
where LHS contains the point-wise evaluation of the compressible topological vector space $g\left(x\right)$.

This leads to a definition of the distribution function, also called generalized function in the literature, which allows to characterize the underlying distribution of the $g\left(x\right)$ and apply calculus methods such as differentiation, multiplication by a smooth function accordingly. Note that the Delta function is replaced with a much more smooth function $\psi\left(x\right)$.
    \label{dfn:mth:ta:dt:1101}
\end{dfn}

\begin{dfn} [Generalized Distribution Function] A useful smooth function as a function of the random process path $x \rightarrow \Xi\left(x\right)$, where $x$ is a random variable, can be defined as
	\begin{equation}
		\centering
		\psi\left(\Xi\left(x\right)\right) = \left(\psi \circ \Xi\right)\left(x\right)
		\label{eq:mth:ta:dt:1089a}
	\end{equation}
leading to the integral of the form
    \begin{equation}
        \centering
         I_{g}\left(\Xi\right) = \int_{\mathds{R}^{n}} g\left(x\right)\psi\left(\Xi\left(x\right)\right)dx
        \label{eq:mth:ta:dt:1089}
    \end{equation} 
    \label{dfn:mth:ta:dt:1099}
\end{dfn}

Integral operator \eqref{eq:mth:ta:dt:1089} can be applied to two kinds of problems. The first class is the scenario when a potential priori knowledge about the underlying distribution of a function $g\left(x\right)$ through $\psi\left(x\right)$ is available. This scenario sounds trivial. In more useful and practical examples, the underlying distribution is unknown. This brings us to the second scenario. The second scenario occurs if $\psi\left(x\right)$ has driven from a kernel of vector space $g\left(x\right)$. This scenario takes the form of generalized homeomorphic relation in \eqref{eq:part2:example_algo_1:gpa:1103}. Then, one can compute the set of linear functional $\Lambda_{i}$, $i\in{\left[1,n\right]}$, that contribute to the final states of $g\left(x\right)$.

\begin{rmk} The most important random process, which we are also interested in compressed sensing, is the joint sampling operator $\Xi\left(x\right)$, where the $x$ is the random sampling variable. The $\psi\left(\Xi\left(x\right)\right)$ in \eqref{eq:mth:ta:dt:1089} is the point mass operator that evaluates every coordinate in $g\left(x\right)$ to find the supports corresponding to the random process path $\Xi\left(x\right)$. Considering the fact that the $\Xi\left(x\right)$ is a stochastic process with a path $x\rightarrow \Xi\left(x\right)$, one can conclude that the $\psi\left(x\right)$ is the event space of the $\Xi\left(x\right)$, or equivalently $\psi\left(x\right)$ contains the potential distribution that $\Xi\left(x\right)$ has taken.
    \label{rmk:mth:ta:dt:1098}
\end{rmk} 

\begin{dfn}[Test Function] \cite[Section 1.13.1]{tao2010epsilon} A test function is any vector space of smooth, compact, and sufficiently continuously differentiable functions $f\in{C^{\alpha}_{c}\left(\mathds{K}\right)}$ in the form of $f:C^{\alpha}_{c}\left(\mathds{K}\right) \rightarrow \mathds{K}$. Usually $\alpha = \infty$ and the test functions are in Schwartz spaces, denoted as a subset of the class of a distribution function $\mathcal{D}\left(\Omega\right)$. 
    \label{dfn:mth:ta:dt:1102}
\end{dfn}

\begin{rmk} Let $X$ be an $n$-dimensional compressible topological vector space lies in a much larger space $\mathds{K}^{d}$. Then, a weak-compact topology defined on an $n$-dimensional subspace of $X$ (resp. weak$^{\ast}$-compact topology defined on separating dual space $X^{\ast}$ of $X$) is compatible with separating semi-norms $f_{n}$ (resp. $f^{\ast}_{n}$). According to the results in \cite{robaeiP7CSKLE}, $n$-dimensional subspace of $X$ is identifiable by sequentially compact subset $K_{n}\subset{X}$. Then, there is $f_{n}\in{C_{c}\left(K_{n}\right)}$ that approximates compressible topological vector space with respect to Riesz extension theorem provided that $K_{n}$ satisfies the requirement in Krein-Milman Theorem \ref{thm:mth_dist_intro_1101} and \ref{thm:mth_dist_intro_1101a}. The approximated subspace $X_{n}$ corresponding to compact subspace $K_{n}$ must contains its extreme points particularly.
    \label{rmk:mth:ta:dt:1101}
\end{rmk}

\begin{exm} Let $f:K_{n}\rightarrow \mathds{R}^{+} \cup \{0\}$ be a family of separating semi-norms as defined in \eqref{eq:mth:ta:ts:1097}. Obviously for an $n$-dimensional sequentially compact compressible topological vector space $X$
\begin{equation}
    \centering
    \|f_{1}\|_{\left(K_{1},\infty\right)} < \|f_{2}\|_{\left(K_{2},\infty\right)} < \cdots < \|f_{n}\|_{\left(K_{n},\infty\right)}
    \label{eq:mth:ta:dt:1091}
\end{equation}

As we discussed in section \ref{m:mth:vcb_fr}, there is subsequence $X_{n}$ that coincides with optimum compactness give by index $n$ over which the sequence of linear functional $\Lambda_{n}$ and semi-norm $f_{n}$ both converge to their limits. Accordingly, there is a pseudo-metric compatible with the given topology indented by
\begin{equation}
    \centering
    d\left(x,y\right) = \sum^{\infty}_{i=1} \frac{c_{i}f_{i}\left(x-y\right)}{1+f_{i}\left(x-y\right)}
    \label{eq:mth:ta:dt:1088}
\end{equation}
for $x,y\in{X}$

Note that the definition in \eqref{eq:mth:ta:dt:1088} coincides with the definition in \eqref{eq:mth_osl_1210}. It is obvious that for the minimal functional $\left|d_{n}\left(x_{l},y_{k}\right)-d\left(x,y\right)\right|\rightarrow 0$. Since $d_{n}\left(x_{l},y_{k}\right)$ uniformly converges over compact subsets to $d\left(x,y\right)$, the $X_{n}$ is the Cauchy subsequence of $X$.
    \label{exm:mth:ta:dt:1101}
\end{exm}

Let $X$ be topological space with separating dual space $X^{*}$. We have already seen that the linear functional $\Lambda\in{X^{*}}$ is lying in the space of continuous functions $C_{c}\left(\Omega\right)$ for open set $\Omega\in{X}$. Let use the sequential weak$^{\ast}$-compact topology for compact continuous function $C_{c}\left(K\right)$ defined on $X^{\ast}$ such that $K_{1}\subset{K_{2}}\cdots\subset{K_{n}}\subset{\Omega}$. Accordingly, a sequential family of compact continuous functions can be defined as $C_{c}\left(K_{1}\right)\subset{C_{c}\left(K_{2}\right)}\cdots\subset{C_{c}\left(K_{n}\right)}\subset{C_{c}\left(\Omega\right)}$. Each of the compact continuous functions $C_{c}\left(K_{i}\right)$ is the Fr\'echet space with a topology defined by separating semi-norms as
\begin{equation}
    \centering
    \|\phi\|_{\left(K^{\ast}_{i},\infty\right)} = \sup_{i\in{\left[1,n\right]}|x\in{K_{i}}}\left|\mathbf{D}^{\alpha}\phi\left(x\right)\right|
    \label{eq:mth:ta:dt:1100}
\end{equation}
\begin{equation}
    \text{and}
    \nonumber
\end{equation}
\begin{equation}
    \centering
    \left|\alpha\right|=\alpha_{1}+\alpha_{2}+\cdots+\alpha_{n}
    \nonumber
    \label{eq:mth:ta:dt:1101}
\end{equation}

Note that \eqref{eq:mth:ta:dt:1100} inherits the uniformly convergence over compact subset as $\phi_{i}$ converges to $\Phi$. $\phi$ and $\Phi$ can be defined as
\begin{equation}
    \centering
    \phi = \langle \Lambda_{i},\psi\left(x\right) \rangle \text{\quad such that \quad} \Lambda_{i}\in{X^{\ast}} \text{, } \left(\psi\left(x\right)\in{\left(X^{\ast}\right)}\right) \hookrightarrow X
    \label{eq:mth:ta:dt:1102}
\end{equation}
\begin{equation}
    \centering
    \Phi = \int_{X^{\ast}} \langle x^{\ast},\psi\left(x\right) \rangle dx^{\ast}
    \label{eq:mth:ta:dt:1103}
\end{equation}
where
\begin{equation}
    \centering
    \langle x^{\ast},\psi \rangle = \int x^{\ast}\psi\left(x\right)dx
    \label{eq:mth:ta:dt:1104}
\end{equation}
\textit{\eqref{eq:mth:ta:dt:1102} defines the dual space $\widehat{X}^{\ast} = \psi\left(x\right)$ in the Lemmas \ref{lm:mth:ta:rr:1103}-\ref{lm:mth:ta:rr:1102}}

We are interested in sifting properties of \eqref{eq:mth:ta:dt:1104} that is
\begin{equation}
    \centering
     \left(\mathbf{G}_{c}\psi\left(\Xi\right)\right)\left(\left(x^{\prime}\right)\right) = \int_{\widehat{X}^{\ast}}\int_{X} \Lambda_{i}\psi\left(\Xi\left(x,\right)\right)d\mu\left(\xi x\right)
     \label{eq:mth:ta:dt:1099}
\end{equation}
\eqref{eq:mth:ta:dt:1099} is the Compressed Sensing Hilbert Schmidt operator recommended in \cite{robaeiP7CSKLE}.

Since $X^{\ast}$ is metrizable, due to the Banach-Agao\u{g}lu theorem, by the Riesz representation theorem, we can define inner regularity $\mu\left(\mathcal{B}\right)$ on Borel set $\mathcal{B}$ corresponding to $K^{\ast}_{n}$ and compatible with weak$^{\ast}$-compact topology defined on $X^{\ast}$ in the Fr\'echet space.
Considering this, we define the distribution $\mathcal{D}_{K}\left(\Omega\right)$ and $\mathcal{D}^{\ast}_{K}\left(\Omega\right)$ in the following definition. $\mathcal{D}^{\ast}_{K}\left(\Omega\right)$ is the topological dual of $\mathcal{D}_{K}\left(\Omega\right)$ for a sequentially compact compressible topological vector space with a separating dual space $X^{\ast}$. 

The last property we are interested in is the differentiation of distribution represented in the next definition.
\begin{dfn}[Differentiation of Distribution] \cite[Section 6.12]{rudin1991functional} Let $\Lambda\in{\mathcal{D}}\left(\Omega\right)$ and $\alpha$ be a multi-indices, the differentiation of distribution takes the form of equality
    \begin{equation}
        \centering
        \|\left(\mathbf{D}^{\alpha}\Lambda\right)\left(\phi\right)\|_{\infty} = \|\Lambda\left(\mathbf{D}^{\alpha}\phi\right)\|_{\infty}
        \label{eq:mth:ta:dt:1107}
    \end{equation}
    for $\phi\in{\mathcal{D}_{K_{n}}\left(\Omega\right)}$.
    \label{dfn:mth:ta:dt:1104}
\end{dfn}

\begin{thm}[CS-KLE and Reflexive Homeomorphism] Let $Y\left(t,\omega\right) = \Gamma\left(\omega\right) X\left(t\right)+e$ to be a noisy measurement of signal $X\in{L^{2}}\left(\mathds{C}^{d}\right)$ filtered by i.i.d process $\Gamma\left(\omega\right)\in{L^{2}\left(\mathds{C}^{M \times d}\right)}$, that is, $\omega\in{\Omega}$ is independently identically distributed random variable. Also, let $Y\in{C\left(X\right)}$ and $X$ be continuous bounded vector spaces with bouneded family of semi-norms. Then, the following inequality defined in $\|\cdot\|_{1,\infty}$ is satisfied 
    \begin{equation}
        \centering
        \|X\left(t\right)\|_{1,\infty}
        \thicksim
        \sup_{\gamma_{i}\left(\omega\right)} \sum_{i\in\left[N\right]} \left|\sqrt{\lambda}\psi\left(t\right)\gamma_{i}\left(\omega\right)\right|
        \label{eq:mth:cs_kle2:kle_rnd:1101}
    \end{equation}
    \label{thm:mth:cs_kle2:kle_rnd:1100}
\end{thm}
Proof has been given in \cite{robaeiP7CSKLE}.
\subsection{Equivalent Locally Convex Space through Fr\'echet Distance Metric and K\"othe sequence}\label{m:mth:fd}
In \cite{robaeiP7CSKLE}, it has been recommended that transition point determined through measuring spectral overlap occurs for an optimum subspace of compressible signal. However, measuring spectral overlap is not possible without knowing the distribution of the actual signal. The two-step approach proposed in \cite{robaeiP7CSKLE} penalizes the performance by wasting resources. Considering locally convex space and a topology defined on $X$ and its dual space $X^{\ast}$, we propose to measure the optimum subspace (resp. $n$-pseudospectrum set) using the Fr\'echet distance metric in \eqref{eq:mth_osl_1211}. For this purpose, we define a sequence space that reveals the internal structure of $X$. Such a sequence needs to carry the structure of the $X$, which means one should derive the sequence space from $Y$ ($\Lambda:X\rightarrow Y$ such that $\Lambda\in{X^{\ast}}$) or  LHS of \eqref{eq:mth:cs_kle2:kle_rnd:1101}. This section explains how Fr\'echet distance metric can measure the optimum subspace dimension of the finite- and infinite-dimensional compressible topological vector spaces. \textit{A potential candidate for sequence space is K\"othe sequence defined such that a distance metric on $X$ (or estimation of $X$) is a Fr\'echet space and induces a topology equivalent to the weak$^{\ast}$-compact topology defined on $X^{\ast}$ of $X$.}

The topology induced on Fr\'echet space can be generated by a translation invariant metric defined in \eqref{eq:mth_osl_1211}. The default value for $c_{i}$ has been given as $2^{-i}$ with respect to dyadic decomposition with side lengths of $2^{i}$, where $i\in{\mathds{N}}$ is the dimension index \cite[Sections 6.12 and 6.13]{rudin1991functional}. The constant $2^{-i}$ decays extremely rapidly as a function of dimension and can only be applied to extremely small size data\footnote{For $i=20$, $c_{20} = 2^{-20}$. Obviously, for practical applications $c_{i\gtrapprox 20}$ will kill all the contributions from $i\gtrapprox 20$ promptly even if they have contained significant amount of energy.}. Obviously, $c_{i}\propto 2^{-i}$ can not be applied to compressible signals. In particular, a topological structure of compressible vector spaces with continuous spectrum extension requires a proper definition of $c_{i}$.

\begin{figure}
	\centering
	\includegraphics[width=3.0in]{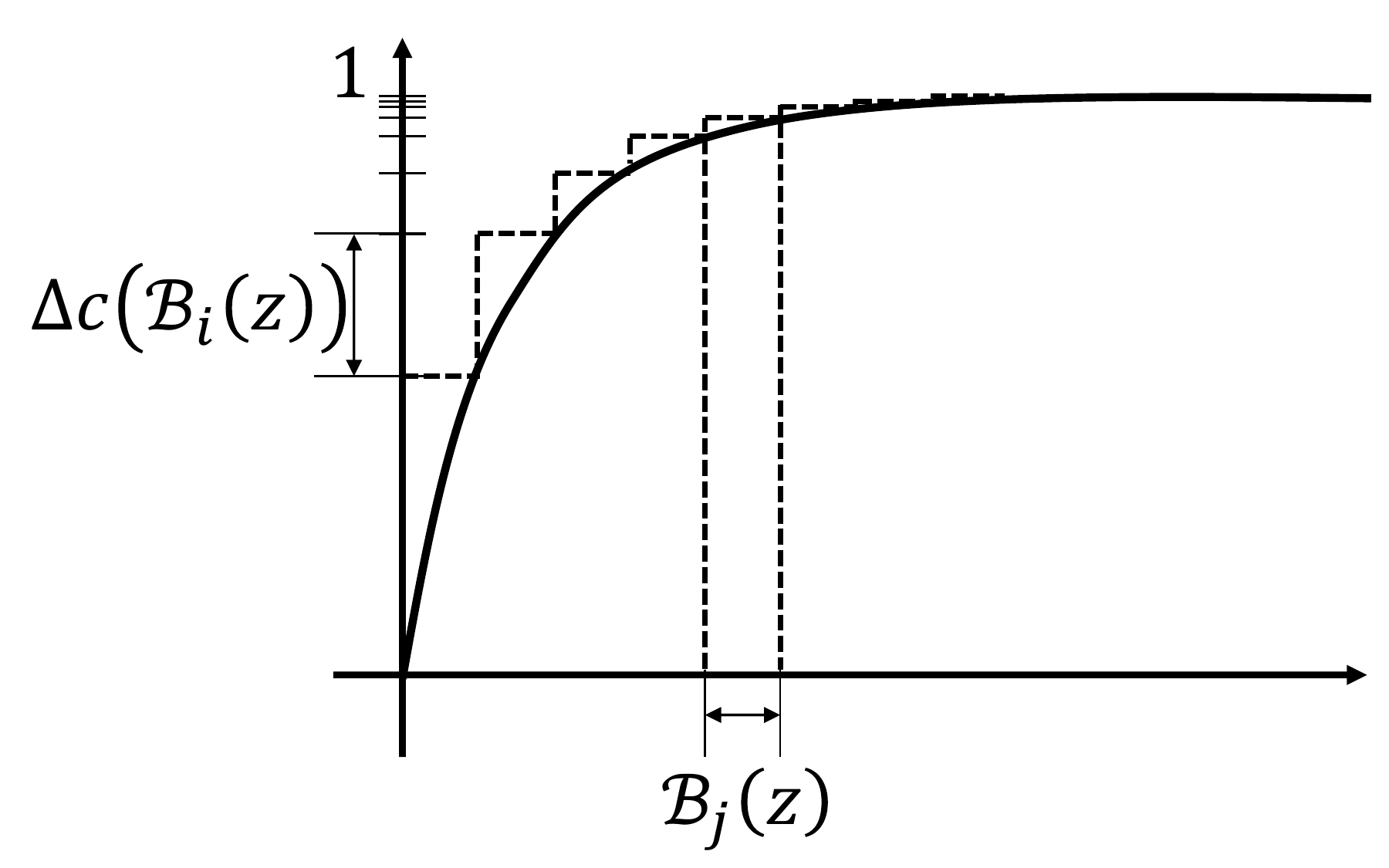}
	\caption{$c\left(z\right)$ and sampled $c_{i}$ for all $i\in{\mathds{N}}$. Sampling is done with respect to Borel set $\mathcal{B}_{i}\left(z\right)$. The dimension of $c_{i}$ grows proportional to $z$.}
	\label{fig:frd:1100}
\end{figure}

In order to define proper sequence $c_{i}$ that reflects the actual dimension growth of compressible topological vector space properly, $c_{i}$ needs to be sampled from continuous dimension function $c\left(x\right)$ that satisfies the following properties:
\begin{enumerate}
	\item $c\left(x\right)$ grows proportional to the dimension of a compressible signal.
	\item $c\left(x\right)$ is in $L^{1}$-space.
	\item $c\left(x\right)$ has a proper Lebesgue sum that converges to its limit uniformly as shown in Fig. \ref{fig:frd:1100}. For this purpose, the Lebesgue sum is defined as
	\begin{equation}
		\centering
		\sum^{\infty}_{i=1} \mathcal{B}_{i}\left(z\right) \Delta c\left(\mathcal{B}_{i}\left(z\right)\right) < \infty
		\label{eq:frd:1100}
	\end{equation}
\end{enumerate}
where $z$ is a continuous variable, and $\{\mathcal{B}_{i}\left(z\right)\}$ is the Borel set defined on $z$. Lebesgue measure has to be normalized by the length of interval, which is denoted by $\Delta c\left(\mathcal{B}_{i}\left(z\right)\right)$ in \eqref{eq:frd:1100}. The decaying power law has been already included in $\Delta c\left(\mathcal{B}_{i}\left(z\right)\right)$. This has been clearly depicted in Fig. \ref{fig:frd:1100}. Item (2) follows directly from (3) and (4). Notice that the definition of $c$ in this way inherits the natural advantage of the Lebesgue measurement. Since $\lim_{i \rightarrow \infty} \Delta c\left(\mathcal{B}_{i}\left(z\right)\right)\rightarrow 0$, then for countably infinite-dimensional and separable compressible space $X$\footnote{$X$ can be optimally decomposed into separable subset $V$ and inseparable subset $U$ as discussed in Theorem \ref{thm:fdos:11010}.}, the Lebesgue measurement on $c$ approaches zeros. As a result, the convergence of the the sum in \eqref{eq:frd:1100} is followed by proportional growth of the $c\left(\mathcal{B}_{i}\left(z\right)\right)$.

\begin{thm}[Vanishing Distribution Derivatives] Let be  $g\in{\mathcal{D}^{\ast}\left(\Omega\right)}$ and $\alpha$ be a multi-index such that $\left|\alpha\right| = \alpha_{1}+\alpha_{2}+ \cdots +\alpha_{n}$. Also, let $c\left(x\right)$ be a function that grows proportional to dimension of compressible topological vector space with bounded Lebesgue sum as in \eqref{eq:frd:1100}. Then, the distribution derivative satisfies
\begin{equation}
	\int_{\Omega}\mathbf{D}^{\alpha}\left(c g\right)\left(x\right)dx = -\int_{\Omega} c\left(x\right)\left(\mathbf{D}^{\alpha}g\right)\left(x\right)dx
	\label{eq:frd:1101}
\end{equation} 
\label{thm:frd:1100}
\end{thm}
\begin{IEEEproof}
\begin{equation}
	\centering
	\int_{\Omega} \left(\mathbf{D}^{\alpha}cg\right)\left(x\right)dx = \int_{\Omega} \left(\mathbf{D}^{\alpha}c\right)\left(x\right)g\left(x\right)dx + \int_{\Omega} c\left(x\right) \left(\mathbf{D}^{\alpha}g\right)\left(x\right)dx
	\label{eq:frd:1102}
\end{equation}
For sufficiently large dimension $X$ and rapidly increasing $c\left(x\right)$, 
that is, $c\left(x\right)$ rapidly reaches its limit, we have 
\begin{equation}
	\begin{split}
	&\int_{\Omega} \left(\mathbf{D}^{\alpha}cg\right)\left(x\right)dx = 0\\
	& \Rightarrow
	\int_{\Omega} \left(\mathbf{D}^{\alpha}c\right)\left(x\right)g\left(x\right)dx = - \int_{\Omega} c\left(x\right) \left(\mathbf{D}^{\alpha}g\right)\left(x\right)dx
	\end{split}
	\label{eq:frd:1103}
\end{equation}
By induction, \eqref{eq:frd:1101} is obtained, and this will end the proof.
\end{IEEEproof}

Distribution derivatives approach zero as $\left(\mathbf{D}^{\alpha}g\right)\left(x\right) \rightarrow 0$, which is equivalent to vanishing of the total variation of compressible signal $X$. Considering this behavior, the optimum subspace of compressible topological vector space $X$ occurs at a Borel set that $d$ vanishes. Setting $\left|\alpha\right| = 2$, \eqref{eq:frd:1101} indicates that the derivative on distribution is equivalent to the derivative on semi-norm $g$. As a result, total variation vanishing on distribution is equivalent to the derivative of semi-norm goes to zeros. This is obvious from Fig. \ref{fig:frd:1100}, where the dimensional growth saturates as the slope of the $c\left(x\right)$ approaches zero as $j \rightarrow \infty$. Consequently, we define the Fr\'echet distance as
\begin{equation}
	\centering
	d = \sum^{\infty}_{i=1}\frac{c_{i}\|D^{\alpha}g_{i}\|_{\left(K_{i},\infty\right)}}{1+\|D^{\alpha}g_{i}\|_{\left(K_{i},\infty\right)}} \text{, \quad} \alpha = 2
	\label{eq:frd:1104}
\end{equation}
where $c_{i}$ and $g_{i}$ are sampled from $c\left(x\right)$ and $g\left(x\right)$ with respect to Borel sets defined on $X$.
\section{Numerical Results}
\subsection{Finite Dimensional Signal}
The numerical evaluation for the locally convex space has been presented in this section. Three classes of signals with respect to their dimensional characteristics have been considered. Finite-dimensional signals, where millimeter-wave channel estimation has been studied. For this purpose, we study the dimension prediction using Fr\'echet distance in \eqref{eq:frd:1104}. Millimeter-wave channel can be categorized as a class of compressible topological vector space with $\rho\rightarrow 0$, i.e., $\frac{n}{M} \ll 1$. For the example in this letter, $\rho \approx 3/2^{12} \approx 7.3*10^{-4}$. In order to study the effect of dimension on the K\"othe sequence purposed in section \ref{m:mth:fd}, we examine two candidates for $c_{i}$, first, sequentially compact Cauchy net $\left(\|\left(X^{\ast}_{i}\right)^{\ast}\|_{1,\infty},\preccurlyeq\right)$, and (2) K\"othe sequence as discussed in section \ref{m:mth:fd}.

In the following, $n$ refers to the optimum subspace estimated using spectral overlap methods with perfect knowledge about underlying distribution, and \textit{$n_{I}$ and $n_{K}$ denote the optimum subspace obtained through inducing topology to RHS of \eqref{eq:mth:cs_kle2:kle_rnd:1101} using $\left(\|\left(X^{\ast}_{i}\right)^{\ast}\|_{1,\infty},\preccurlyeq\right)$ and K\"othe sequence.} The central carrier frequency is set to $28$GHZ with a bandwidth of $0.1$GHz and $3$GHz for narrow-band and wide-band, respectively. The bandwidth is uniformly divided into $32$ subchannels, and steering vectors are generated for the subchannel using the same angles of departure and arrival. In order to measure optimum subspace and beamfroming gain, we estimate the millimeter-wave channel from compressed sensing measurement vector $\overline{y}_{N_{B}}$ of subchannel $N_{B}$ as
\begin{equation}
	\centering
	\overline{y}_{N_{B}} =\Phi\Psi\overline{h}_{v}+\overline{e}
	\label{eq:nmr:fds:1100}
\end{equation}
where $\overline{h}_{v}$ is the Kronecker-vectorized compressible millimeter-wave subchannel. In \eqref{eq:nmr:fds:1100} $\Psi$ is the DFT dictionary matrix, $\Phi = \left[\phi_{0}\vert\vert\phi_{1}\vert\vert\cdots\vert\vert\phi_{M-1}\right]^{T}$, and $\overline{e}=\left[\overline{e}_{0}\vert\vert\overline{e}_{1}\vert\vert\cdots\vert\vert\overline{e}_{M-1}\right]^{T}$ are concatenated sampling matrix and noise vector (vertical concatenation is denoted by $\vert\vert$). Consequently, for multiple subcarriers, MMV is obtained as $\left[\overline{y}_{0}\vert\vert\overline{y}_{1}\vert\vert\cdots \vert\vert\overline{y}_{N_{B}-1}\right]_{MN_{r,rf}\times N_{B}}$, where $N_{B}$ is the total number of coherent bandwidth blocks.

\textbf{\textit{Narrow Band Channel (No Beam Squint Between Subchannels) -}} Given the multiple compressed sensing measurement vectors, results show that both $c_{i} \coloneqq \left(\|\left(X^{\ast}_{i}\right)^{\ast}\|_{1,\infty},\preccurlyeq\right)$ and K\"othe sequence predict the optimum dimension $n$ accurately. The accuracy of beamforming gain estimation indicates the accuracy of the topology induced by $\|\left(X^{\ast}_{i}\right)^{\ast}\|_{1,\infty}$ in the approximation stage, and also the accuracy of the topology induced by the distance metric \eqref{eq:frd:1104}.

\begin{figure*}
	\centering
	\subfloat[]{\includegraphics[width=3.0in]{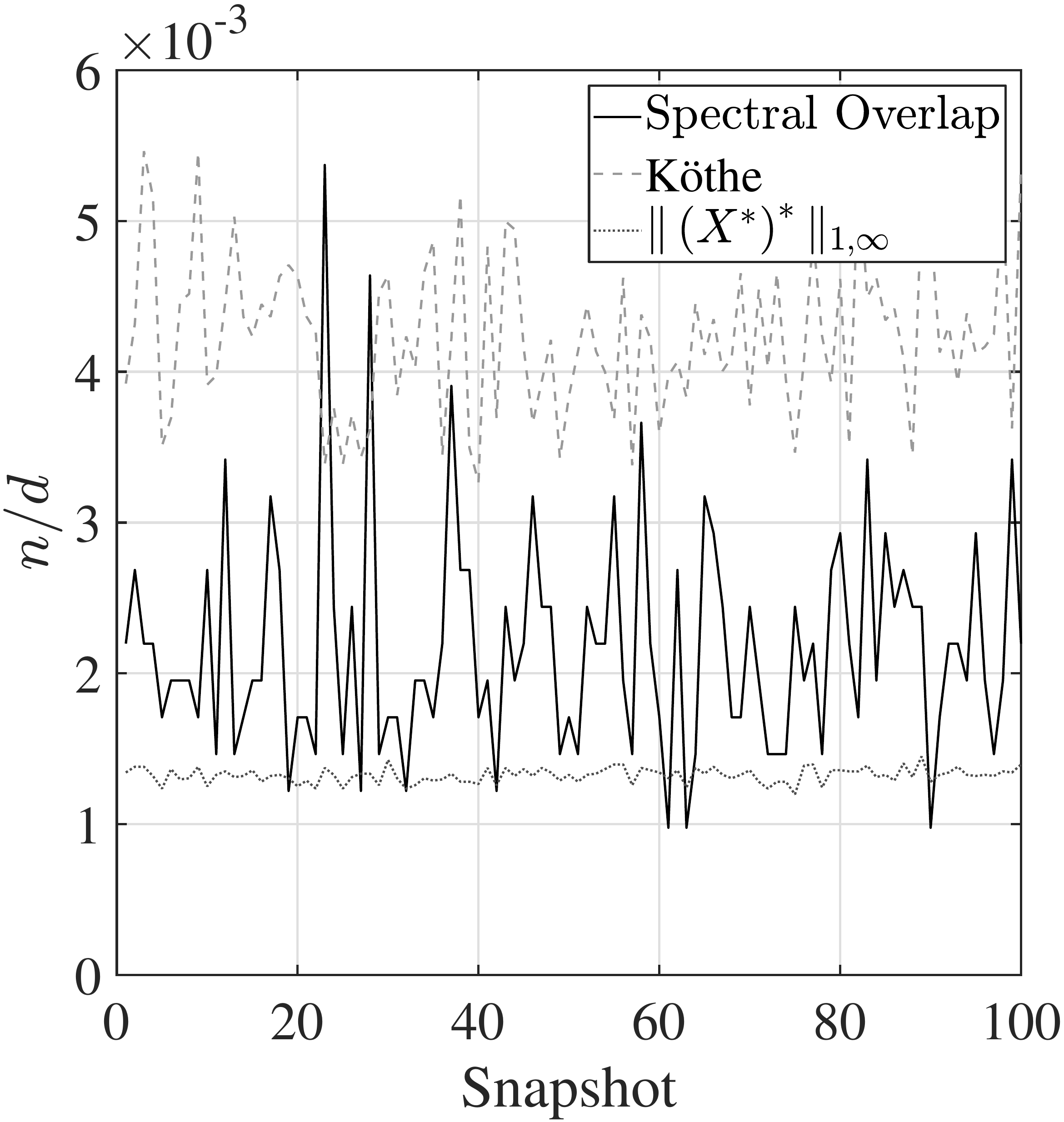}\label{fig:nmr:fds:1100_a}}
	\hspace{4mm}
	\subfloat[]{\includegraphics[width=3.15in]{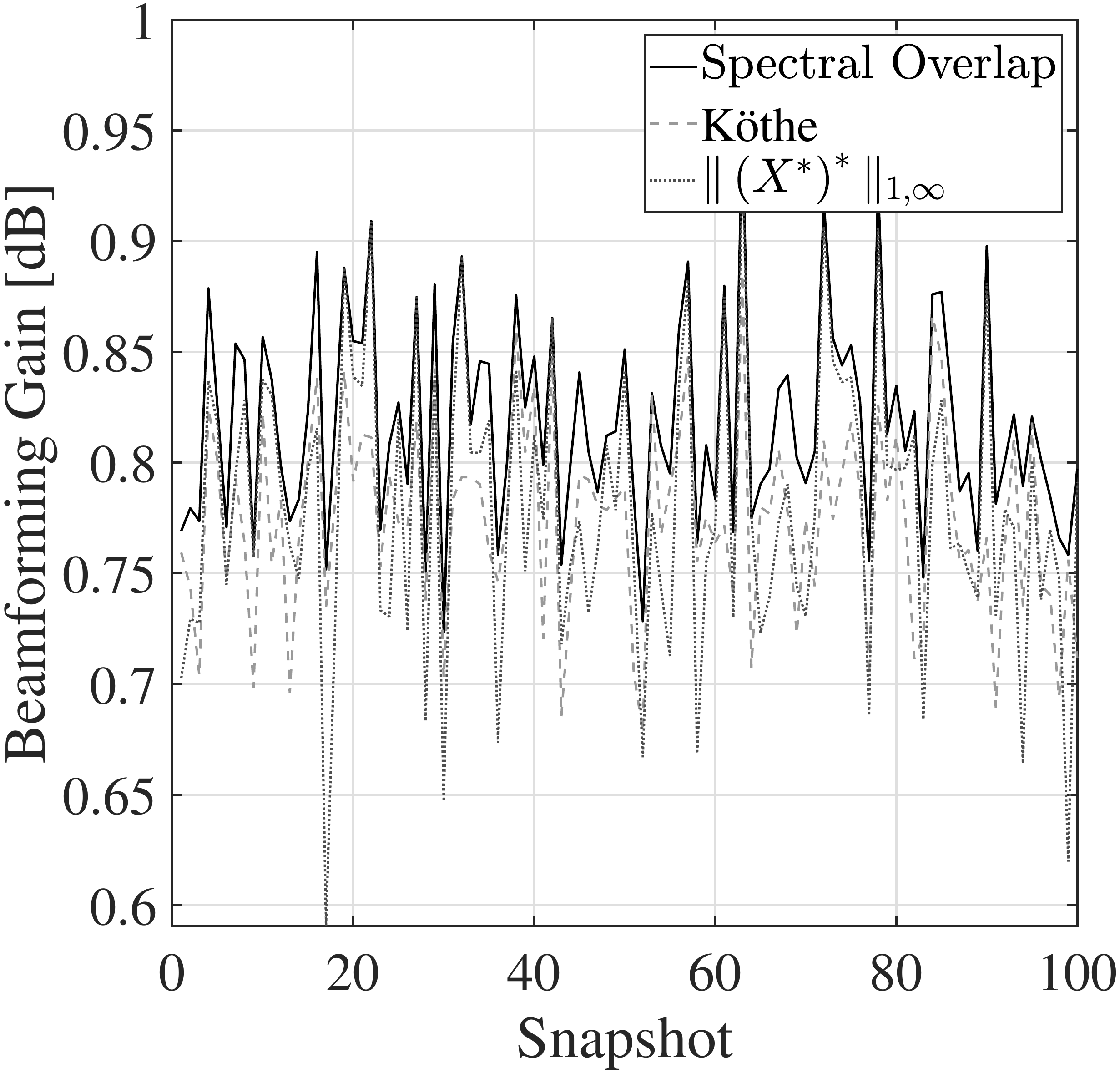}\label{fig:nmr:fds:1100_b}}
	\caption{Optimum dimension $n$ measured for the millimeter-wave directional channel without beam-squint, narrowband $100$MHz bandwidth is divided to $32$ subchannels $100$ trials, (a) optimum dimension ratio for $100$ trials, (b) beamforming gain in [dB].}
	\label{fig:nmr:fds:1100}
\end{figure*}

In conclusion, observation shows that for finite-dimensional MMV with partial overlap between the support sets, $n_{I}$ and $n_{K}$ provide the lower and upper bounds for optimum dimension $n$ almost surely. Fig. \ref{fig:nmr:fds:1100} shows the optimum dimension ratio and corresponding beamforming gain (in [dB]) for $100$ trials. We obverse the beamforming gain with $\left(\|\left(X^{\ast}_{i}\right)^{\ast}\|_{1,\infty},\preccurlyeq\right)$ and K\"othe sequences follows the optimum subspace $\rho$ of the spectral overlap method with prefect knowledge about the channel.
\begin{figure*}
	\centering
	\subfloat[]{\includegraphics[width=3.0in]{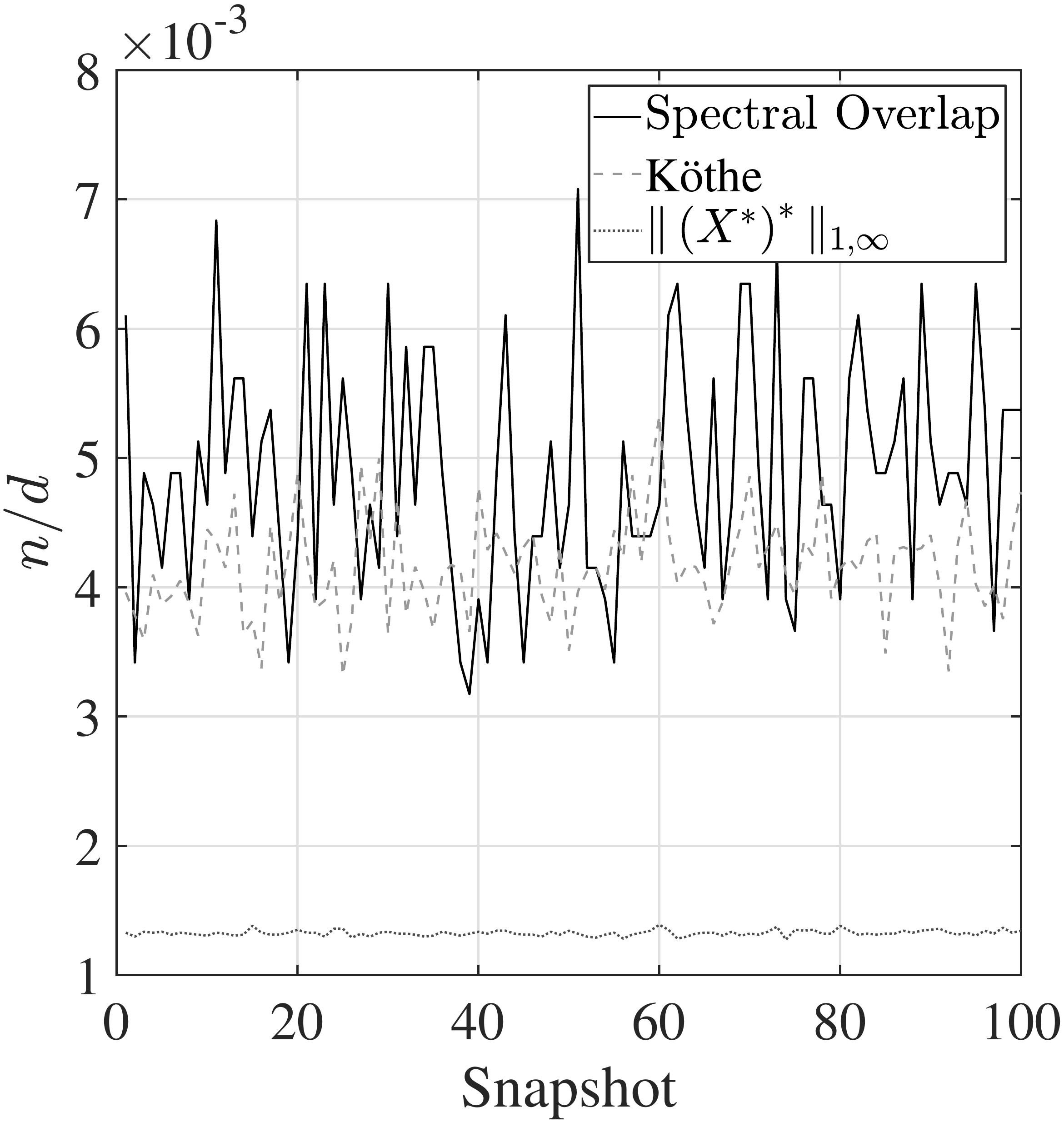}\label{fig:nmr:fds:1101_a}}
	\hspace{4mm}
	\subfloat[]{\includegraphics[width=3.05in]{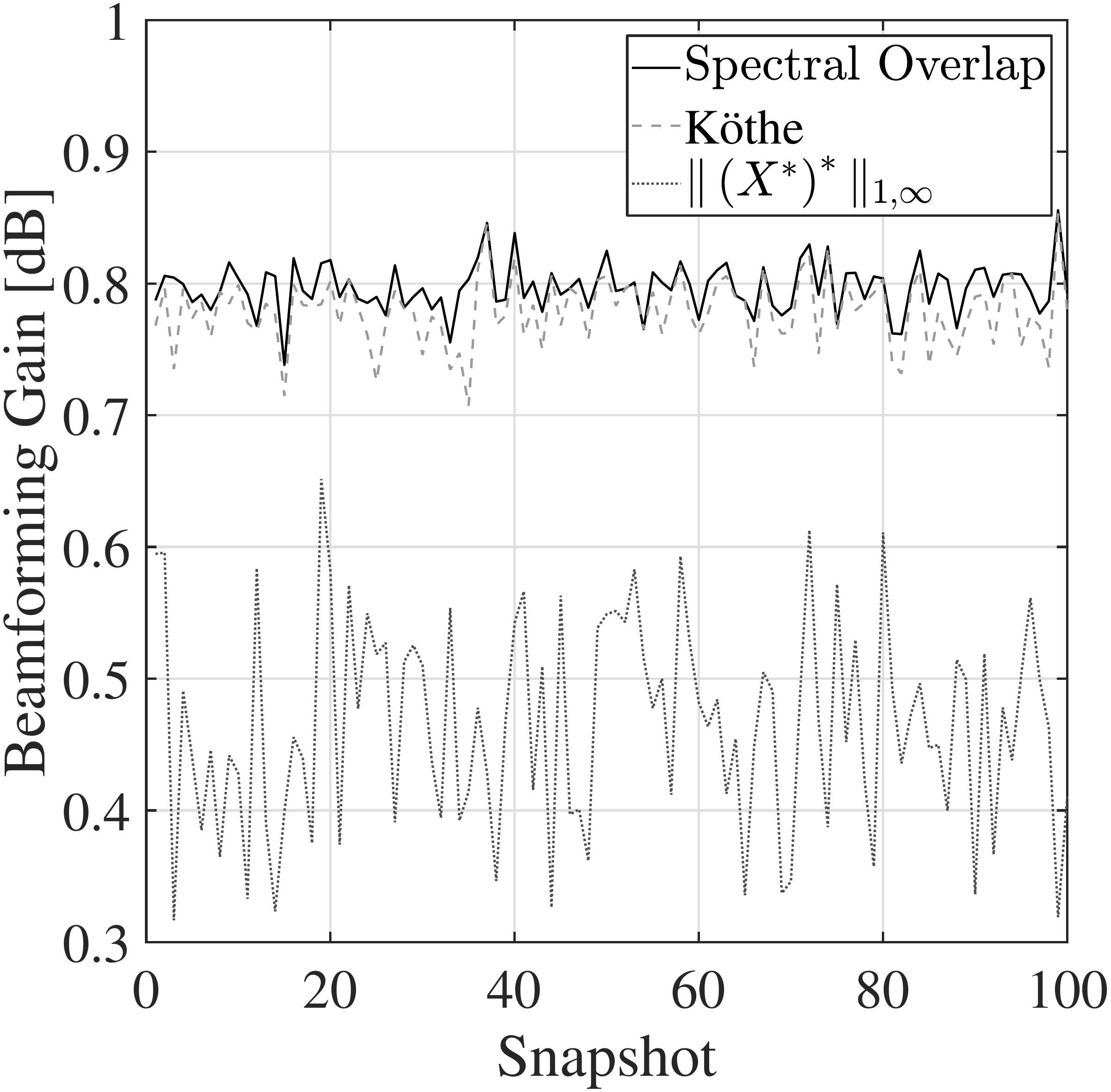}\label{fig:nmr:fds:1101_a}}
	\caption{Optimum dimension $n$ measured for the millimeter-wave directional channel with beam-squint, wideband $3$GHz bandwidth is divided to $32$ subchannels and $100$ trials, (a) optimum dimension ratio, (b) beamforming gain in [dB].}
	\label{fig:nmr:fds:1101}
\end{figure*}

\textbf{\textit{Wide-Band Channel (With Beam Squint Between Subchannels) -}} Under beam-squint impact, the support set corresponding to beam squint has been widened by a factor of $2$ on average. Since the underlying topology has been defined for SMV, the $n_{I}$ and $n_{K}$ are the same as the narrow-band channel discussed above. In other words, $n$ get approaches to $n_{K}$, and the beamforming gain predicted by the Fr\'echet distance metric with respect to K\"othe sequence uniformly approaches the perfect spectral overlap method\footnote{By uniform approach, we refer to sequentially convergence of the topologies predicted by distance metric \eqref{eq:frd:1104} to Theorem \ref{thm:mth:cs_kle2:kle_rnd:1100} as a function of dimension.} Fig. \ref{fig:nmr:fds:1101} shows the optimum dimension ratio and beamforming gain for the wide-band channel with beam squint effect. Obviously, the K\"othe sequence provides a more accurate topology than the $\left(\|\left(X^{\ast}_{i}\right)^{\ast}\|_{1,\infty},\preccurlyeq\right)$. 

\textbf{\textit{Conclusion -}} The distance metric \eqref{eq:frd:1104} induces a topology that almost perfectly matches the topology $\|\left(X^{\ast}\right)^{\ast}\|_{1,\infty}$ proposed in Theorem \ref{thm:mth:cs_kle2:kle_rnd:1100}. We notice that the transition point predicted by the distance metric matches the transition point predicted by the spectral overlap method. Remember that the spectral overlap method measures the spectral overlap between the actual and estimated signals. Consequently, the achievable beamforming gain almost perfectly matches the beamforming gain predicted by the spectral overlap method at the neighborhood of the transition point. In this way, numerically, we show that \textit{the topology induced by the Fr\'echet distance metric in \eqref{eq:frd:1104} matches perfectly the topology induced by the CS-KLE relation in \ref{thm:mth:cs_kle2:kle_rnd:1100}.}
\subsection{Semi-Infinite Dimensional Signal}
\begin{figure*}
\centering
\subfloat[]{\includegraphics[width=2.0in]{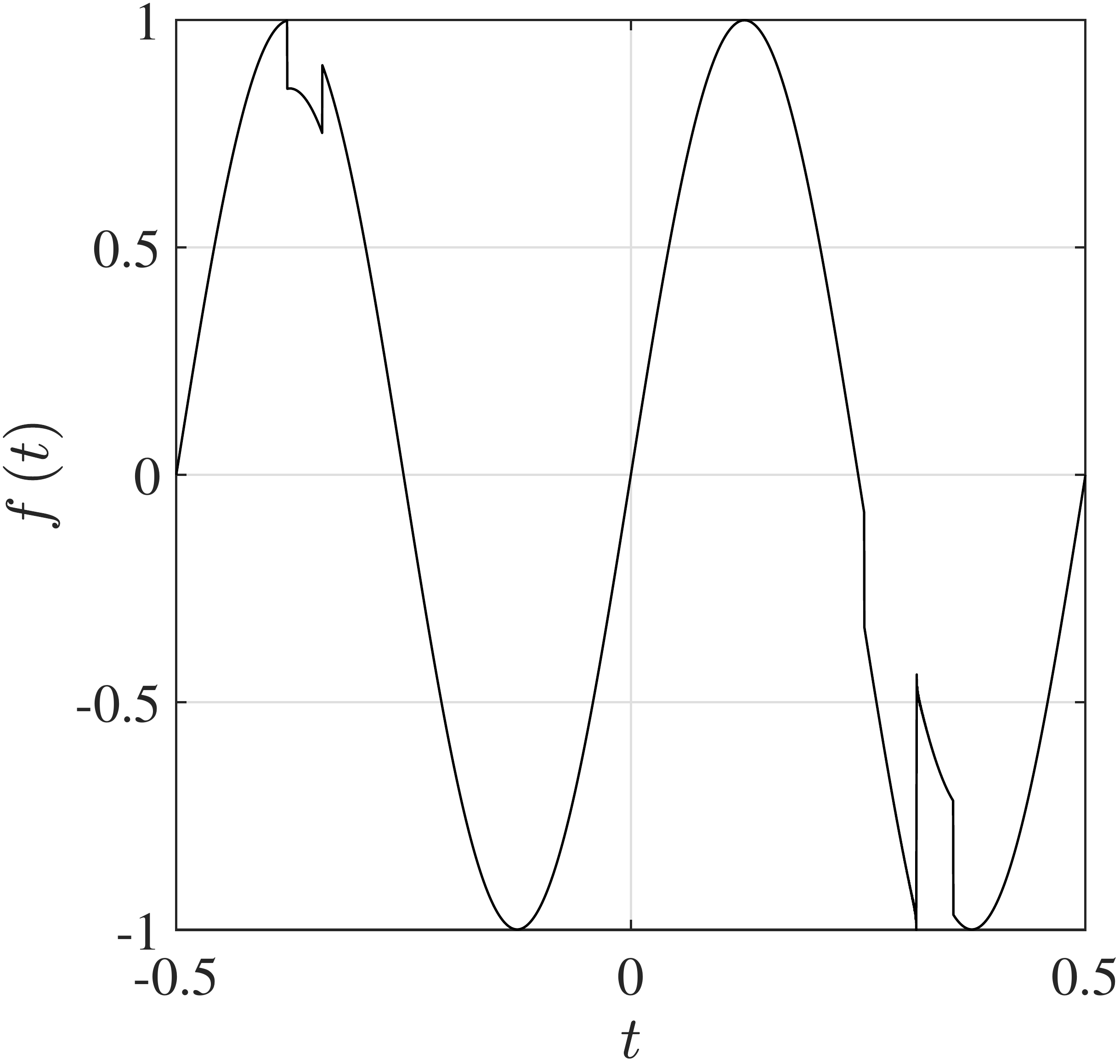}\label{fig:semiinf:1100_a}}
\hspace{4mm}
\subfloat[]{\includegraphics[width=2.0in]{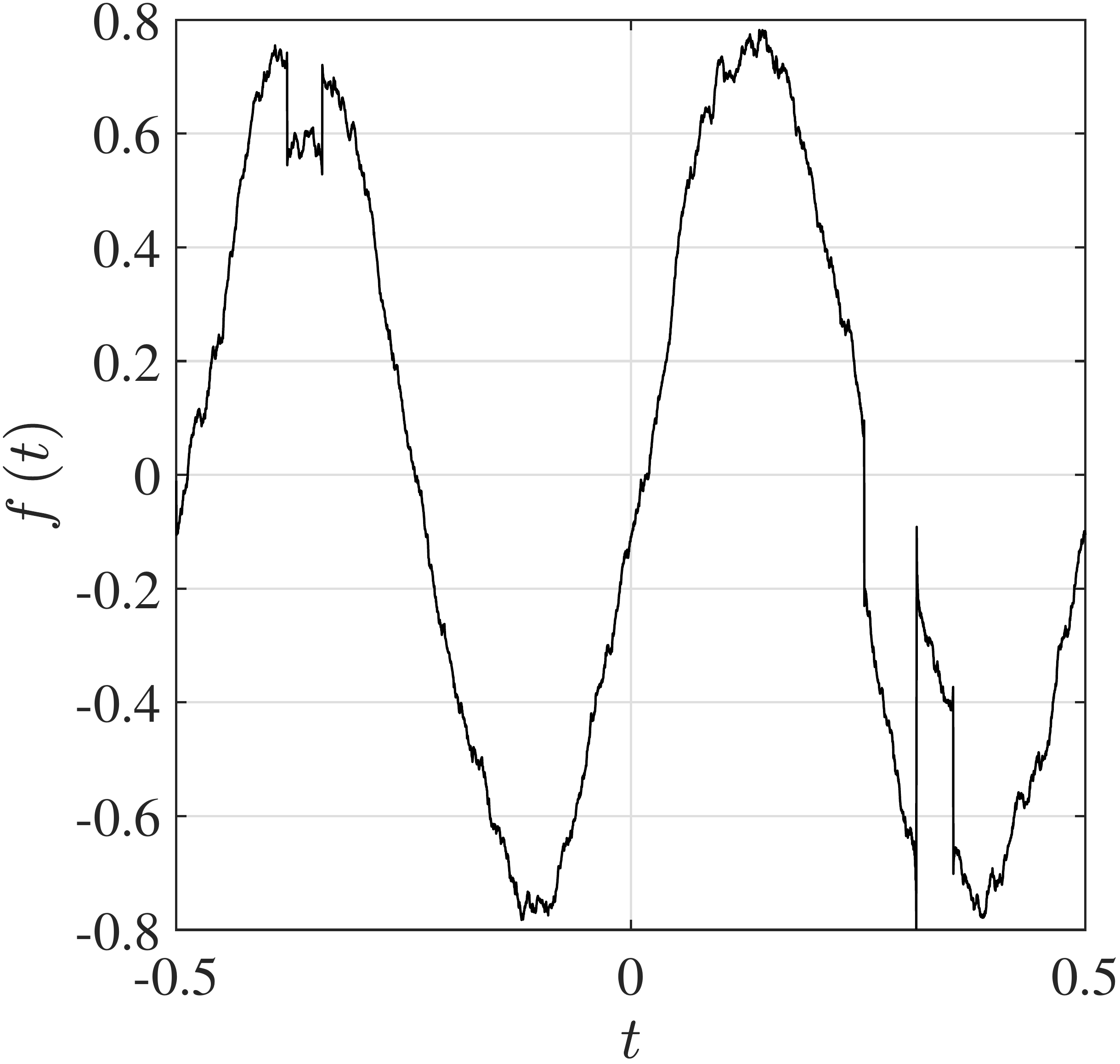}\label{fig:semiinf:1100_b}}
\hspace{4mm}
\subfloat[]{\includegraphics[width=2.0in]{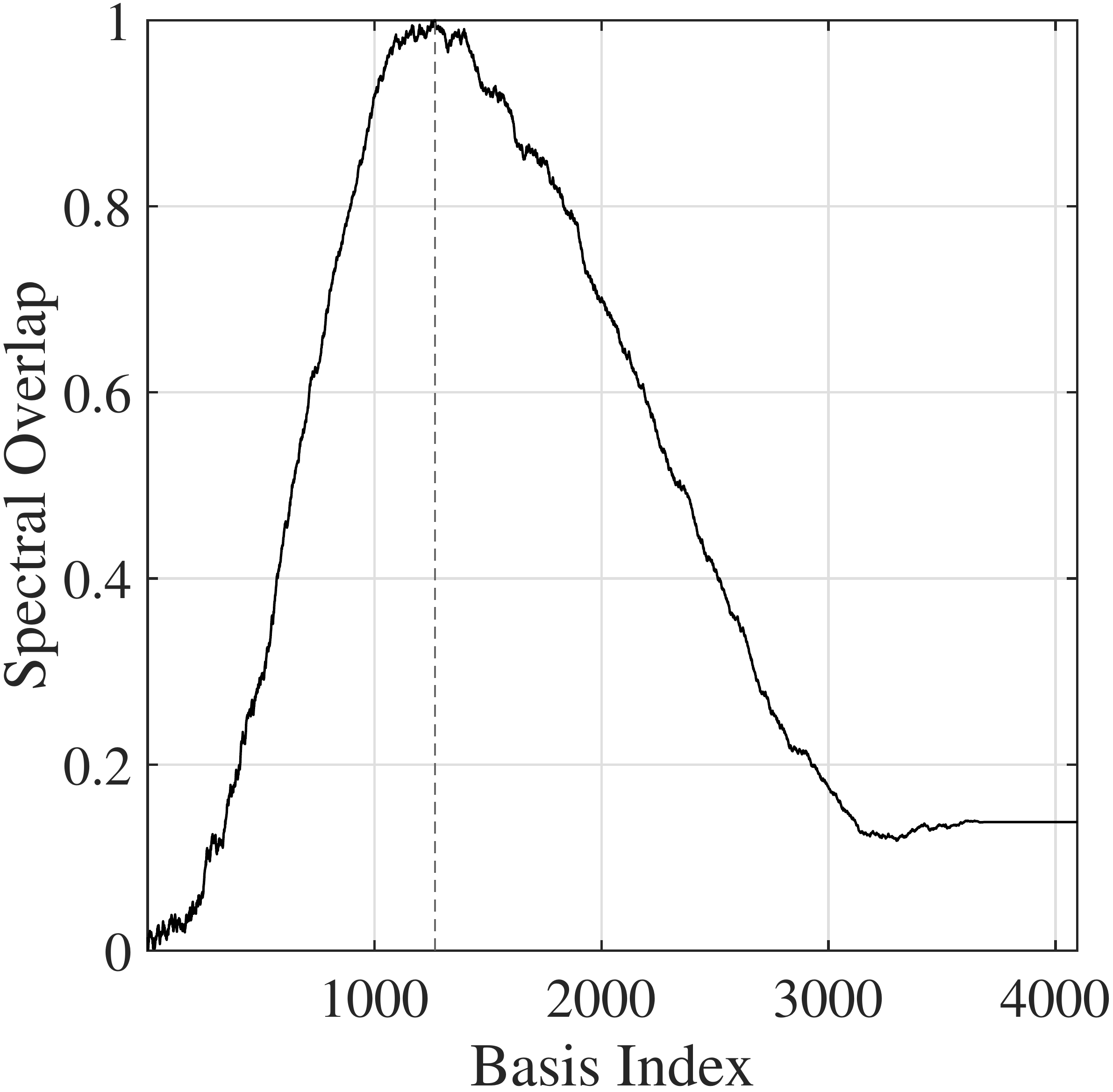}\label{fig:semiinf:1100_c}}
\caption{Semi-infinite dimensional signal generated using $512$ iterations, (a) original signal, (b) CS-KLE reconstruction, (c) spectral overlap with $n = 1276$.}
\label{fig:semiinf:1100}
\end{figure*}

In this section, we examine the spectral continuum of the semi-infinite dimensional signal in Fig \ref{fig:semiinf:1100_a}. We observed that the spectral of the $f\left(t\right)$ spreads up to $n = 1276$ where it normalized spectral overlap reaches the maximum of $1$. First, $f\left(t\right)$ is estimated using CS-KLE relation in Theorem \ref{thm:mth:cs_kle2:kle_rnd:1100}. Then, for derived $\left(\|X^{\ast}_{i}\|_{\infty},\preccurlyeq\right)$ in LHS, $c_{i}$ is derived as K\"othe sequence space. Accordingly, the Fr\'echet distance metric has been computed. This problem is a class of compressible topological vector spaces where the compressibility factor $\rho$ moves toward $d$, where $d$ can go to infinity. However, unlike the infinite-dimensional signal, semi-infinite dimensional signal contains finitely many high frequency components, leading to Gibb's effect. Applying Fr\'echet metric to $f\left(t\right)$, we computed $n_{K} = 1100$, leading to optimum subspace dimension precision of $\%13.9$.

The result above indicated that provided proper $c_{i}$ sequence space, the Fr\'echet distance metric induces the same topology as the $\left(\|X^{\ast}_{i}\|_{1,\infty},\preccurlyeq\right)$ in the RHS of Theorem \ref{thm:mth:cs_kle2:kle_rnd:1100}.
\subsection{Infinite Dimensional Signal}
\begin{figure*}[t!]
\centering
\subfloat[]{\includegraphics[width=2.0in]{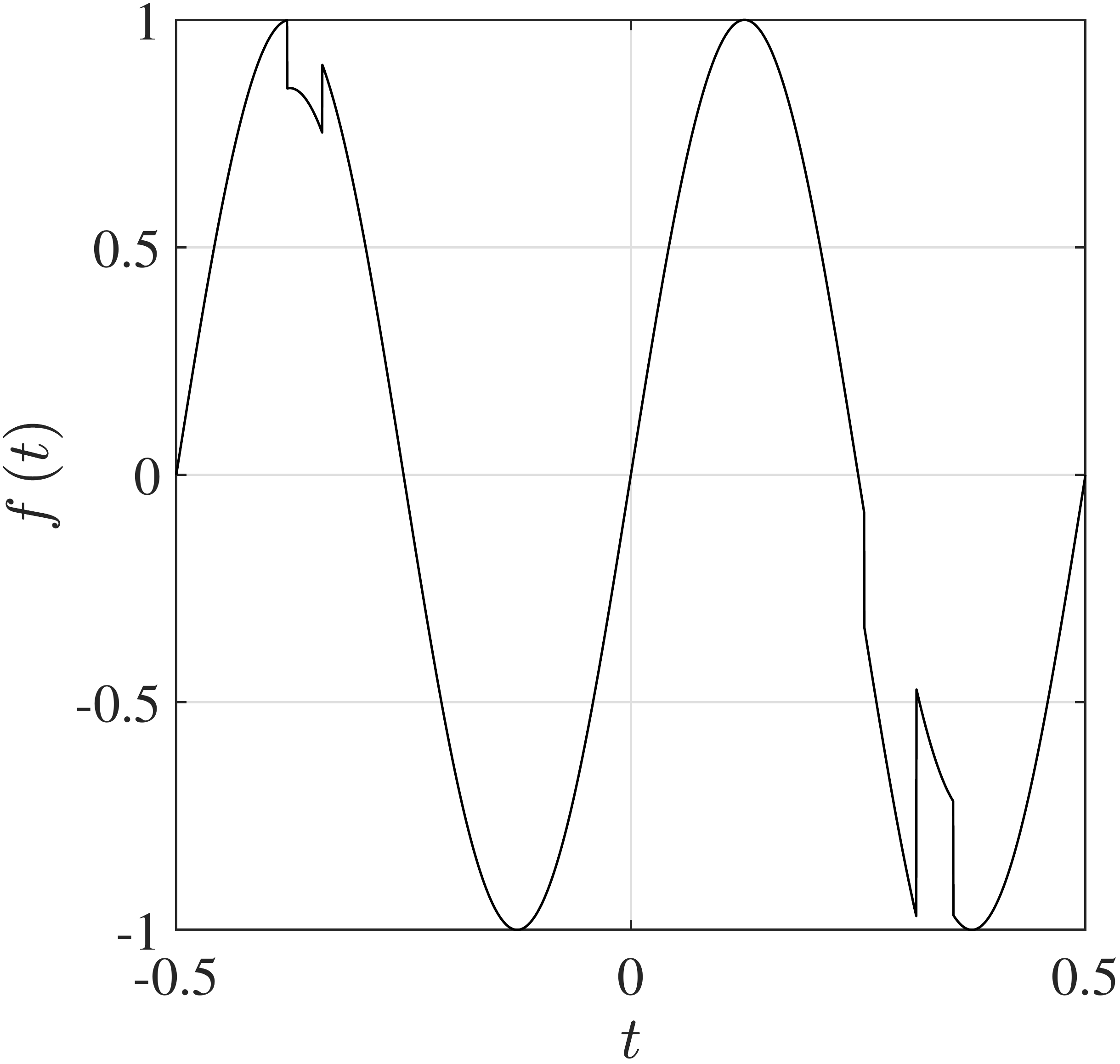}\label{fig:infi:1100_a}}
\hspace{4mm}
\subfloat[]{\includegraphics[width=2.0in]{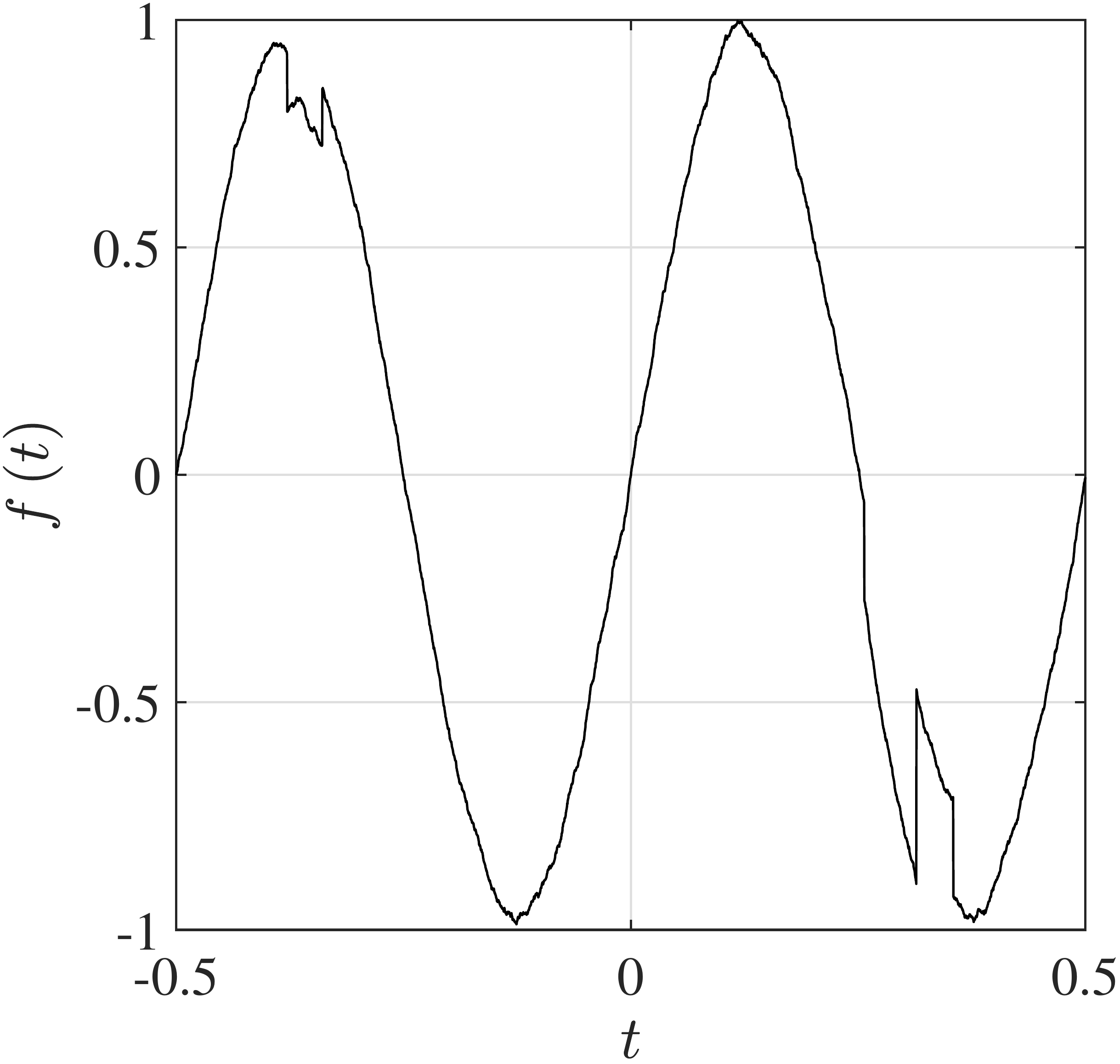}\label{fig:infi:1100_b}}
\hspace{4mm}
\subfloat[]{\includegraphics[width=2.0in]{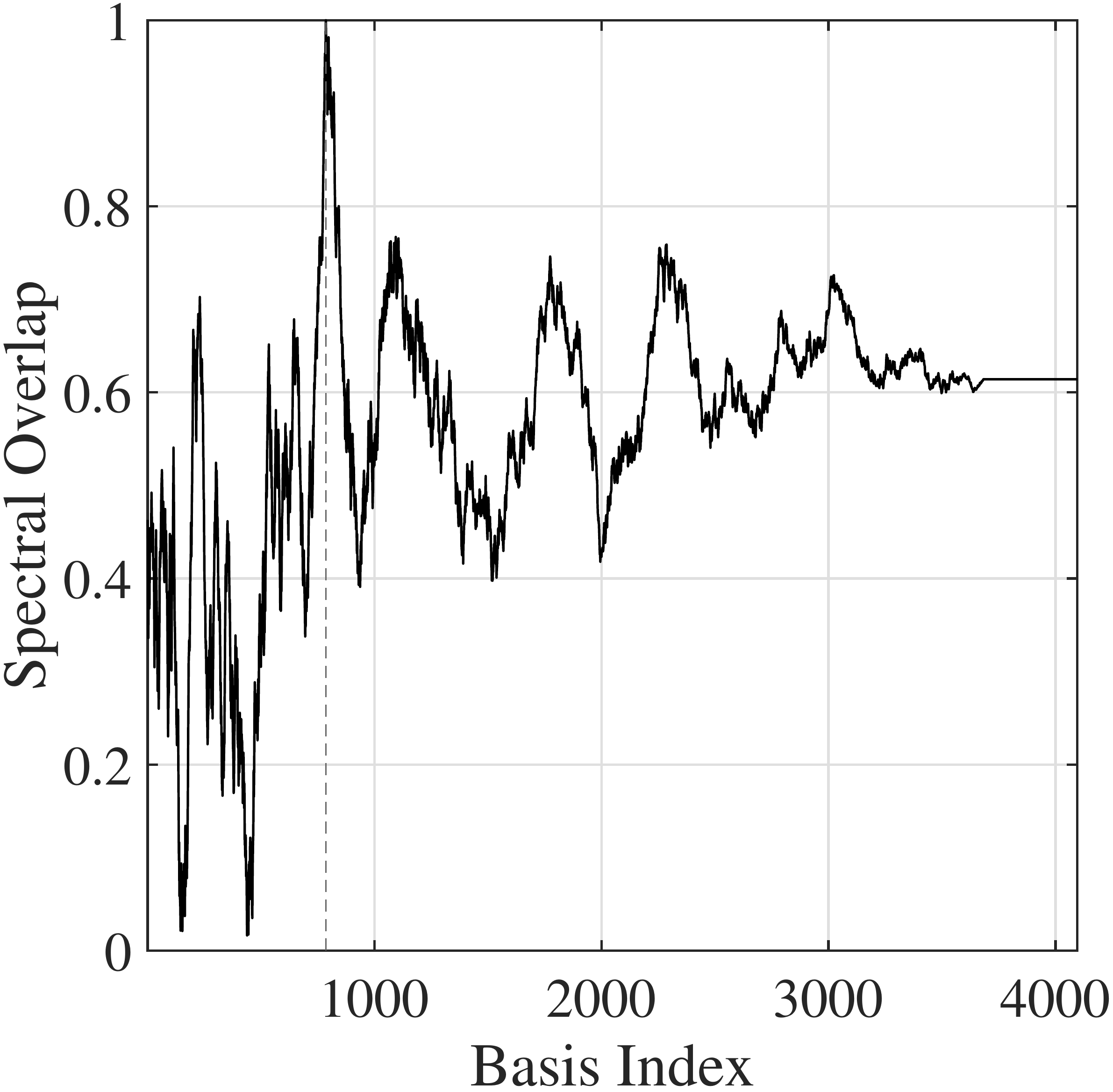}\label{fig:infi:1100_c}}
\caption{Infinite dimensional counterpart of the signal in Fig. \ref{fig:semiinf:1100}, (a) original signal, (b) CS-KLE reconstruction, (c) spectral overlap with $n=786$.}
\label{fig:infi:1100}
\end{figure*}

This section applies Fr\'echet distance metric to infinite-dimensional signal to predict the pseudospectrum set that provides the continuous spectrum of an infinite-dimensional signal. We use the same compressible vector space examined in \cite{robaeiP7CSKLE}. Fig. \ref{fig:infi:1100_a} shows infinite dimensional signal. Obviously, the signal of interest contains the step-wise change in the signal requiring infinite iteration of Fourier transform. We have measured $n=786$ using spectral overlap with perfect knowledge about the underlying distribution. Fig. \ref{fig:infi:1100_b} shows the reconstructed signal. Defining $c_{i}$ K\"othe sequence for the RHS of the Theorem \ref{thm:mth:cs_kle2:kle_rnd:1100}, and applying the Fr\'echet distance metric, the optimum subspace $n_{K}=795$ and spectral overlap of $0.92$ are obtained. The dimension estimation precision is about $\frac{795-786}{786}\times 100 = \%1.1$. Fig. \ref{fig:infi:1100_c} demonstrates the spectral overlap generated for the signal reconstructed in the Fr\'echet space.

\begin{figure*}[t!]
\centering
\subfloat[]{\includegraphics[width=2.0in]{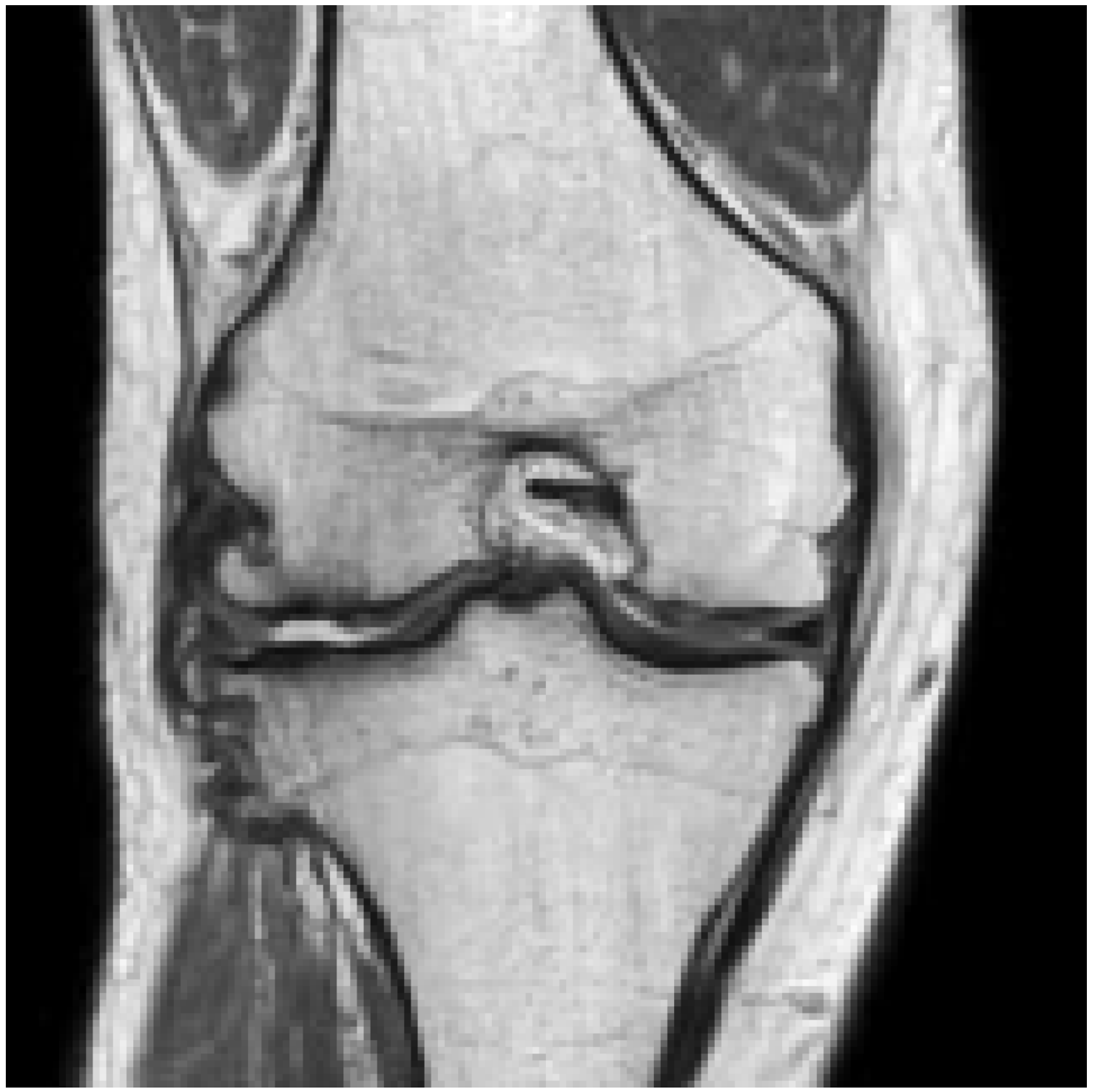}\label{fig:infi:1102_a}}
\hspace{4mm}
\subfloat[]{\includegraphics[width=2.0in]{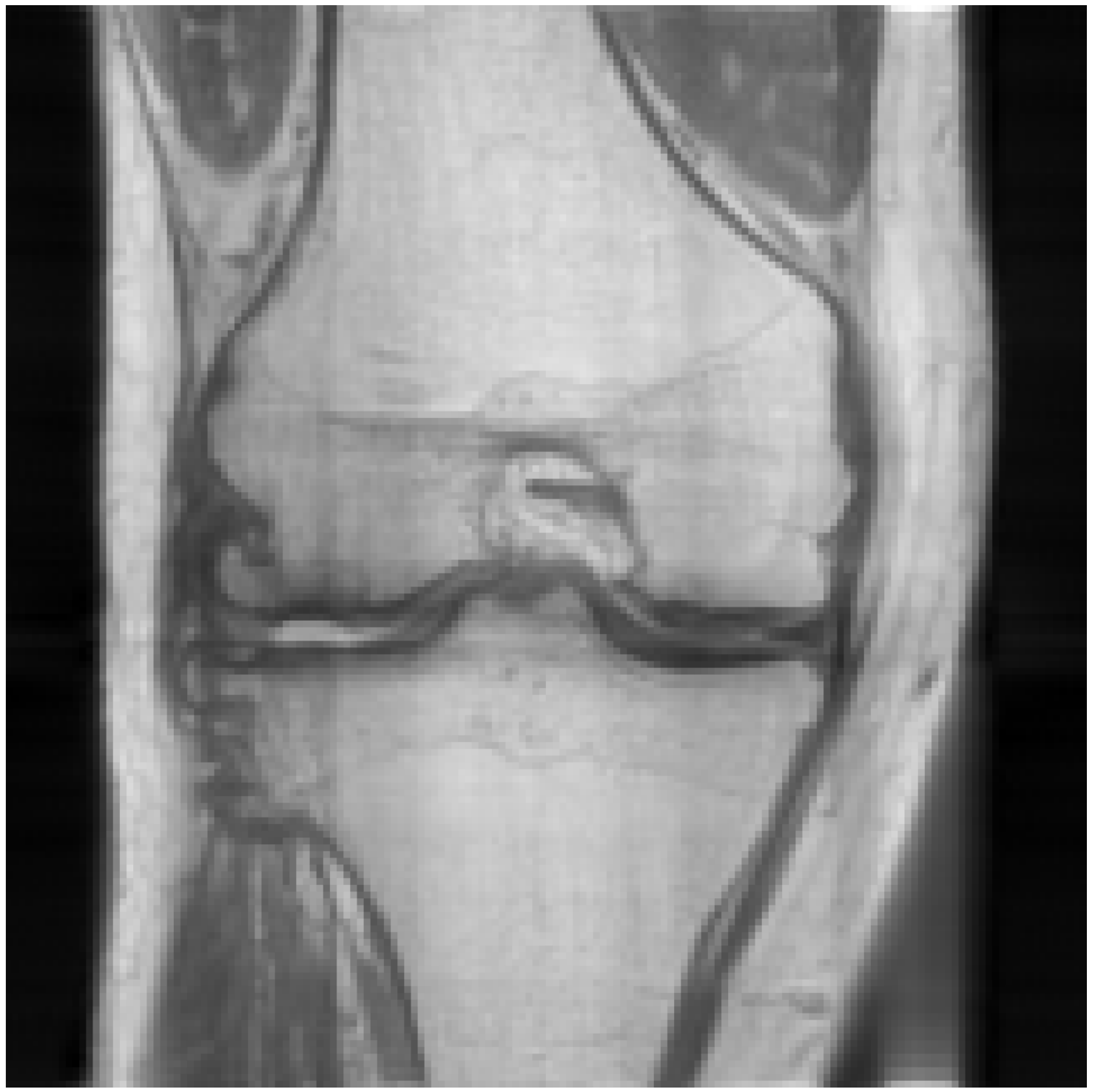}\label{fig:infi:1102_b}}
\hspace{4mm}
\subfloat[]{\includegraphics[width=2.0in]{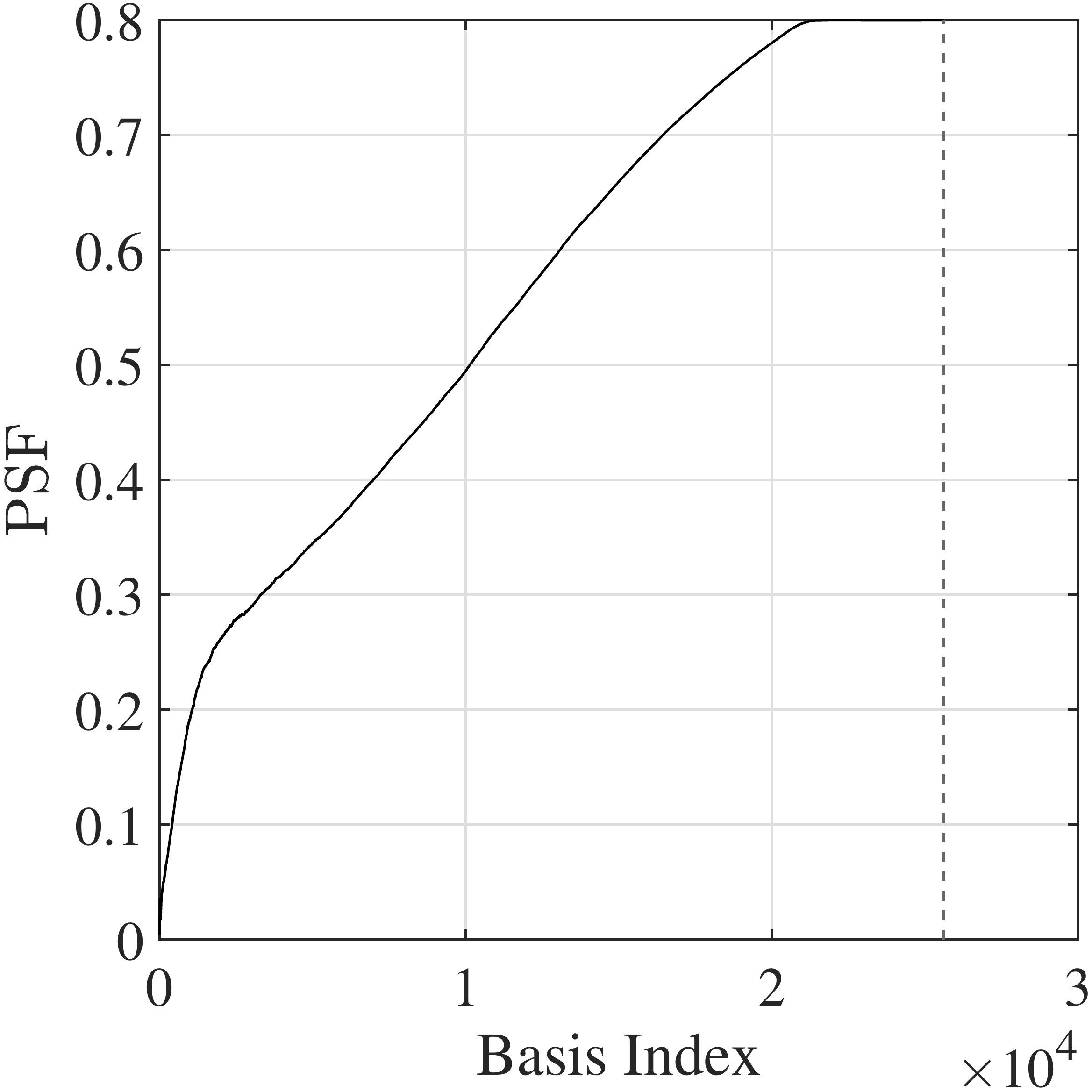}\label{fig:infi:1102_c}}
\caption{MRI reconstruction, (a) original MRI image, (b) CS-KLE reconstruction, (c) PSF with $n=25594$.}
\label{fig:infi:1102}
\end{figure*}

Finally, we compute $c_{i}$ and apply the Fr\'echet distance metric to the MRI image in Fig. \ref{fig:infi:1102_a}, with CS-KLE reconstruction in Fig. \ref{fig:infi:1102_b}. We observe that for the image, the optimum subspace may be latent significantly due to high frequency components. Using spectral overlap measured in the form pf Point Spread Function (PSF) in Fig. \ref{fig:infi:1102_c}, we have obtained $n=25594$ with almost perfect spectral overlap of $0.9998$. For topological space induced by the Fr\'echet distance metric, we obtained $0.8$ spectral overlap in the neighborhood of $n_{K}= 21430$. Obviously, the dimension estimation precision is about $16.3\%$. Since optimum subspace dimension obtained through the Fr\'echet distance metric is smaller than the estimation via spectral overlap, $n_{K}$ should be considered as the lower bound for the image in this case.

We conclude that the topology induced by the Fr\'echet distance metric is compatible with the topology induced by RHS of Theorem \ref{thm:mth:cs_kle2:kle_rnd:1100}.
\section{Discussion and Conclusion}\label{dis:disc}
Studying the compressibility of the random process by introducing topology allows for studying the uniform convergence and asymptotic behavior of the macrostates that a system can occupy. This approach provides an extremely powerful tool to evaluate the internal structure of compressible topological vector space, such as spectral distribution, diagonalization, continuous spectral extension, the power decay rate, and so many other properties (equivalent topologies induced by sequence spaces). 

In particular, diagonalization of compressible topological vector space with respect to random sampling operators was our focus in this work. We have applied conventional theories in functional and stochastical analysis to predict the optimum subspace of continuous compressed sensing problems without priori knowledge about the underlying distribution. We have defined the required definitions and theories to formulate the behavior of the compressible random processes in locally convex space. For this purpose, the fundamental Theorem \ref{thm:mth:ta:h:1099} as the generalized reflexive homeomorphism has been proved, for which we provided two examples. We observed that for a random process generated by a random process path, the Daniell-Kolmogorov extension theorem in stochastical analysis agrees with the Hahn-Banach extension theorem and its generalized counterpart, the Riesz extension theorem, in functional analysis.

The main obstacle to developing a topological analysis of the compressible random processes was the non-trivial null space. According to the null space property of the compressed sensing, the null space of compressible signals should not be too tight. In order to overcome the null space property, we introduced absorbing null space based on the Minkowski functional. Based on the fundamental property of the Lebesgue measure, which assigns zero to the subset with a small measure, we proved the null space can be defined as an absorbing and connected topological structure. 

We have proved that the weak-compact and weak-$^{\ast}$-compact topologies can be obtained for compressible topological vector space $X$, its dual space $X^{\ast}$, and double dual space $\left(X^{\ast}\right)^{\ast}$ with respect to the Hahn-Banach and Banach-Agao\u{g}lu theorems. We have noticed that semi-infinite and infinite-dimensional signals can be modeled by continuous spectrum approximation represented by $n$-pseudospectrum. We proved that the $n$-pseudospectrum had to satisfy the closed convex hull requirement and contain its extreme points in order for $X$ to converge to its limit with respect to the minimal linear function obtained through the Riesz extension theorem. 

Given $n$-pseudospectrum the optimum subspace $X_{n}$ is the optimum restricted representation of the compressible topological vector space $X$. Finally, by analyzing compressible random process in Fr\'echet space, the equivalent topological of $X$ obtained as K\"othe sequence. Considering the equivalent topology in the given sequence space, the Fr\'echet distance had been obtained over which the optimum dimension of compressible topological vector space was defined.

In order to evaluate the proposed topological analysis in locally convex space, we examined three categories of signals, (1) finite-dimensional millimeter-wave directional channel, (2) semi-infinite dimensional signal with Gibb's effect, and (3) vector spaces with local infinite-dimensional fluctuations. We have shown that approximated $n$-pseudospectrum space perfectly matches the theoretical spectral overlap method with perfect knowledge about the underlying distribution.

The extension of the proposed topological analysis of the compressibility to multiple measurement vectors is an open problem.
Similar to conventional convexity, the intersection preserves the $\sigma$-algebra. That is, the intersection of an arbitrary number of $\sigma$-algebra is also $\sigma$-algebra. If generated $\sigma$-algebras also contain the bases of the topological vector space, then their intersection still contains the bases for the resultant locally convex space. The idea of the intersection of $\sigma$-algebra is helpful when we construct a new locally convex space from given multiple measurement vectors. The resultant $\sigma$-algebra contains the smallest possible $\sigma$-algebra for the new locally convex space generated from multiple measurement vectors. The resultant $\sigma$-algebra is required, but it is insufficient and only determines the lower bound for the $\sigma$-algebra. While the upper bound is for $\sigma$-algebra given by the Minkowski summation of the convex bodies, the optimum $\sigma$-algebra sets of multiple measurement vector is not determined. We summarize the problem in the following question.

\begin{quote}
What is the optimum $\sigma$-algebra for the locally convex space of multiple measurement vectors?
\end{quote}

The problem can be addressed using Zorn's Lemma, where we can find the minimal linear functional for the new locally convex space. However, it looks like that similarly to the single measurement vector, multiple measurement vector can be categorized as an absorbing problem. We believe that absorbing methods are the natural approach to developing growing subspaces for the compressible random processes without priori knowledge of underlying distributions.
%
\bibliographystyle{IEEEtran}
\bibliography{references1.bib,references2.bib}

\end{document}